\documentclass[prb,aps,showpacs,groupedaddress,superscriptaddress,twocolumn,toc=flat]{revtex4-2}

\usepackage[colorlinks=true,linkcolor=blue,urlcolor=blue,citecolor=blue]{hyperref}

\usepackage[utf8x]{inputenc}
\usepackage{color}
\usepackage{bbm} 

\usepackage{amsfonts,amsmath,amssymb,stmaryrd}

\usepackage{graphicx}
\usepackage{subfigure}  
\usepackage{bbm} 
\usepackage{hyperref}
\usepackage{epsfig}
\usepackage{mathrsfs}
\usepackage{verbatim}
\usepackage{centernot}
\usepackage{ulem}
\usepackage{longtable}
\usepackage{nicefrac}
\usepackage[framemethod=tikz]{mdframed}

\usepackage{array}

\usepackage{cancel,ifthen}
\newcommand{\cmnt}[2][NoInPuT]{\ifthenelse{\equal{#1}{NoInPuT}}{}{{\color{red}\sout{#1}}} {\color{blue} #2}}

\usepackage{bm} 

\LTcapwidth=\textwidth

\begin{document}
\normalem       

\title{Competing instabilities at long length scales in the one-dimensional  Bose-Fermi-Hubbard model at commensurate fillings}

\author{Janik Sch\"onmeier-Kromer}
\affiliation{Department of Physics and Arnold Sommerfeld Center for Theoretical Physics (ASC), Ludwig-Maximilians-Universit\"at M\"unchen, Theresienstr. 37, M\"unchen D-80333, Germany}
\affiliation{Munich Center for Quantum Science and Technology (MCQST), Schellingstr. 4, D-80799 M\"unchen, Germany}

\author{Lode Pollet}
\affiliation{Department of Physics and Arnold Sommerfeld Center for Theoretical Physics (ASC), Ludwig-Maximilians-Universit\"at M\"unchen, Theresienstr. 37, M\"unchen D-80333, Germany}
\affiliation{Munich Center for Quantum Science and Technology (MCQST), Schellingstr. 4, D-80799 M\"unchen, Germany}
\address{Wilczek Quantum Center, School of Physics and Astronomy, Shanghai Jiao Tong University, Shanghai 200240, China}

\begin{abstract}

We study the phase diagram of the one-dimensional  Bose-Fermi-Hubbard model at unit filling for the scalar bosons and half filling for the $S=1/2$ fermions using quantum Monte Carlo simulations. The bare interaction between the fermions is set to zero. A central question of our study is what type of interactions can be induced between the fermions by the bosons, for both weak and strong interspecies coupling. We find that the induced interactions can lead to competing instabilities favoring phase separation, superconducting phases, and density wave structures, in many cases at work on length scales of more than 100 sites. Marginal bosonic superfluids with a density matrix decaying faster than what is allowed for pure bosonic systems with on-site interactions, are also found.
\end{abstract}


\maketitle

\section{Introduction}
Due to their clean and fully controllable yet versatile setup, quantum gases in optical lattices have proven to represent an ideal candidate to realize a quantum simulator for classically incomputable many-body problems in condensed matter theory~\cite{Bloch2008,Lewenstein2007}. For monoatomic bosonic gases trapped in a three-dimensional (3D) optical lattice the theoretically predicted \cite{Jaksch1998,Oosten2001} quantum phase transition from a Mott insulator to a superfluid was experimentally proven to exist \cite{Greiner2002}. 

The interplay between bosons and fermions is ubiquitous in nature. 
In conventional superconductors phonons mediate an attractive interaction between the electrons. 
Mixtures of $^3$He and $^4$He caught attention already in the 1940s and showed such effects as a very low solubility and the Pomeranchuk effect~\cite{Pomeranchuk1949,Zharkov1960,DeBruynOuboter1960}. 
 Every Bose-Fermi system has however its own types of interactions and characteristics, and requires a separate study.
 In the field of ultracold atoms, several degenerate Bose-Fermi mixtures have been cooled to degeneracy over the past 20 years~\cite{Truscott2001, Schreck2001, Hadzibabic2002, Roati2002, Tey2010, Lous2017, Juergensen2012, Roy2017, DeSalvo2017}. Induced interactions between the fermions mediated by the bosons leading to a fermionic pair superfluid attracted theoretical interest shortly after the first condensates were experimentally realized~\cite{Houbiers1997, Heiselberg2000, Viverit2000, Viverit2002}. In more recent years, experimental efforts focused on the interactions between a bosonic and fermionic pair superfluid~\cite{Ferrier2014, Delehaye2015, Fava2018, Wu2018}.
 Experimental loading of a  fermionic $^{40}$K and bosonic $^{87}$Rb mixture in an optical lattice~\cite{Guenter2006} can lead to stronger localization and interaction effects such as phase separation, spin or charge density waves and supersolids ~\cite{Albus2003, Buechler2003, Buechler2004, Mathey2004, MatheyWang2007, Mathey2007, Pollet2006, PolletKollath2008, Suzuki2008, EnssZwerger2009, Titvinidze2008, Mering2008, Mering2010, Orignac2010, Anders2012, Bukov2014, Bilitewski2015}.

In this work we examine the ground state phase diagram of the one-dimensional Bose-Fermi-Hubbard model at unit filling for the scalar bosons and half filling for the $S=1/2$ fermions. The bare fermions are taken to be free particles. As the system size and inverse temperature increase, the induced interactions grow stronger. Nevertheless,  those induced interactions leading to pair flow~\cite{Kuklov2003} are expected to be weak  for weak inter-species coupling~\cite{Batrouni2021}. In this Bardeen-Cooper-Schrieffer regime the pair size is very large, and therefore very large system sizes are required.
 For stronger inter-species coupling, localization effects are stronger and a competition between instabilities towards  superfluids and density wave structures is expected. It turns out that the renormalization flow can still extend to 100s of lattice sites, with misleading behavior on intermediate length scales. 
The purpose of this work is hence to map out the physics of induced interactions and competing instabilities for the one-dimensional Bose-Fermi-Hubbard model. It is one of the few sign-positive models for mixtures where unbiased numerical simulations can be carried out.

\section{Model and Phase Diagram}
\label{sec:model}
%
\begin{figure}[!htb]
        \begin{tabular}{ll}
        \includegraphics[width=0.5 \columnwidth]{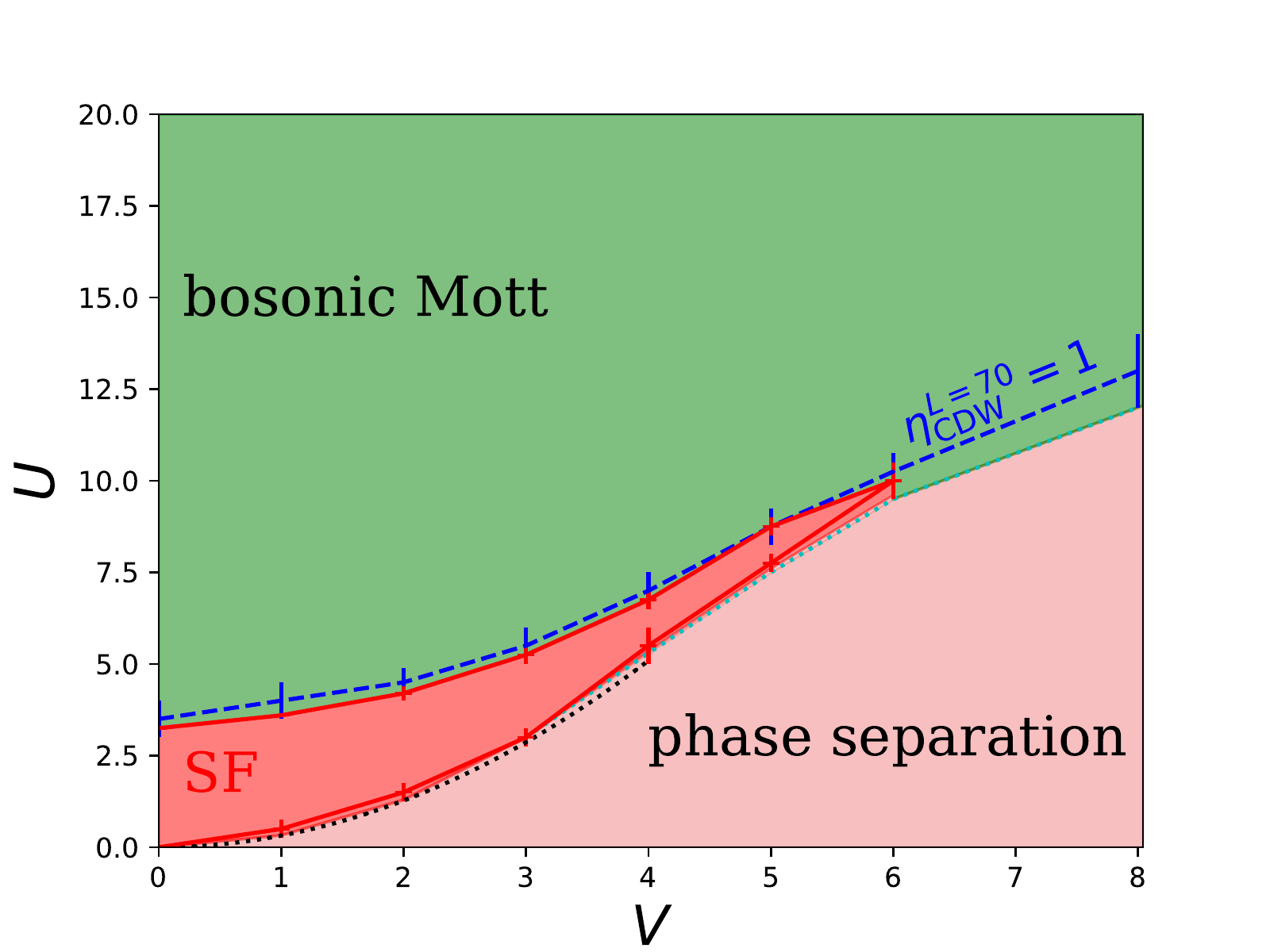} &
        \includegraphics[width=0.5 \columnwidth]{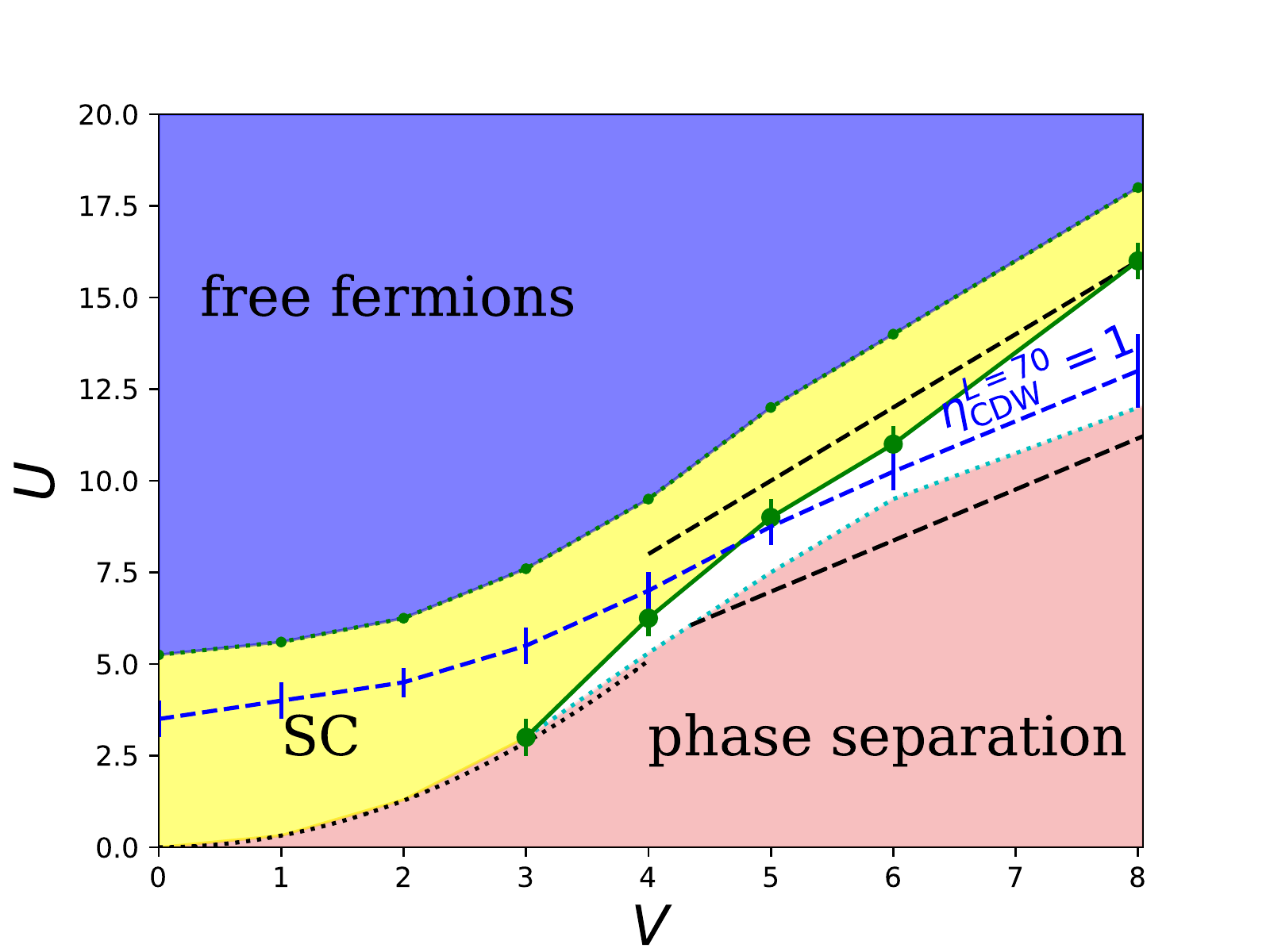} 
        \end{tabular}
        \caption{  \label{fig:phasediagram} Phase diagram of the model Eq.~\ref{eq:Hamiltonian} at unit bosonic and fermionic half filling with all hopping amplitudes set to 1, as obtained from quantum Monte Carlo simulations for $V=1,2,3,4,5,6$ and $8$. Lines are a guide to the eye. Left (bosons): The red area demarcates the area of bosonic superfluidity; the green area is the uniform bosonic Mott inslulator, and phase separation is shown in light red. For comparison with the right panel, we show the blue dashed line which corresponds to $\eta_{\rm CDW} = 1.0$ ({\it i.e.}, the exponent controlling the (fermionic) charge density wave (CDW) correlation function) on a length scale $L=70$, with lower values than 1 above the line. Right (fermions): Dominant superconducting fluctuations (SC) are found in the yellow area.  The dotted green line is the crossover on our length scales towards quasi free fermionic behavior (purple area).  In the white area Luttinger liquid behavior is seen with dominant charge correlations above the blue dashed line. The dotted and lower dashed black lines are estimates for phase separation; the upper dashed line is a leading order strong coupling argument separating uniform systems from ones with charge density waves.  Those lines are only informative, but are approached asymptotically for larger values of $V$. The cyan dotted line is where the numerically observed boundary of phase separation. Further explanations are given in a separate paragraph in Sec.~\ref{sec:model} in the text. 
  }
\end{figure}

%
We study  the one-dimensional Bose-Fermi-Hubbard model with on-site interactions,
%

\begin{eqnarray}
H & = & -  t_{\rm F} \sum_{\langle i,j \rangle, \sigma = \uparrow, \downarrow} \left( c_{i,\sigma}^{\dagger} c_{j, \sigma} + {\rm h.c.}\right) \nonumber \\
{} & {} & - t_{\rm B} \sum_{\langle i,j \rangle } \left( b_{i}^{\dagger} b_{j} + {\rm h.c.}\right) + \frac{U}{2}  \sum_i n^{\rm B}_i(n^{\rm B}_i - 1 ) \nonumber \\
{} & {} & + V  \sum_i n^{\rm B}_i (n^{\rm F}_{i, \uparrow} + n^{\rm F}_{i, \downarrow} ),
\label{eq:Hamiltonian}
\end{eqnarray}
where $b^{\dagger}_{i}$ creates a soft-core boson on site $i$ and $c^{\dagger}_{i,\sigma}$ a hard-core boson on site $i$ with spin $\sigma = \uparrow, \downarrow$. Particles can hop between nearest neighbors  $\langle i,j \rangle$; the hopping amplitudes are  $t_{\rm F}$ for the hard-core bosons and $t_{\rm B}$ for the soft-core bosons. Those amplitudes are chosen equal, and set as the energy unit,  $t_{\rm B} = t_{\rm F} = t = 1$. We work at unit filling for the soft-core bosons and half filling for the hard-core bosons.  The on-site density-density interactions are constant over the lattice. Its amplitude for the intra-species soft-core bosonic interaction is $U >0$, and the amplitude for the inter-species interactions is $V$, whose sign is irrelevant at half filling. There is no bare intra-species interaction between the hard-core bosons. The lattice spacing is set to one, $a=1$, and the length of the chain is $L$. The Jordan-Wigner transformation between hard-core bosons and fermions requires an odd number of fermions if we use periodic boundary  conditions; we will use the language of bosons and fermions in this sense below. The model is simulated using a straight-forward extension of the worm algorithm~\cite{Prokofev1998qmc} presented in Ref.~\cite{Pollet2007,sadoune_efficient_2022}.

The parameters are chosen such that there is no apparent small parameter, {\it i.e.}, in a regime where numerics are necessary. 
Contemporary Density Matrix Renormalization Group (DMRG) studies put a strong cutoff on the bosonic occupation number in order to keep the local Hilbert space tractable and focused on superfluid-insulator transitions~\cite{Avella2019, Avella2020, GuerreroSuarez2021}.
Our main result is summarized in the phase diagram shown in Fig.~\ref{fig:phasediagram}, valid in the thermodynamic limit. However, monitoring the competing instabilities that develop on length scales consisting of hundreds of sites can not be read off in full from this phase diagram. This competition is explained in the detailed analysis in the sections below. As we will see, the simulations are notoriously hard, and, in certain cases, autocorrelation times that exceed one million Monte Carlo steps are observed, indicative of strong metastabilities and competing phases, perhaps in the vicinity of first order transitions. This  inevitably leads to a few uncertainties in the phase diagram, which can only be resolved if better algorithms are devised and better computer hardware is available. \\

To set the ideas, we briefly explain the phase diagram shown in  Fig.~\ref{fig:phasediagram}.
The phase diagram is dominated by two big areas: for $V \gg U$ the system separates into a bosonic and a fermionic system, and for $U \gg V$ the bosons form a uniform $n=1$ Mott insulator above which the fermions are quasi free. We focus on the channel-like region in between. 
 The red solid line demarcates the area of uniform bosonic superfluidity. It extends to remarkably large values of $U$ and $V$ 
and probably closes in a cuspy way for a value of $V$ close to $V=6$. Mesoscopic superflow is found for $V=8, U=14$ as well but, extrapolating our results, it vanishes around $L \approx 500$. In the tip of this region,  the bosonic superfluid is marginal with a density matrix decaying faster than what is allowed for the pure bosonic system. We found no indications of bosonic supersolid behavior ({\it i.e.}, concomitant bosonic superfluidity and density waves breaking the lattice symmetry). 
The dotted black line is the weak coupling argument for phase separation (see Sec.~\ref{sec:weak_coupling}), which is found to exist everywhere in the lower parts of the phase diagram. The dashed black lines are the leading order strong coupling predictions (see Sec.~\ref{sec:strong_coupling})  separating phase separation, a structure with density wave character, and a uniform system. We suspect that phase separation extends up to the black and cyan dotted lines.
In the fermionic sector, the fermions are insulating below the full green line and show pair flow above and to the left of it, {\it at least up to the system sizes that we can simulate}. Above (below) the dashed blue line the decay of the charge density wave (CDW) correlations is slower (faster) than for free fermions. Whether the charge density waves can spontaneously break the lattice symmetry is impossible to say based on our system sizes for $V \le 6$, but for $V=8$ we see some strong indications of that for $U=13$. The green dotted line indicates a crossover scale where, {\it on our length scales}, the up and down particles are so weakly coupled on top of a uniform bosonic Mott insulator that they can be considered quasi-free. For exponentially low temperatures, pair flow is expected everywhere in the upper part of the phase diagram.

The remainder of the paper is structured as follows.  In Sec.~\ref{sec:analytics} we present analytical arguments in the limiting cases of the phase diagram such as weak and strong inter-species coupling, arguments based on bosonization, and the type of induced interactions in the random phase approximation for various parameter regimes. This is followed by Sec.~\ref{sec:worm} where we highlight some specifics of our quantum Monte Carlo algorithm, and we list in Sec.~\ref{sec:order_parameters} all relevant quantities that are computed in the simulations and used for analyzing the phases. 
In Sec.~\ref{sec:scans} we systematically go through the phase diagram for various values of $V$ and discuss the obtained Monte Carlo results. In particular, renormalization flows as a function of the system size allow one to monitor the competing instabilities, and infer the structure of the phase diagram.  Note that in this paper we refer to a renormalization flow always as a flow in the system size, unlike in the setting of the renormalization group. We conclude in Sec.~\ref{sec:conclusion}.

\section{Simple Analytical considerations in limiting cases} \label{sec:analytics}

\subsection{Weak interspecies coupling, $V \le 2t$ }\label{sec:weak_coupling}

For weak values of $U$ the bosons are well described by a Luttinger liquid with a linear spectrum. To set the ideas, we can for $U \le 2.5$ use the Bogoliubov approximation, which is very accurate  for this range of $U$-values~\cite{Ejima2012}. 
Integrating out the Bogoliubov quasiparticles results in  a total action consisting of the one for the bare fermions (including a shift to $\mu_F$ by $n_0 V$ which we do not mention because we stay at fermionic half filling) and an induced part $\exp(-S)$ with $S$ given by
\begin{equation}
S = \frac{n_0 V^2}{2} \int_0^{\beta} d\tau_1  d\tau_2 \sum_{i,j,\sigma,\sigma'} n_{i}^{\sigma}(\tau_1) D^0(i-j, \tau_1 - \tau_2) n_{j}^{\sigma'}(\tau_2).
\end{equation}
Here, $n_0$ is the quasi-condensate and the kernel $D^0(x,\tau)$ is given by
\begin{equation}
D^0(x,\tau) = \int \frac{dk}{2\pi} e^{ikx} \frac{e^{E_k \tau} + e^{E_k(\beta - \tau)}}{e^{\beta E_k} - 1} \frac{\epsilon_k}{E_k}.
\end{equation}
Here, $E_k$ is the Bogoliubov dispersion, and $\epsilon_k \ge 0$ is the bare bosonic dispersion shifted by $2t$. This kernel is peaked at $\tau = 0$ (and $\tau = \beta$) in imaginary time. The instantaneous limit can be taken when the bosons are fast compared with the fermions.   The velocity of the bare fermions at half filling ($k_F = \pi/2$) is $v_F = 2t \sin (k_F) = 2t$. The instantaneous approximation is hence reasonable for low values of $V$ when $U$ is not too small; in particular, close to the bosonic Mott transition it is valid. In the static approximation we have
\begin{equation}
D^0_{\rm static}(x) = \int \frac{dk}{2\pi} e^{ikx} \frac{2  \epsilon_k }{\sqrt{ \epsilon_k^2 + 2 \epsilon_k n_0 U} }.
\end{equation}
This induced interaction is spin agnostic, attractive on-site ($x=0$), but repulsive for $x > 0$ and scaling as $\sim 1/x^2$, which is short-range in 1D but can be rather large for next and next-next-nearest neighbor interactions. If we nevertheless only keep the local part, then we arrive at an attractive Hubbard model whose phase diagram is known~\cite{Giamarchi2007}. In particular, we expect dominant superconducting pair fluctuations, especially for weak $V$ in this BCS-like regime, but the pairing gap is exponentially weak in the induced interaction and scales exponentially in  $\sqrt{V^2/U}$.
Note that this approach neglects the back-coupling of the fermions on the bosons, {\it i.e.}, underestimates CDW structures, which, given the non-local form of the static interaction could still be important. 
Furthermore, the induced pair-superfluidity is in competition with a tendency to phase separate, which at the mean-field level, is found when $V^2 > U$~\cite{Viverit2000, Buechler2003, EnssZwerger2009}. At $U= 0$ the bosons occupy a number of sites scaling as $L^{1/3}$, the fermions occupy the rest. In  the thermodynamic limit coupling free bosons with spinless fermions via $V$ is hence always unstable, with a separation line given by $U = V^2 / \pi$. However, in case of induced interactions, the previous arguments do not immediately apply and the criterion for phase separation can be written as $\partial \mu_\sigma / \partial n_{\sigma} <  0$~\cite{Viverit2000}, which is numerically hard to use however. Since the gain in energy due to quasi-condensation is rather small in practice, we expect that the criterion found above is a very good upper bound and it is indicated by the dotted black line in Fig.~\ref{fig:phasediagram}. Simulations in close proximity to phase separation are difficult, and we saw no indications of being able to significantly improve on the behavior $U = V^2 / \pi$ (which is also close to the bosonization prediction, see Sec.~\ref{sec:bosonization}).\\

The bosons undergo a transition from a Luttinger liquid to a Mott insulator, which, for $V=0$, is found at $U=3.25(5)$~\cite{Kromer2014}. Turning on $V$, and having established that the bosons are fast with respect to the fermions, the renormalization flow of the bosons is understood to be already strong on the system sizes that we can simulate, whereas the fermions remain nearly free. Neglecting the backaction of the bosons on the free fermions, the bosonic Mott transition shifts upwards quadratically with $V^2$ and proportionality factor $D(\epsilon) = \sqrt{1 - (\epsilon/2t)^2}/(\pi t)$ at $\epsilon = 0$. 
This prediction for the location of the  bosonic to superfluid Mott insulator is a reasonable approximation for low values of $V \ll 1$. As can be seen in Fig.~\ref{fig:phasediagram}, it is even not too bad at $V=2$ but slightly overestimates the critical point because not all fermions can be considered non-interacting at the length scales of strong bosonic renormalization.
When the bosons are deep in the uniform Mott insulator, we can approximate their dispersion by $E(k) = \Delta + \frac{k^2}{2m^*}$ where $\Delta$ is the gap (of order $U$ in the pure bosonic system) and $m^*$ the effective mass for particle and hole excitations. The kernel is now $D^{\rm Mott}(x,\tau) = \int \frac{dk}{2\pi} e^{ikx} e^{-\Delta \tau}$ to leading order in the ground state. For $\Delta \gtrsim 2t$ we hence cannot expect to see any induced interaction, {\it i.e.}, the bosonic and fermionic systems decouple and the bosons remain in a uniform $n=1$ Mott insulator and the fermions are quasi-free (this is the upper part of the phase diagram in Fig.~\ref{fig:phasediagram}). For $\Delta \ll 2t$ induced interactions remain possible. However, in the weak $V$ limit and recalling the exponential weakness of the pairing gap, we can only see such pairing fluctuations when $\Delta$ is {\it very} close to 0, which in turn requires very big system sizes to distinguishes pair flow from two correlated superfluids.

\subsection{Strong interspecies coupling, $V \gg 4t$ }\label{sec:strong_coupling}

\begin{figure}[h]
\includegraphics[width=\linewidth]{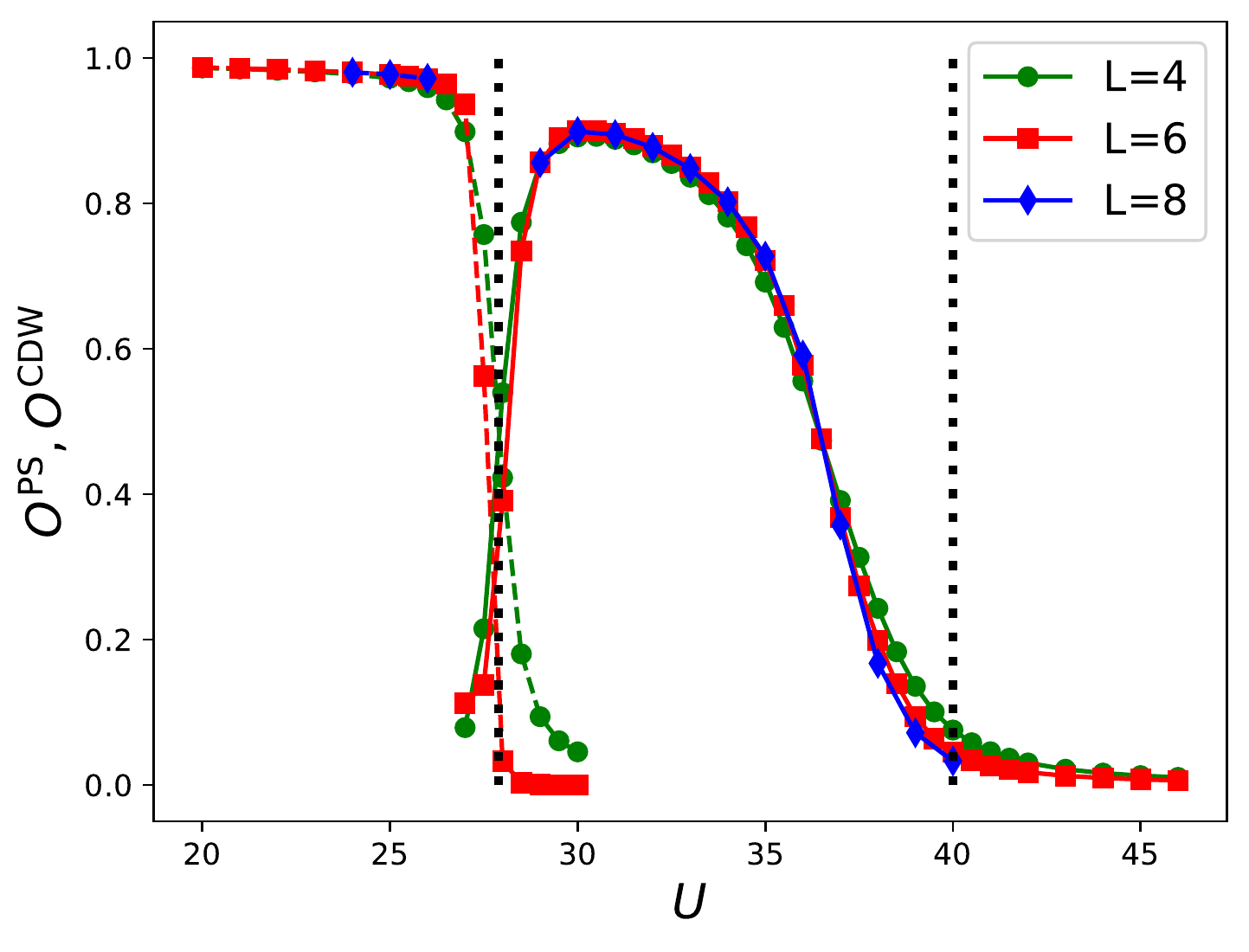}
\caption{\label{fig:Lanczos_V20} 
Results for the phase separation ($O^{\rm PS}$, dashed lines in the left part of the figure) and charge-density wave ($O^{\rm CDW} = S^{\rm CDW}(k = \pi)/L$, solid lines in the right part of the figure) from exact diagonalization (Lanczos calculations) for $V=20$. The black dotted lines are the strong coupling predictions separating a regime of phase separation (left) from a CDW phase (middle) and a uniform phase (right). Note that the calculations did not converge for $L=8$ in the range $26 < U < 29$.
}
\end{figure}

\begin{figure}[h]
\includegraphics[width=\linewidth]{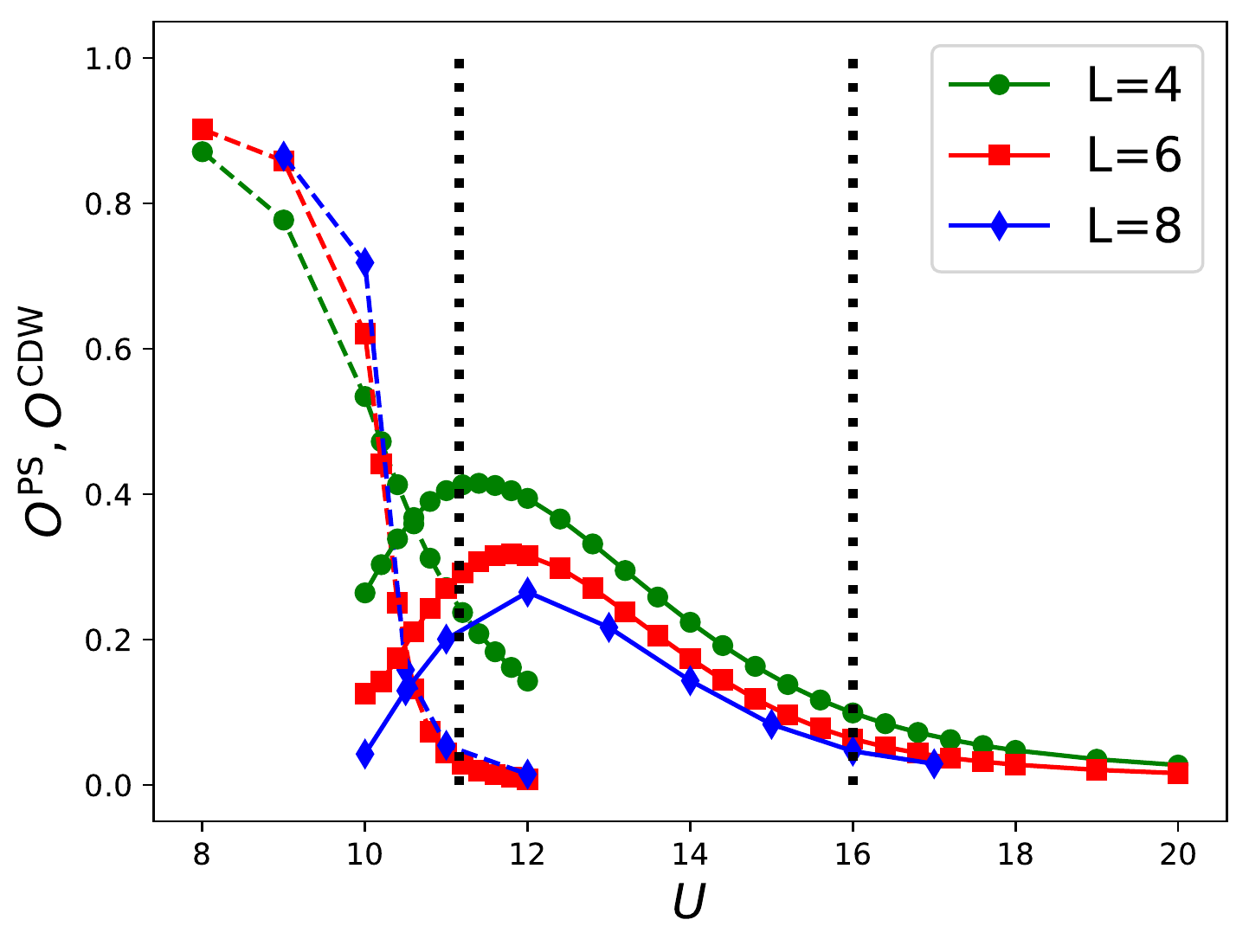}
\caption{\label{fig:Lanczos_V8} 
Same as in Fig.~\ref{fig:Lanczos_V20} but for $V=8$.
}
\end{figure}

\begin{figure}[h]
\includegraphics[width=\linewidth]{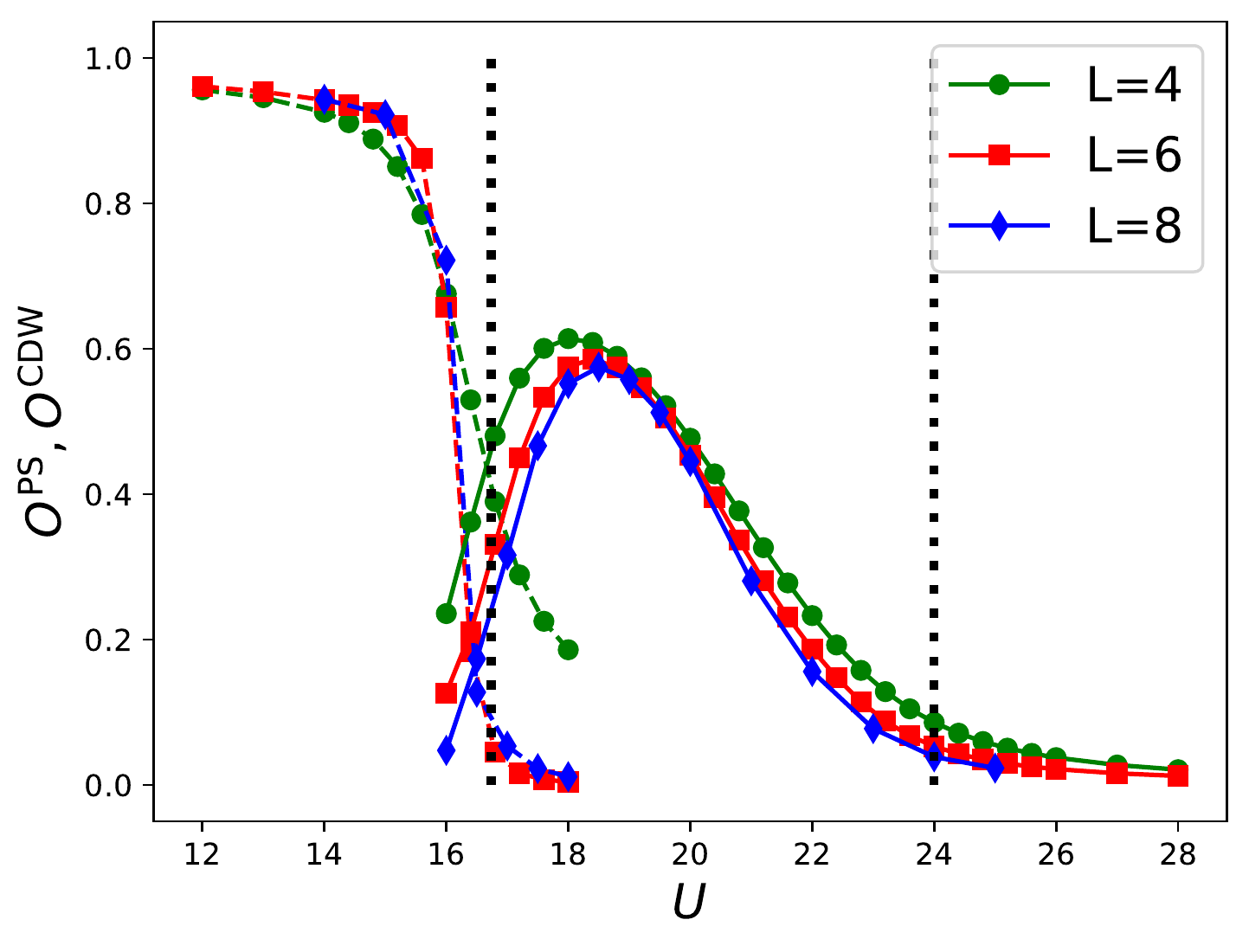}
\caption{\label{fig:Lanczos_V12} 
Same as in Fig.~\ref{fig:Lanczos_V20} but for $V=12$.
}
\end{figure}

Since the system phase separates when $U \ll V$, we focus on the regime where both couplings are strong,  $U,V \gg 4t$.
We can map the system onto an effective  model as follows. 
First, turning off the hopping amplitudes, we consider product states that minimize the potential energy subject to our filling constraints. 
We can fill $M$ sites with 2 bosons, $N$ sites with one fermion and one boson, and $L - M -N$ sites with up and down fermions. Sites which contain more than 2 particles per site are higher in energy and are discarded.
The energy of this configuration is $E_g = M U + N V$, subject to the constraint of particle number conservation, $2M + N = L$. 
Eliminating $N$, we minimize $E_g = M  (U - 2 V)$. 
Hence, for  $V < U/2$ (this is the upper black dashed line in Fig.~\ref{fig:phasediagram}) we expect that every site contains one boson ({\it i.e.}, we have a homogeneous bosonic Mott insulator with unit density) and one fermion on average. As there is no interaction between the fermions and the potential energy between bosons and fermions is constant (this case is equivalent to the Hartree approximation), the fermions delocalize on top of the Mott insulator and behave as (nearly) free fermions. From the arguments in the previous paragraph we expect this behavior when the bosonic gap is large, while pair flow is expected for a small bosonic gap.
For $V > U/2$ we must have that $M= L/2$, {\it i.e.}, half of the system is doubly occupied with bosons and the other half with fermionic doublons (molecules). Even though all configurations are equally likely, we expect that the kinetic terms will select either the fully phase separated regime or a phase with charge density wave order.   

Second, we analyze the effect of the hopping terms on those ground states to see which phase can lower its energy the most, and is hence preferred when we lower $U$ away from $U = 2 V$. 
In second order perturbation theory there is only a diagonal virtual exchange term (no flips are possible) 
described by a $J_z$ term
\begin{equation}
J_z = \frac{6 t_{\rm B}^2}{U} -  \frac{2 t_{\rm B}^2}{2V - U } - \frac{2 t_{\rm F}^2}{2V}.
\end{equation}

This predicts a transition to the phase separated regime at  $V/U = \frac{10 + \sqrt{52}}{24} \approx 0.717$ (this linear behavior is the lower black dashed line in Fig.~\ref{fig:phasediagram}).

The reliability of these arguments was checked with exact diagonalization using a Lanczos method. We computed a naive measure for phase separation, $O^{\rm PS}$, as the overlap of (one of) the ground states with fully phase separated product states. This is an underestimation of the phase separated states but for large $V$ and $U$ it is believed to be very accurate, consistent with the strong coupling argument. In the phase-separated regime we also expect the ground state to be $L-$fold degenerate. CDW order was probed by the staggered structure factor of the CDW correlation functions, as introduced in Sec.~\ref{sec:order_parameters}, but divided by the number of sites. The ground-state of a system with CDW order should be two-fold degenerate in the thermodynamic limit. For $V=20$, we see in Fig.~\ref{fig:Lanczos_V20} that the strong-coupling prediction is quantitatively accurate. Furthermore, the CDW order parameters for $L=6$ and $L=8$ lie on top of each other. Hence, finite size effects are tiny. For $V=8$ however (this is the largest value of $V$ for which we have obtained quantum Monte Carlo results), the Lanczos results are far off from the thermodynamic limit, as can be seen in Fig.~\ref{fig:Lanczos_V8}: the peak position of the CDW correlator shifts to larger values of $U$, and the signal diminishes with $L$.  We also observe a tiny energy difference between the ground state energy and the energy of the first excited state in this parameter regime. Based on the Lanczos results, we can envisage two scenarios: (i) the CDW correlations survive in the thermodynamic limit (as for $V=20$) but with a very tiny amplitude, or (ii) there is no lattice symmetry breaking and we instead find a Luttinger liquid with dominant CDW powerlaw correlations. Also a combination is possible, where (ii) is found close to the phase separation regime and (i) for larger values of $U$. Also the location of the phase separation boundary could shift to larger values of $U$ as we increase $L$.
As we will see below, the quantum Monte Carlo results for substantially bigger $L$ are likewise unable to differentiate between these scenarios. Especially for $U=14$ we will see strong competing instabilities at work up to 100s of lattice sites. The quantum Monte Carlo simulations for $V=8$ show the strongest tendency for CDW order at $U=13$. 

Finally, For $V=12$ (see Fig.~\ref{fig:Lanczos_V12}), the Lanczos results suggest that true CDW order is stable for $U \gtrsim 19$, but  probably not for smaller values of $U$ where we would again expect dominant CDW power-law behavior.

\subsection{Intermediate coupling, $V \sim 3t-4t$  }

\begin{figure}[h]
\includegraphics[width=\linewidth]{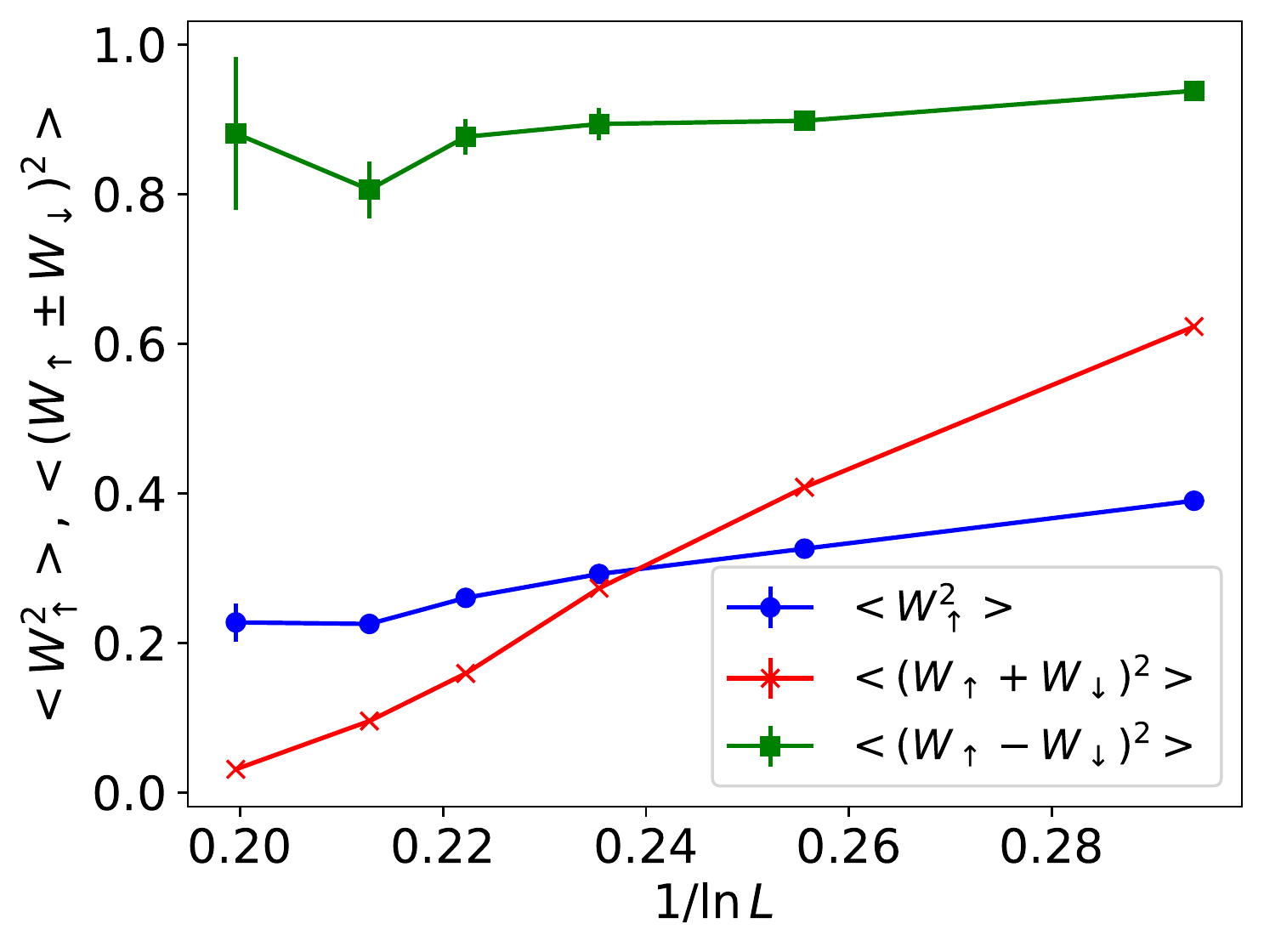}
\caption{\label{fig:wind_superconductor} 
Winding number squared for the up fermions (blue), the pair flow channel (green), and the counter flow channel (red) as a function of inverse system size $L = \beta$ for $V=3, U=4.5$. In the counter-flow channel one clearly sees the renormalization flow towards 0 while in the pair-flow channel the signal stays constant. When the counter flow is 0, the winding numbers squared of the individual components should be 1/4 of the value seen for the pair-flow.
}
\end{figure}

\begin{figure}[h]
\includegraphics[width=\linewidth]{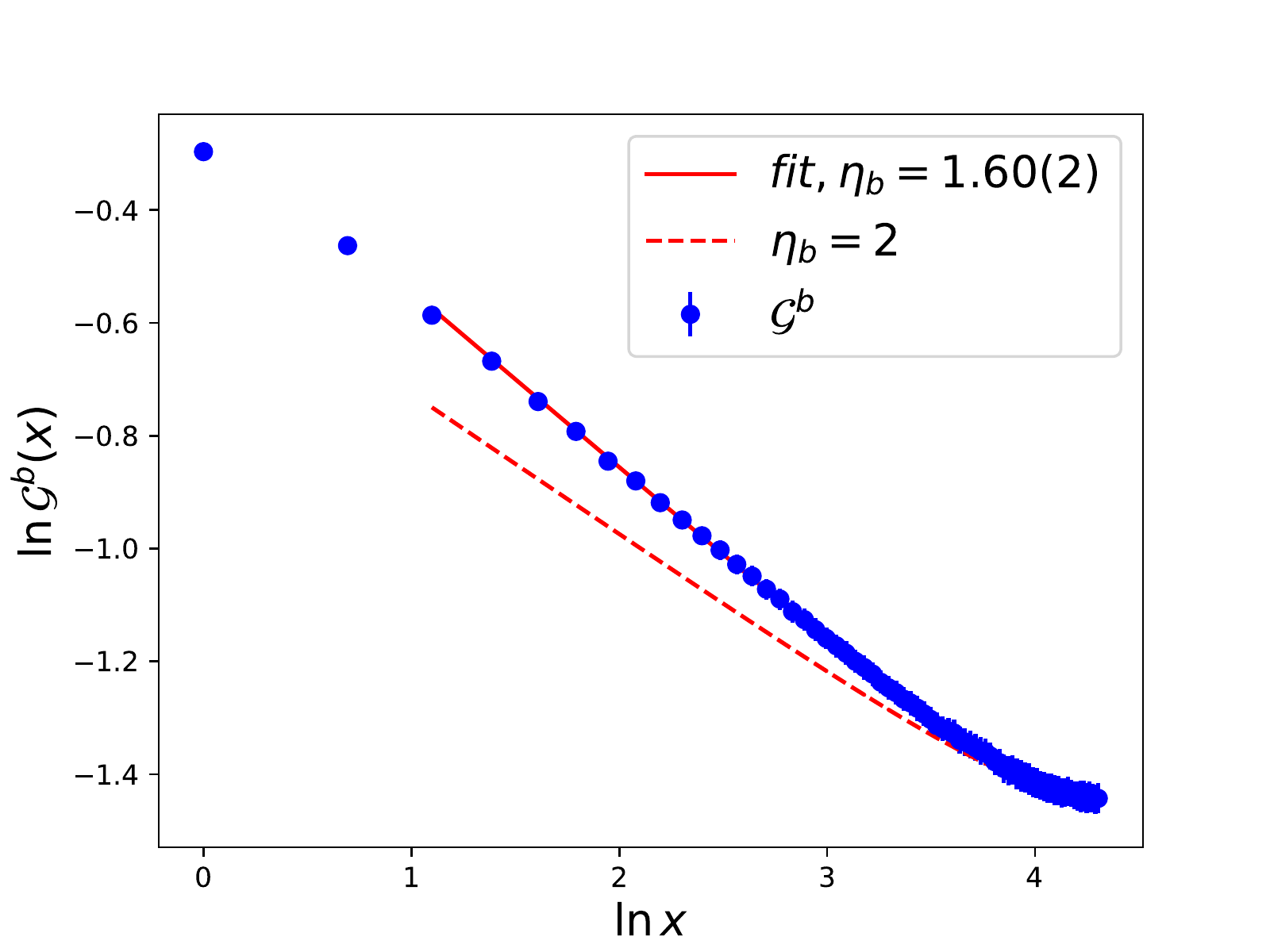}
\caption{\label{fig:wind_marginal_sf} 
The equal-time bosonic density matrix $\mathcal{G}^b(x) = \left< b^{\dagger}(x)b(0) + {\rm h.c.} \right>$ and corresponding fit with the cord function $\sim d(x \vert L)^{-1/(2\eta_b)}$ showing a powerlaw decay with exponent $\eta_b \ll 2$ for $V=4, U=6, L = \beta=150$. The dashed line shows the powerlaw with $\eta_b =2$. This powerlaw was seen for all smaller system sizes as well.  
For $U=6.5$ we found $\eta_b = 1.55(2)$. 
}
\end{figure}

In this regime, we expect no simple analytical arguments to hold and give only a few comments on the phase diagram and/or the most surprising results here.
Unlike for weak interspecies coupling, we can in this regime often monitor the renormalization flow till the order is complete. In Fig.~\ref{fig:wind_superconductor} we see how a full pair superflow has developed for $V=3$ and $U = 4.5$ on a system size $L \sim 200$. 

The extension of the bosonic superfluid to these large values $V=4$ is remarkable. For $V=4$, $U=6$ the bosonic winding numbers (indicative of their superfluid density) remain constant as a function of system size, hinting at a superfluid. Fitting the equal time bosonic density matrix with a powerlaw $d^{\frac{1}{2 \eta_B}}(x \vert L)$ we extract that $\eta_B = 1.60(2)$ as is shown in Fig.~\ref{fig:wind_marginal_sf}. Recalling that we have commensurate bosons at unit filling, the fact that this Luttinger parameter is less than 2 and does not flow to zero with system size is paradoxical (see however the next section).


\subsection{Considerations based on bosonization}\label{sec:bosonization}

In this section we review the results of Ref.~\cite{MatheyWang2007}, in which abelian bosonization for each component separately was applied. The fermion degrees of freedom were subsequently cast in the standard spin and charge channels. If we assume that the spin sector is always massive thanks to the induced attractive interactions, which is supported by the Monte Carlo data except for the free fermion case (which we found when $U \gg V$), then the bosonization analysis bears a strong resemblance to the spinless case. In Ref.~\cite{MatheyWang2007} it was assumed that the bosons are fast compared to the fermions. Lattice commensurabilities, which would lead to a non-linear and hence non-tractable problem, were analyzed to lowest order by looking at their scaling dimension only. 

The predictions of the theory of Ref.~\cite{MatheyWang2007} are that the single-particle bosonic density matrix decays as a powerlaw,
\begin{equation}
\mathcal{G}^b(x) \sim x^{-\frac{1}{2 \eta_b}},
\end{equation}
where $\eta_b$ ($K_\epsilon$ in the notation of  Ref.~\cite{MatheyWang2007}) differs from the bare bosonic Luttinger parameter $K_b$ and is also different from the the exponent governing the decay of the density-density correlation function, whose $2k_b$ component decays as
\begin{equation}
C^{\rm bb}(x) \sim \cos(2 k_b x) x^{-2  \eta_{\rm bb}}.
\end{equation}
In the notation of Ref.~\cite{MatheyWang2007}, $K_{\delta} = \eta_{\rm bb}$. The momentum $k_b = \pi$ is set by the bosonic filling factor.
The transition to the bosonic Mott insulator is found for $K_{\delta} = 2$. Note that $K_{\delta} > K_{\epsilon} > K_b$ (which allows the behavior seen in Fig.~\ref{fig:wind_marginal_sf} with $\eta_b < 2$).  In the massive spin sector, the fermionic single-particle density matrix and the SDW correlator are exponential, but the CDW correlator has a powerlaw behavior as
\begin{equation}
C^{\rm CDW}(x) \sim \cos(2k_f x) x^{-\eta_{\rm CDW}}.
\end{equation}
In the notation of Ref.~\cite{MatheyWang2007}, $\eta_{\rm CDW} = K_{\beta} + K_{\sigma} \to K_{\beta} $. If the full form of the SDW correlator is needed, it is $\eta_{\rm SDW} = K_{\beta} + K^{-1}_{\sigma}$.
The $2k_f$ component of the singlet correlation function has a powerlaw $K_{\beta} + K_{\gamma}^{-1}$, but the signal to noise ratio on this quantity in the Monte Carlo simulations is too low for this to be useful to us. 
The polaron operators mentioned in Ref.~\cite{MatheyWang2007} are not implemented.
We will take the pragmatic view that the generic form of the correlation functions above holds in the non-massive phases in order to analyze our simulations. However, Ref.~\cite{MatheyWang2007} has only limited power in predicting the phase diagram. It gives a good account of the phase separation (also quantitatively for low values of $V$),  and a criterion for the bosonic Mott transition (through $K_\delta=2$) but it systematically predicts singlet pairing for the fermions for our microscopic parameters.  For our parameters the commensurabilities are such that lattice symmetry breaking is possible, and that insulating CDW structures in the bosonic sector are also possible, which is incompatible with their assumptions. 


\section{Algorithm} \label{sec:worm}

We employ the worm algorithm~\cite{Prokofev1998qmc} based on the update scheme of Ref.~\cite{sadoune_efficient_2022}  to perform quantum Monte Carlo simulations. The three species are implemented as three different layers. The inter-species coupling then corresponds to a density-density coupling between the layers. For the fermions, we employ two types of worms: the standard algorithm, which samples the single-particle Green function for each component separately, and a two-worm algorithm necessary to be ergodic in the superconducting phases. The latter is implemented in the particle-particle or the particle-hole channel in such a way that  the worm creation operator of the spin-up particles moves around in space-time synchronously  as the worm creation operator for the spin-down particles, but not necessarily on the same site. In the particle-hole channel, it is the worm annihilation operator for one of the species that moves together with the worm creation operator of the other species. With these updates the algorithm is ergodic for superconducting phases made up of two fermions.  Simulations would face an exponential barrier however in case of a molecular superfluid  phase consisting of the pairing between a hard-core boson  and a soft-core boson because we do not allow configurations with the simultaneous presence of their respective worm operators. Such phases would certainly be important for other parameter choices than the ones we considered.
Nevertheless, the simulations are often hard for large system sizes and inverse temperatures. We also employed simulated annealing strategies, in which we gradually lower the temperature, and also quench strategies, in which we take a stable and converged configuration in the, say, superconducting phase and then quench the parameters to a nearby point in parameter space where the simulations are harder. 
We typically scale system size and inverse temperature as $L = \beta$, as is justified for Luttinger liquids. Test simulations done for $V \le 4$ with a larger aspect ratio did not change the physics.

\section{The various order parameters} \label{sec:order_parameters}

The Monte Carlo simulations provide us with a plethora of thermodynamic quantities.  We examined in total 20 different quantities, which must behave consistently before we can be sure of the phase that we observe.
\begin{itemize}
\item The superfluid densities of the individual bosons $\rho_{s,b} = \frac{L \left< W_b^2 \right>}{t \beta}$, and 'fermions' (simulated as hard-core bosons), $\rho_{s,\sigma} = \frac{L \left< W_\sigma^2 \right>}{t \beta}$. Here, $W_b$ and $W_\sigma$ are the winding numbers of the bosons and the fermions, respectively~\cite{Pollock1987}. We also looked at the paired superfluid density $\rho_{\text{PSF}} = \frac{L\langle (W_{\uparrow} - W_{\downarrow})^2\rangle}{t\beta}$ and the counter-rotating superfluid density $\rho_{\text{SCF}} = \frac{L\langle (W_{\uparrow} + W_{\downarrow})^2\rangle}{t\beta}$ in order to probe the system for pairing correlations between the fermions~\cite{Kuklov2003}.
\item The equal-time density matrices $\mathcal{G}^b(x) = \left< b^{\dagger}(x) b(0) + {\rm h.c.} \right>$ and $\mathcal{G}^\sigma(x) = \left< c_{\sigma}^{\dagger}(x) c_{\sigma}(0) + {\rm h.c.} \right>$ of the bosons and the fermions respectively. For the fermions, the Jordan-Wigner string between sites $0$ and $x$ must be inserted in the Monte Carlo measurements since the fermions are simulated as hard-core bosons~\cite{Girardeau1960}. Its Fourier transform yields the momentum distribution $n^{\sigma}(k) = \sum_x e^{ikx} \mathcal{G}^\sigma(x)$. Sign-positive statistics remain possible with periodic boundary conditions when the number of up spins and the number of odd spins are both odd.
\item The equal-time 4-point correlation functions $\mathcal{G}^{pp}(x) = \left< c_{\uparrow}^{\dagger}(x) c_{\downarrow}^{\dagger}(x) c_{\uparrow}(0) c_{\downarrow}(0)  + {\rm hc}   \right>$ and $\mathcal{G}^{ph}(x) =  \left< c_{\uparrow}^{\dagger}(x) c_{\downarrow}(x) c_{\uparrow}(0) c^{\dagger}_{\downarrow}(0) + {\rm hc} \right> $. These probe the Green functions of molecules seen as a local Cooper pair of two fermions with opposite spin. Due to the attractive induced interactions, $\mathcal{G}^{ph}(x)$ should always decay exponentially in the thermodynamic limit.
\item For grand-canonical simulations we computed the compressibility of the bosons, $\kappa_{b} = \beta( \left<N_b^2 \right> - \left< N_b \right>^2)/L $ where $N_b$ is the total number of bosons in the configuration.
\item For canonical simulations, we fine-tune the chemical potential $\mu_b$ such that $\left <N_b \right> = L$. The fermionic chemical potentials are always $\mu_\sigma = V$ to ensure that $\left< n_{\sigma} \right> = 1/2$.
\item In canonical simulations we also compute the bosonic Green function at zero momentum as a function of imaginary time $\mathcal{G}^b(p=0, \tau)$ in order to measure the bosonic single particle gap in the insulating phases. The Green function $\mathcal{G}^b(x=0, \tau)$ was also computed for very low values of $V$ in order to extract the speed of sound~\cite{Prokofev1998qmc}.
\item We compute the connected static density-density correlation functions $C^{\rm bb}(x) = \left<n_b(x) n_b(0) \right> - 1$,  $C^{\rm CDW}(x) = \left< (n_{\uparrow} + n_{\downarrow})(x)  (n_{\uparrow} + n_{\downarrow})(0)  \right> - 1$ and $C^{\rm SDW}(x) = \frac{1}{4} \left< (n_{\uparrow} - n_{\downarrow})(x)  (n_{\uparrow} - n_{\downarrow})(0)  \right> $ for bosons and fermions in the charge density wave (CDW) and spin density wave (SDW) channels. We also compute the correlator $C^{\rm MDW}(x) = \left< M(x) M(0)  \right> $, where $M(x) = 2 n_b(x) - n_{\uparrow}(x) - n_{\downarrow})(x)$ is the molecular density wave correlator.
\item We compute the corresponding structure factors $S^{\rm bb}(k) = \sum_j e^{ikj} C^b(j)$, $S^{\rm CDW/SDW} (k) = \sum_j e^{ikj} C^{\rm CDW/SDW}(j)$.
\end{itemize}

As our bare fermions are non-interacting, their $n_{\sigma}(k)$ shows a Fermi surface with a jump of 1 at $k= \pm k_F= \pm \frac{\pi}{2}$. For finite $V$ and increasing $L$, this jump decreases. In the thermodynamic limit, the jump must be zero as there are no quasi-particles in one dimension, only collective excitations. The jump is hence a good diagnostic for how strongly the renormalization flow from the non-interacting fermions into the Luttinger liquid (or a massive phase) has developed. \\

We always expect a spin gap because of the attractive induced interactions in the thermodynamic limit. The limit $\lim_{q \to 0} \pi S^{\rm SDW}(q)/q$ should hence approach 0. In practice, any value below 1 for finite $L$ is seen as evidence for a spin gap that will fully develop as $L \to \infty$. For a system with rotational spin invariance, one must have that $\lim_{q \to 0} \pi S^{\rm SDW}(q)/q = 1$. This is what we hence see for free fermions. Likewise, the limit $\lim_{q \to 0} \pi S^{\rm CDW}(q)/q$ of the CDW structure factor, captures the non-oscillating part of the CDW correlation function. \\

Since it is possible to break a discrete symmetry in 1D, true CDW order is possible. It is seen as $C^{\rm CDW}(x)$ oscillating with $C \cos(\pi x)$ for large $x$ and a constant $C > 0$.
This implies that $S^{\rm CDW}(k = \pi)$ grows linearly with $L$. True CDW order for the fermions must imply true CDW order for the bosons, and vice versa, due to the nature of the boson-fermion coupling.  
Strong CDW correlations are also possible in the form of quasi-long range order, when the oscillating $2k_F$  part of the $C^{\rm CDW}(x)$ correlation function decays as $\sim \vert x \vert ^{-2+ \alpha}$. For $\alpha > 0$ this implies a diverging susceptibility. The density remains uniform in this case, just as the bosonic density remains uniform.
In order to treat the periodic boundary conditions adequately, we must use the cord function $d(x \vert L) = \sin (\pi x/L)$ in the argument of the correlation functions, cf.~\cite{Pollet2006}. Near half the system size, a power-law decay of the cord function is hard to discern from asymptotic (near-)constant behavior. Given our system sizes of the order  $L=100$ we are therefore often unable to discriminate between lattice symmetry breaking with a low value of $C$ and a slowly decaying powerlaw. 
 In practice we attempted both fits, which are often equally good, and used a cutoff value of $C \approx \frac{1}{L}$ for true long range order for at least $L \ge 70$. 
 We will refer to algebraic CDW order when the decay of the CDW correlations is described by a powerlaw but slower than for free fermions, {\it i.e.}, $\eta_{\rm CDW} < 1$.  \\

\section{Parameter scans} \label{sec:scans}

In the sections below we systematically plot for different values of  $U$ and $L = \beta$ in the canonical ensemble the $Z$-factors and winding numbers squared in the bosonic, counter-flow and pair-flow channels for $L=30,50$ and $70$. This is done for the values $V=1,2,3,4,5,6$ and $8$. Additional plots and additional system sizes are provided in case interesting physics happens. The bosonic superfluid density, single-particle density matrix and the properties of the fermionic single particle matrices converge typically the fastest. Counter-flow and especially pair-flow properties converge slower, and the large-distance charge density wave properties are usually hard to determine precisely because of the low values involved.
Our resolution on  the $V$ parameter is 1, as already mentioned,  and it is $0.5$ on the  $U$  parameter in the weak and moderately strong coupling regimes. For $V=6$ and $V=8$ the resolution on $U$ is $1$.  This means that the error bars in the phase diagram must be of the order $U/2$. We do not attempt to discuss the nature of the phase transitions or determine their precise location for two reasons.
First, although the data for the (bosonic) superfluid properties are suggestive of a Kosterlitz-Thouless renormalization flow, we do not know the value of the jump in $\eta_b$ at the transition. 
We take a pragmatic view and when we observe a typical scaling of the superfluid densities as $\sim 1/ \ln(L/L_0)$ with $L_0$ comparable to the lattice spacing, then we know that we are close to the critical point. 
Second, computing quantities like $\eta_{\rm CDW}$ is known to be hard, already for systems with fewer flavors of particles~\cite{Kuehner2000, Clay1999}. When fitting the CDW correlation function, there  are uncertainties about how to choose the fitting interval, the influence  of a non-oscillating term $1/r^2$ (which is not always negligible) next to the oscillating first harmonic which has an unkown powerlaw decay, and the influence of the periodic boundary conditions. 
It is argued that a low momentum analysis of the corresponding static structure factor (see Sec.~\ref{sec:order_parameters}) gives a better signal to noise ratio~\cite{Clay1999, Sengupta2002, Sandvik2004}. Although we also take this route here, the drawback in this approach is that there are cases where the curve bends down for low momenta. Increasing the system size adds lower momentum points to the curve, which in turn tends to bend the curve further down; {\it i.e.}, the extrapolation to zero momentum can be misleading for small system sizes.

\subsection{Scan of phase diagram at $V=1$}

\begin{figure}[!htb]
        \begin{tabular}{ll}
        \includegraphics[width=0.5 \columnwidth]{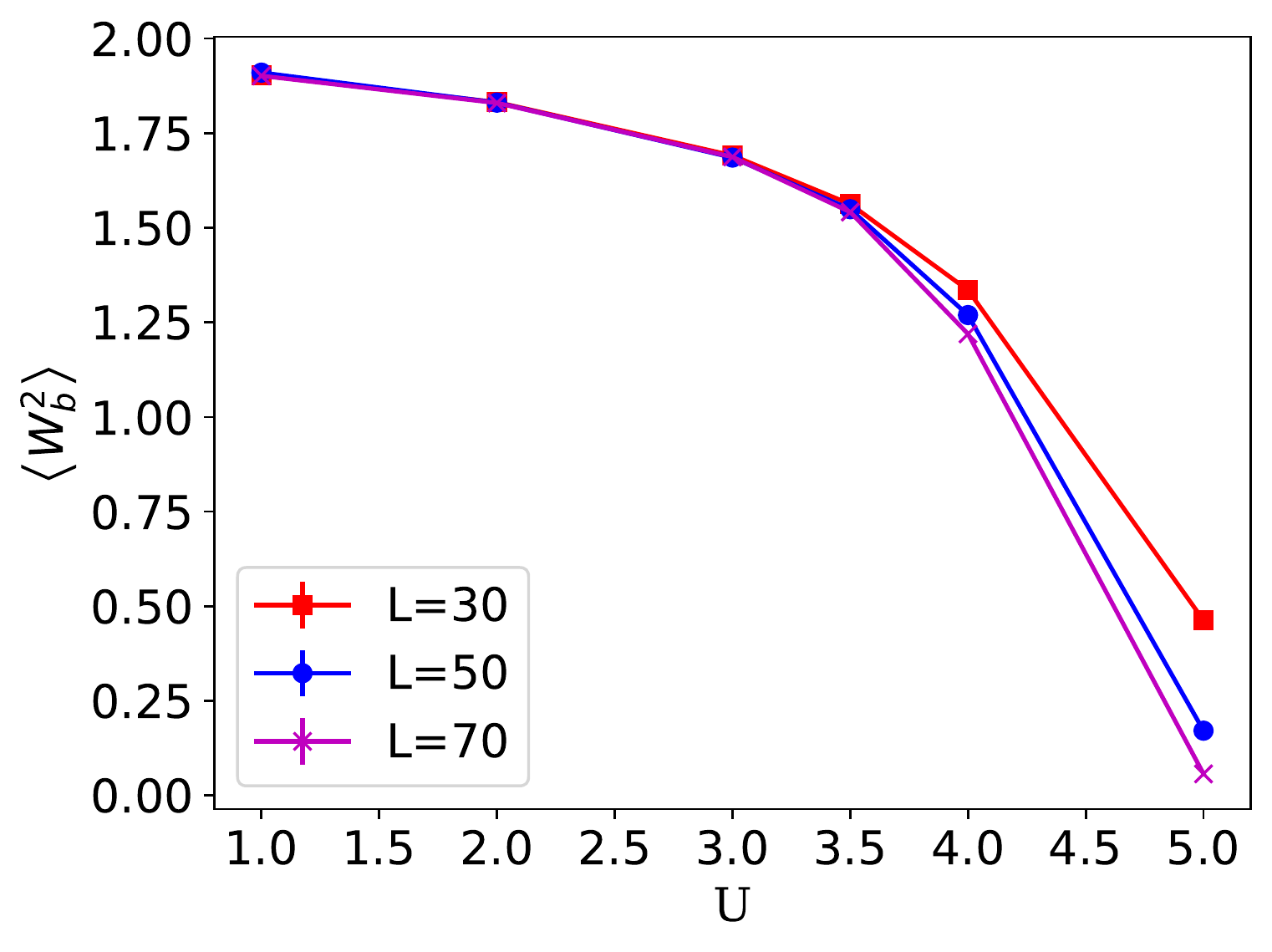} &
        \includegraphics[width=0.5 \columnwidth]{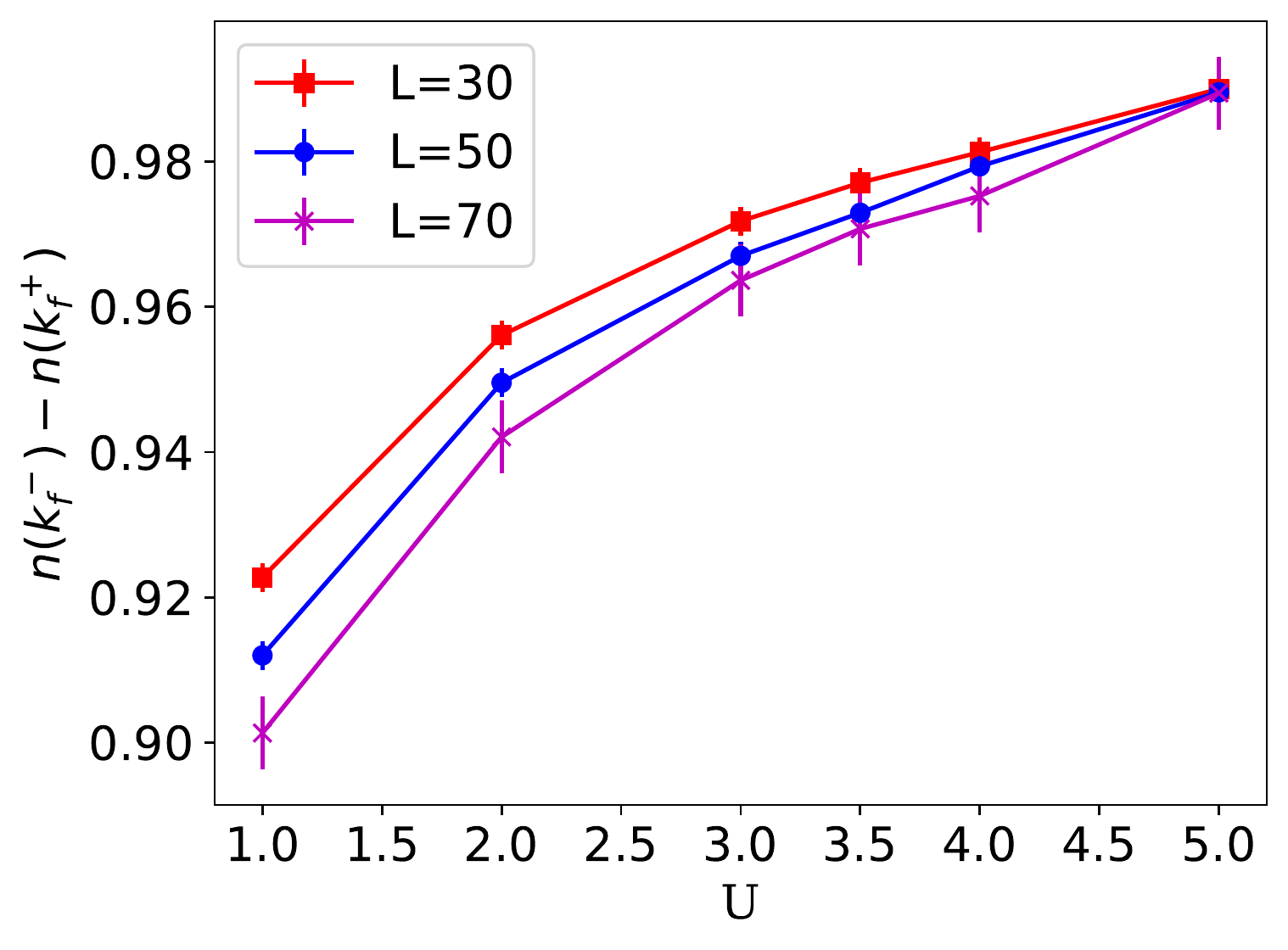} 
        \end{tabular}
        \caption{  \label{fig:V1_app} Left: Bosonic winding number squared for $V=1$ as a function of $U$ for different $L=\beta$; the bosonic Mott transition is expected at $U = 3.57(5)$. 
        Right: The jumps in fermionic occupation number  at $k=k_F$ ({\it i.e.}, the $Z$-factor)  for the same system, showing that the fermions remain almost free on these system sizes.       }
\end{figure}

\begin{figure}[!htb]
        \begin{tabular}{ll}
        \includegraphics[width=0.5 \columnwidth]{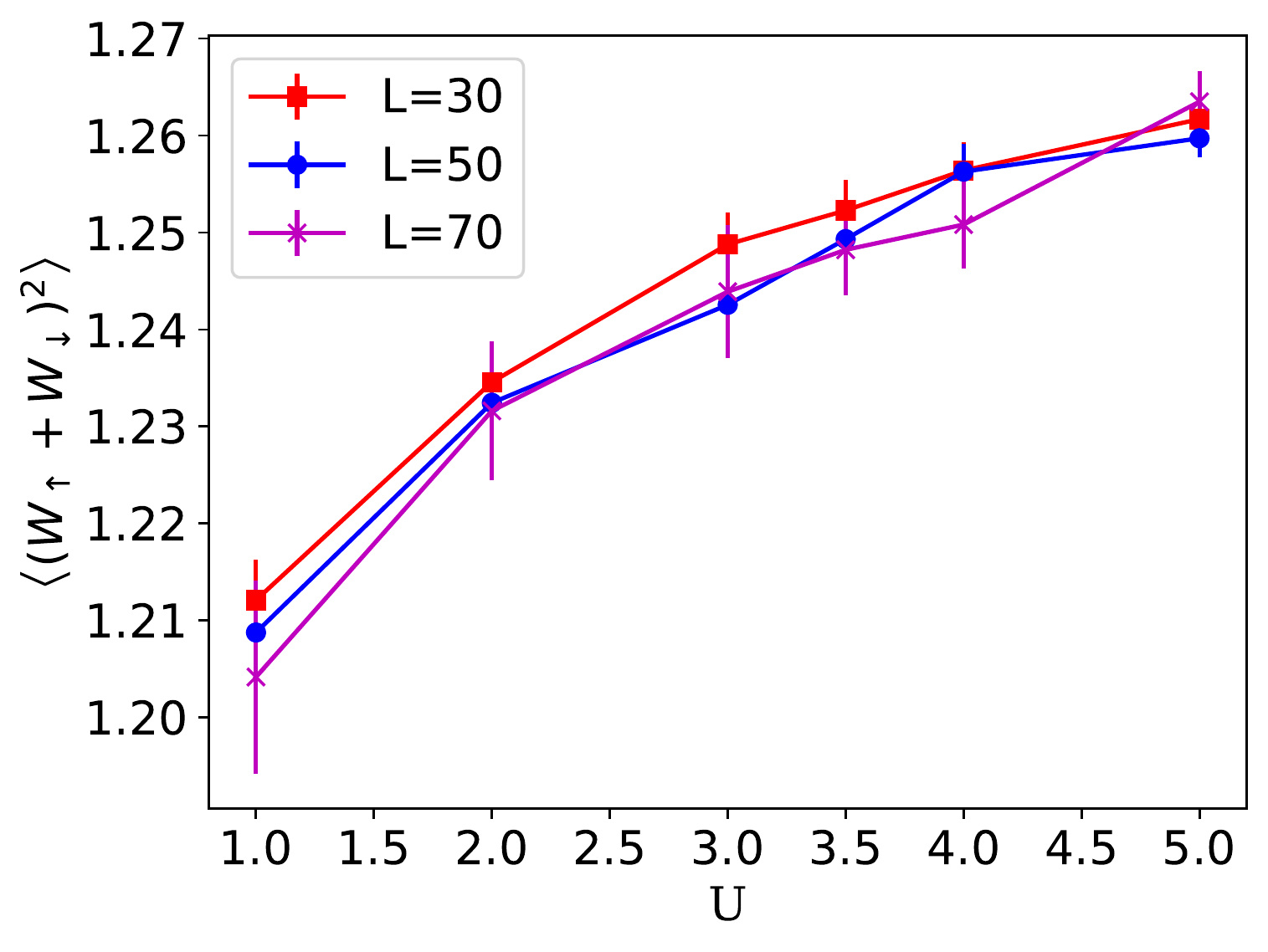} &
        \includegraphics[width=0.5 \columnwidth]{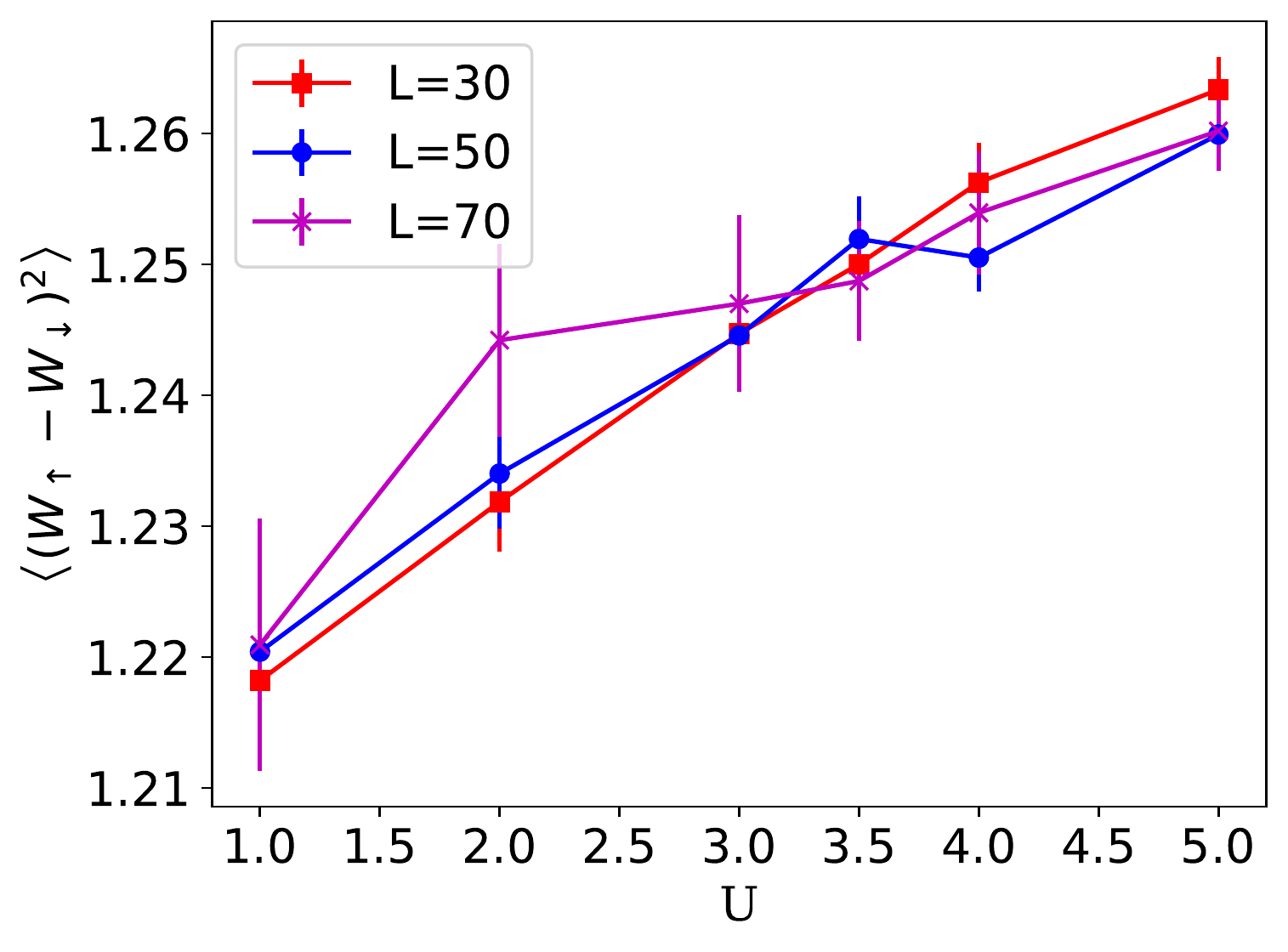}
        \end{tabular}
        \caption{Left: Winding numbers squared in the counter-flow  channel for $V = 1$; Right: Winding numbers squared in the pair-flow channel for $V= 1$.  
                 \label{fig:V1_spf_scf} }
\end{figure}

\begin{figure}[!htb]
        \begin{tabular}{ll}
        \includegraphics[width=0.5 \columnwidth]{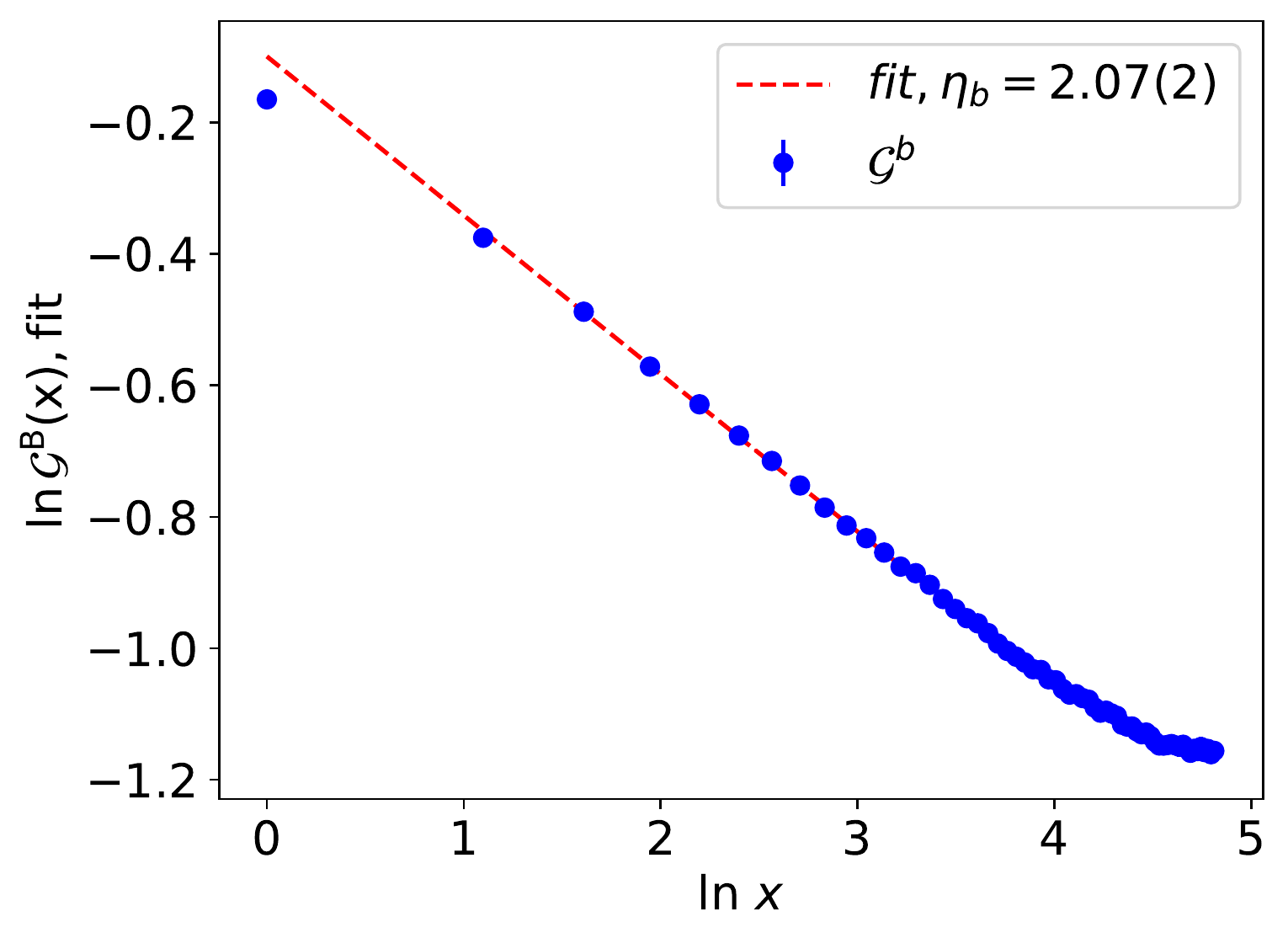} &
        \includegraphics[width=0.5 \columnwidth]{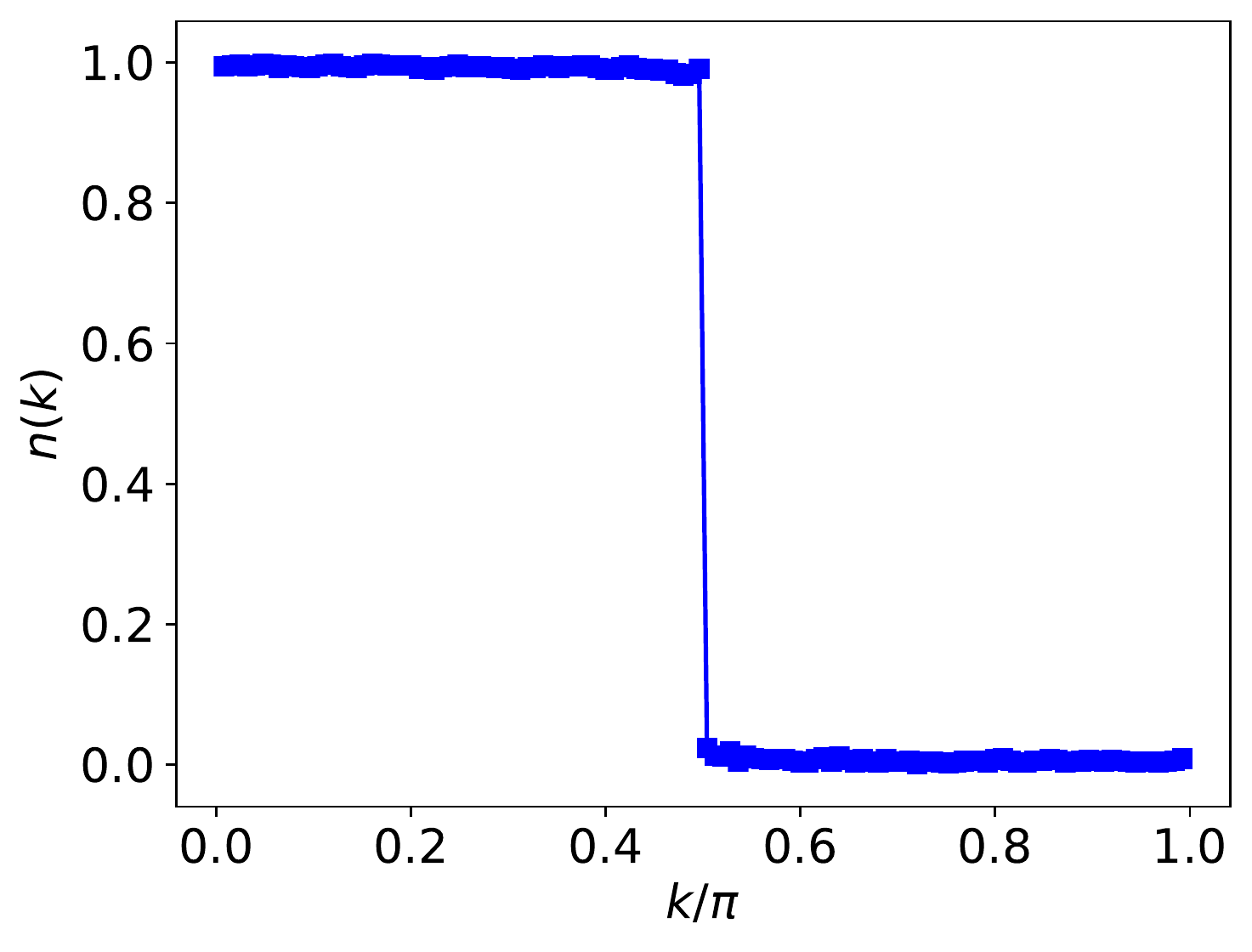}
        \end{tabular}
        \caption{Left: Bosonic equal-time density matrix and the fit $f(x) = a - \rm{log} (d(x | L))/(2 \eta_b)$ for a system with parameters $V = 1, U = 3.5, L = \beta =250$.  The extracted power law is  $\eta_b=2.07(2)$, which, to a very good approximation for $V=1$, can be compared with the bare Luttinger parameter $K_b$ and which predicts a Kosterlitz-Thouless transition to the Mott insulator (MI) at $\eta_b \approx K_b^{\rm MI} = 2$.
        Right: Momentum occupation of the fermions for the same system, showing behavior very close to the one of free fermions. \label{fig:V1_U3.5_L250_app} }
\end{figure}

In case of very weak inter-species coupling, the induced interaction is likewise very weak with fermionic $Z$-factors above 0.9 for $U \ge 1$ and $L < 70$, and increasing with $U$ (see Fig.~\ref{fig:V1_app}). It is therefore a good approximation to consider the bosonic and fermionic subsystems as quasi  uncoupled on these system sizes, {\it i.e. }, the bosons can be described by the standard Luttinger theory and the fermions remain nearly free. We indeed see that the powerlaws in the CDW and SDW channels are $\eta_{\rm CDW} \approx 1$ and $\eta_{\rm SDW} \approx 1$, and the difference between the counter and pair superfluid densities (Fig.~\ref{fig:V1_spf_scf}) is no more than one percent, indicating very weakly coupled superfluids of up and down spins.
We show in Fig.~\ref{fig:V1_U3.5_L250_app} the bosonic density matrix very close to the Mott transition for $V= 3.5$ for a system size $L=\beta=250$ as well as the fermionic occupation numbers. We see for $U > 3.5$ the typical flow of the bosonic winding numbers towards the insulating state for $L=\beta=30,50$ and 70 in the left panel of Fig.~\ref{fig:V1_app}, and this flow is also seen in the powerlaw decay of $\mathcal{G}^b(x)$. As the fermions remain nearly free on these system sizes, this supports the claim that we expect the bosonic Mott transition at $U^{\rm crit} ( V= 1) =  U^{\rm crit} ( V= 0) + \frac{1}{\pi} V^2 = 3.25(5) + \frac{1}{\pi}  =3.57(5)$, where we took the critical value for the superfluid to Mott insulator transition for the Bose-Hubbard model from Ref.~\cite{Kromer2014}. In short, the system behaves as expected for $V=1$. \\


\subsection{Scan of phase diagram at $V=2$}

\begin{figure}[!htb]
        \begin{tabular}{ll}
        \includegraphics[width=0.5 \columnwidth]{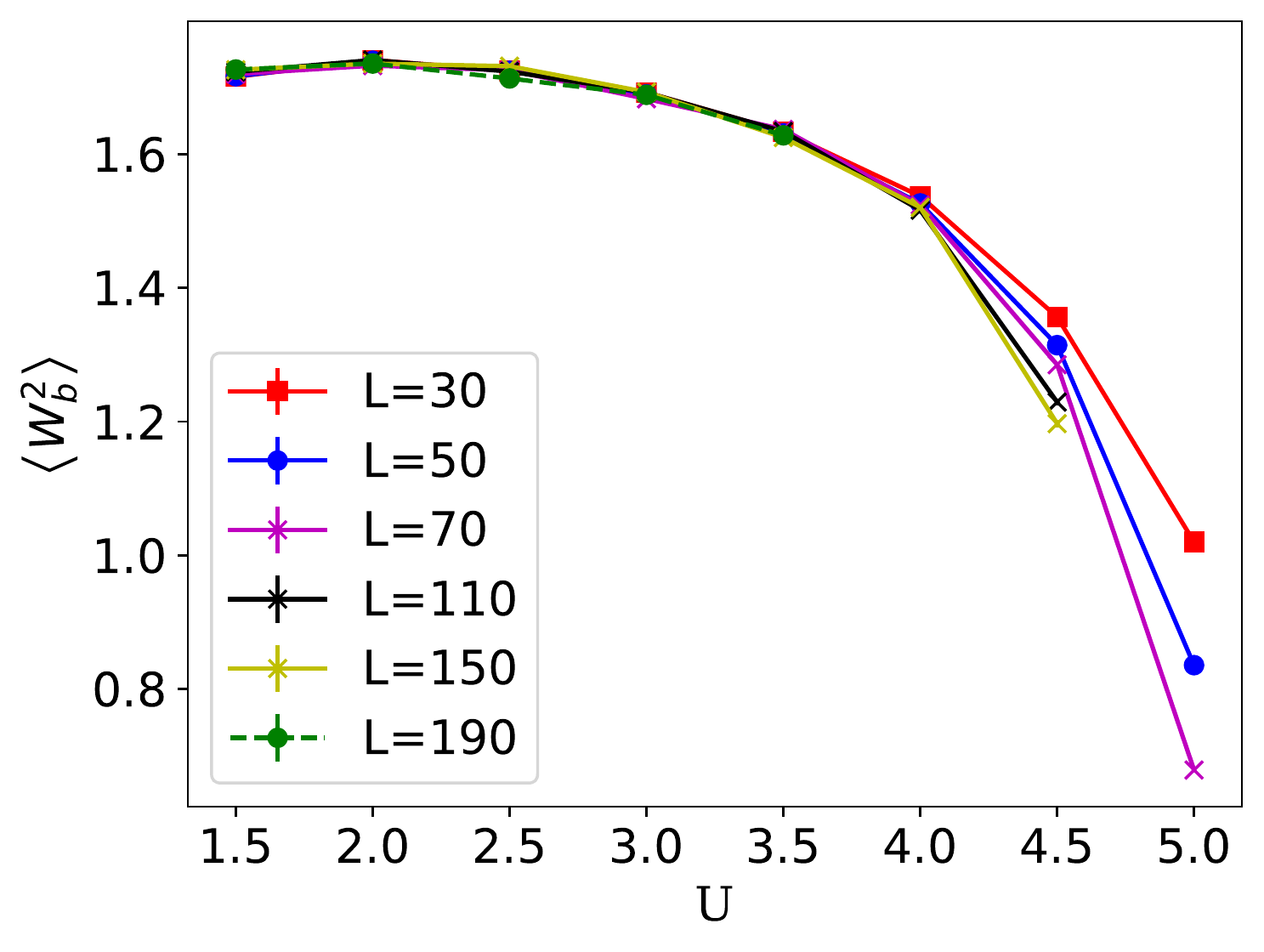} &
        \includegraphics[width=0.5 \columnwidth]{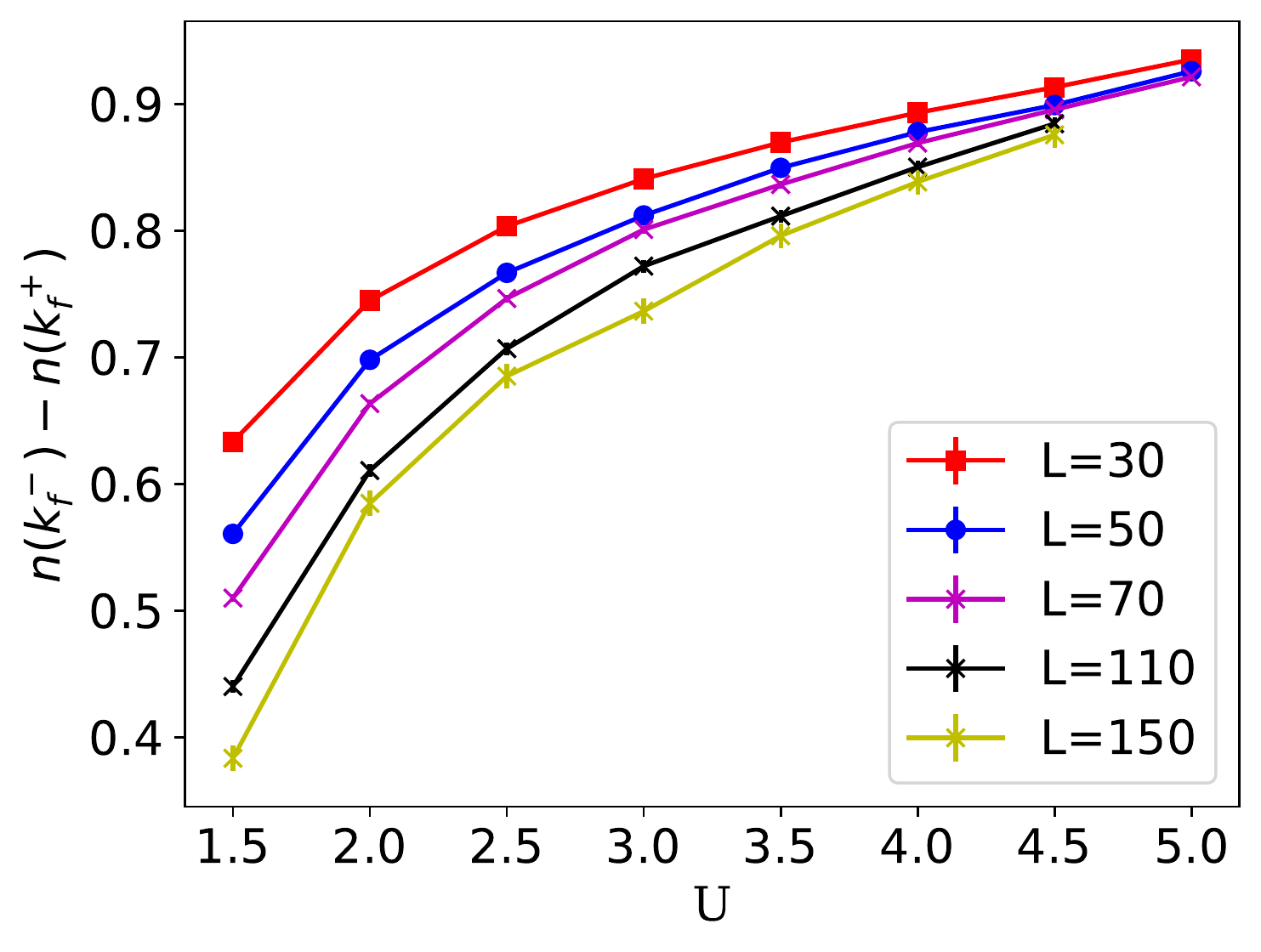} 
        \end{tabular}
        \caption{  \label{fig:V2_app} Left: Bosonic winding number squared for $V=2$ as a function of $U$ for different $L=\beta$.  Phase separation is seen for $U \lesssim 1$  but not for $U=1.5$; the bosonic Mott transition is expected at $U = 4.3(2)$. 
        Right: The jumps in fermionic occupation number  at $k=k_F$ ({\it i.e.}, the $Z$-factor)  for the same system, which are seen to range from 0.4 to 0.9.     }
\end{figure}

\begin{figure}[!htb]
        \begin{tabular}{ll}
        \includegraphics[width=0.5 \columnwidth]{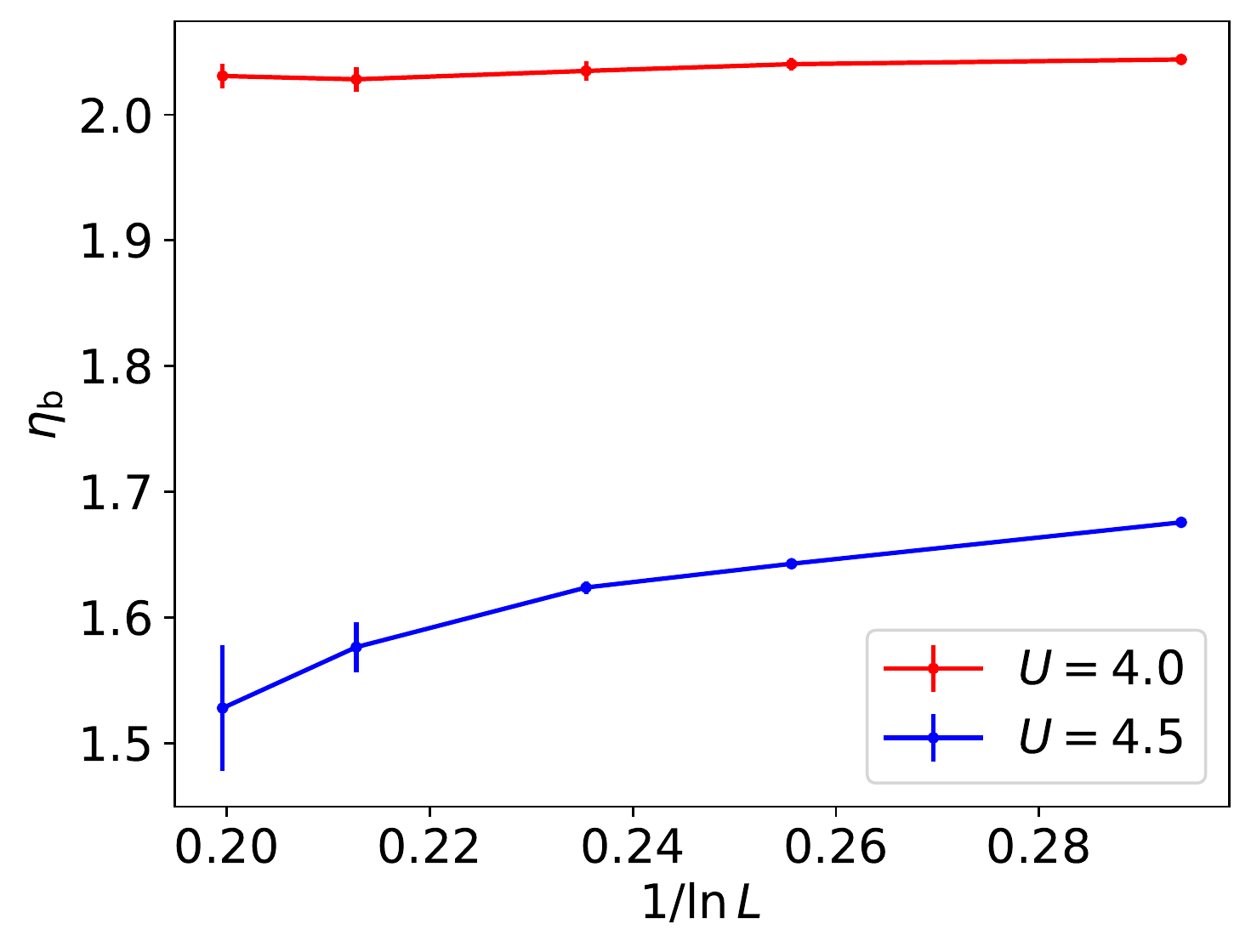} &
        \includegraphics[width=0.5 \columnwidth]{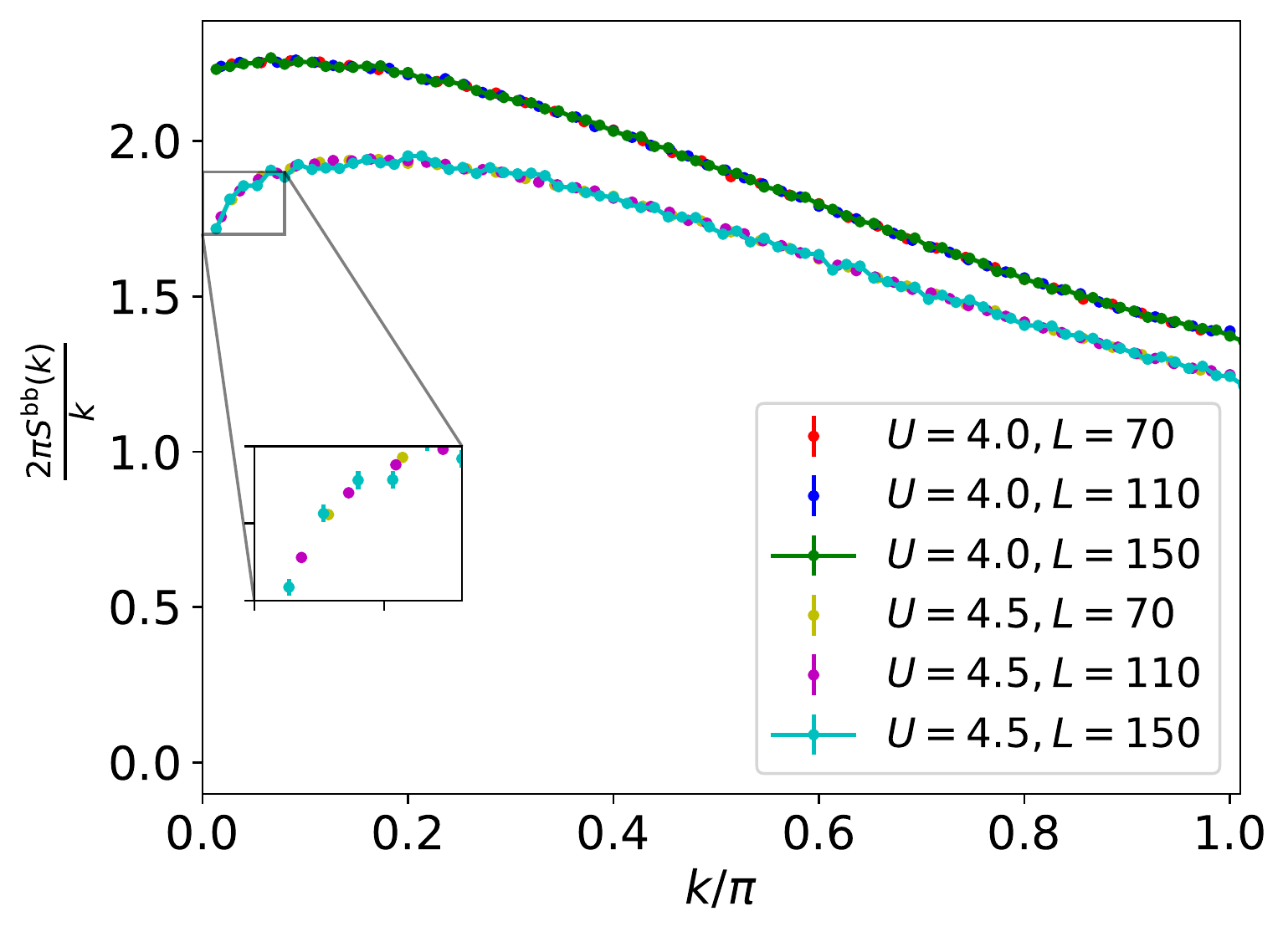} 
        \end{tabular}
        \caption{  \label{fig:V2_sf_bos_transition} Left: The powerlaw $\eta_b$ governing the decay of the bosonic single particle density matrix for $U=4$ and $U=4.5$ and $L=30, 50, 70, 110$ and $150$. On the superfluid side, convergence is reached, on the insulating side, the quantity flows to 0. According to the bosonization theory~\cite{MatheyWang2007}, the transition is expected for a non-universal value slightly below 2.
        Right: Scaled bosonic structure factor close to the superfluid to Mott insulator transition for $U=4$ and $U=4.5$ for $L = 70, 110, 150$. On the superfluid side, $U=4$, the extrapolated value for $k \to 0$ is larger than 2 and the structure factors show convergence in $L$. On the insulating side, $U = 4.5$, the extrapolated value for $k \to 0$ is smaller than 2 and increasing $L$ leads to a stronger curvature for low $k$. The extrapolated value at $k=0$ scales to 0 in the thermodynamic limit. 
                 }
\end{figure}

\begin{figure}[!htb]
        \begin{tabular}{ll}
        \includegraphics[width=0.5 \columnwidth]{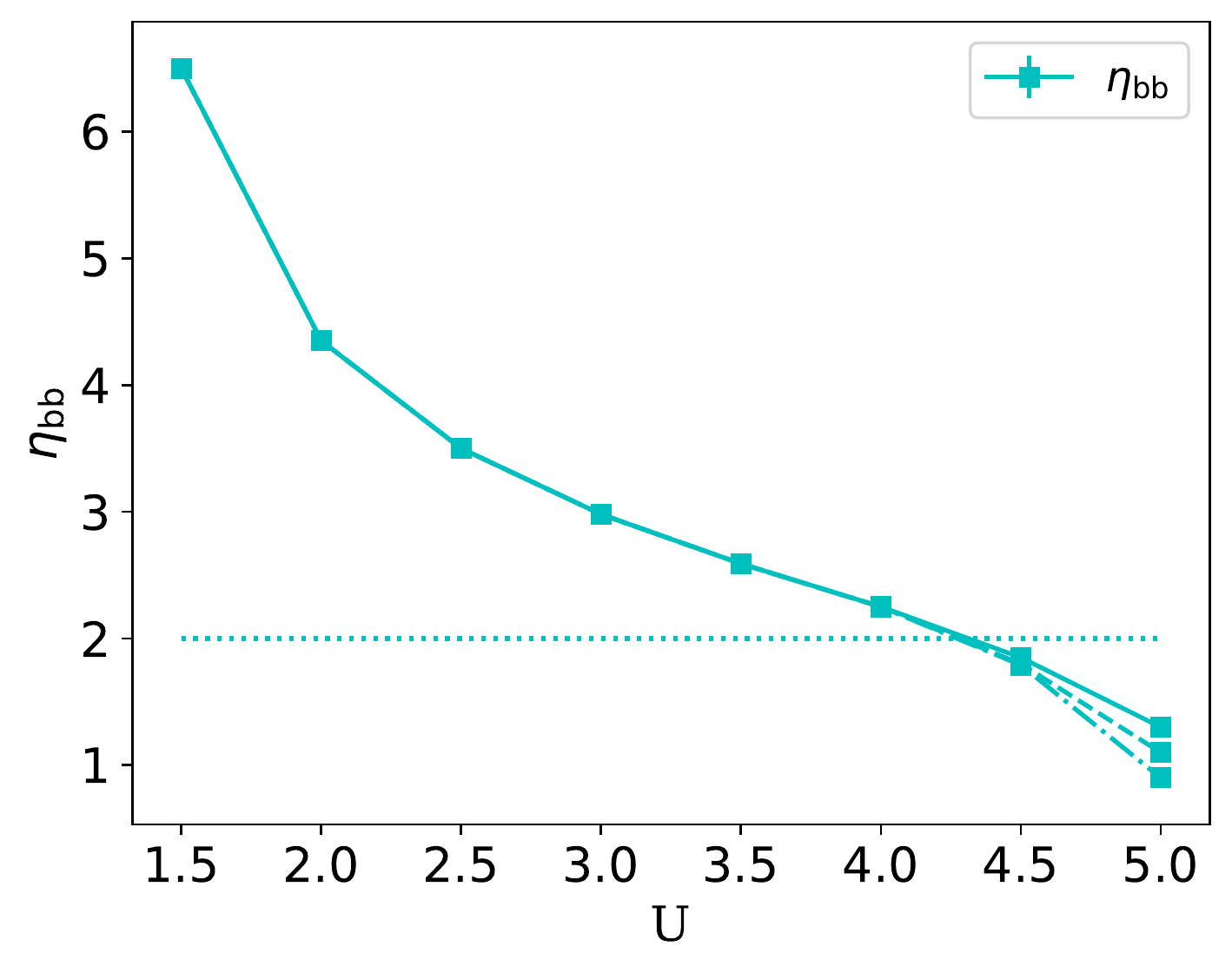} &
        \includegraphics[width=0.5 \columnwidth]{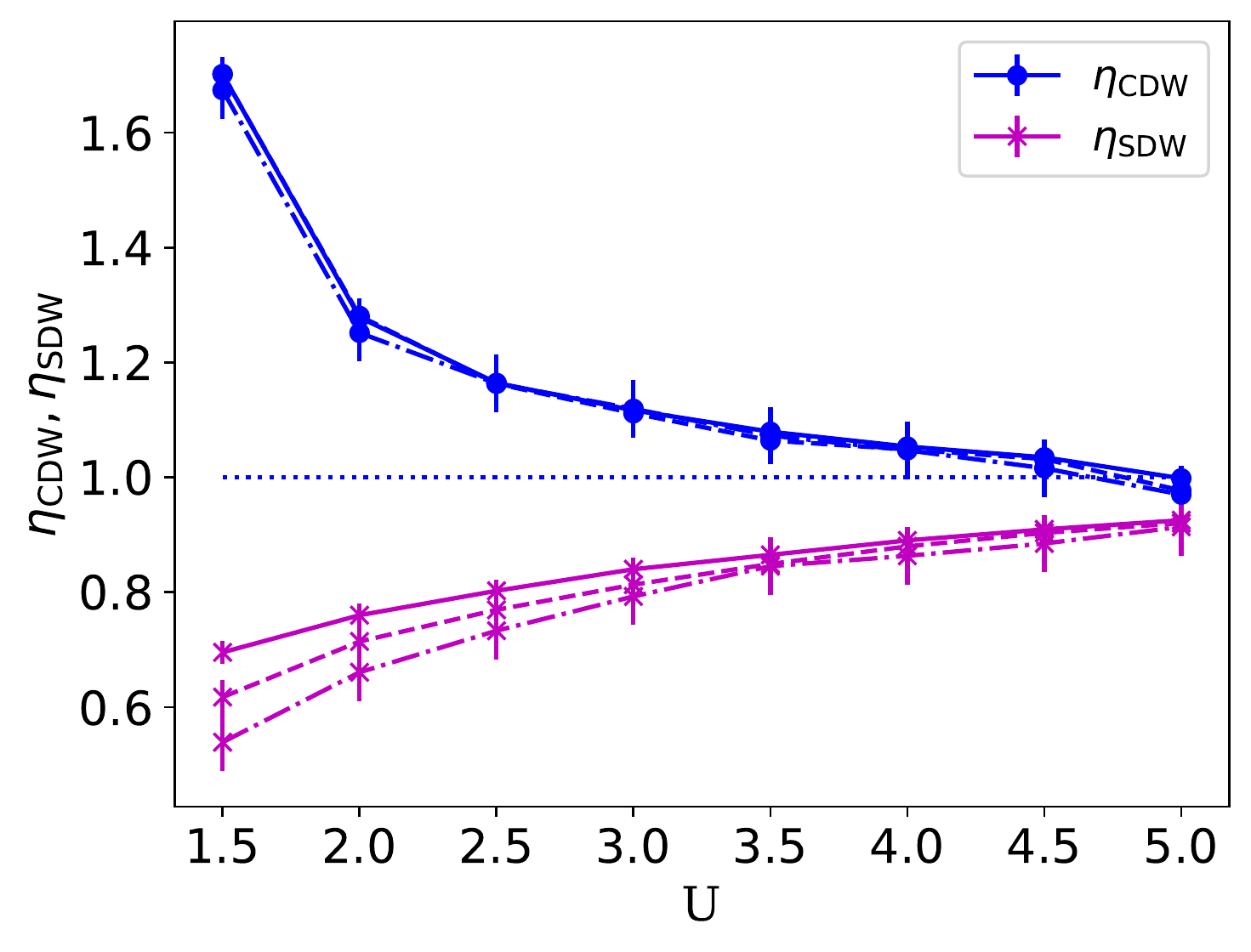}
        \end{tabular}
        \caption{Left: The parameter $\eta_{\rm bb}$ as obtained from the bosonic static structure factor, $\lim_{k \to 0} \frac{ 2 \pi S^{\rm bb}(k)}{k}$. Bosonization predicts a transition from a Luttinger liquid to a Mott insulator when $\eta_{\rm bb} = 2$ (indicated by the dotted line).
        Right: The parameters $\eta_{\rm CDW}$ (blue) and $\eta_{\rm SDW}$ (purple) for $V = 2$ and $L=30, 50, 70$ (top to bottom) obtained from their respective static structure factors.  The dotted line is the value of $\eta_{\rm CDW}$ for free fermions. The parameter $\eta_{\rm SDW}$  should scale to 0 owing to the presence of a spin gap for attractive induced interations.
                \label{fig:V2_Lutt}}
\end{figure}

\begin{figure}[!htb]
        \begin{tabular}{ll}
        \includegraphics[width=0.5 \columnwidth]{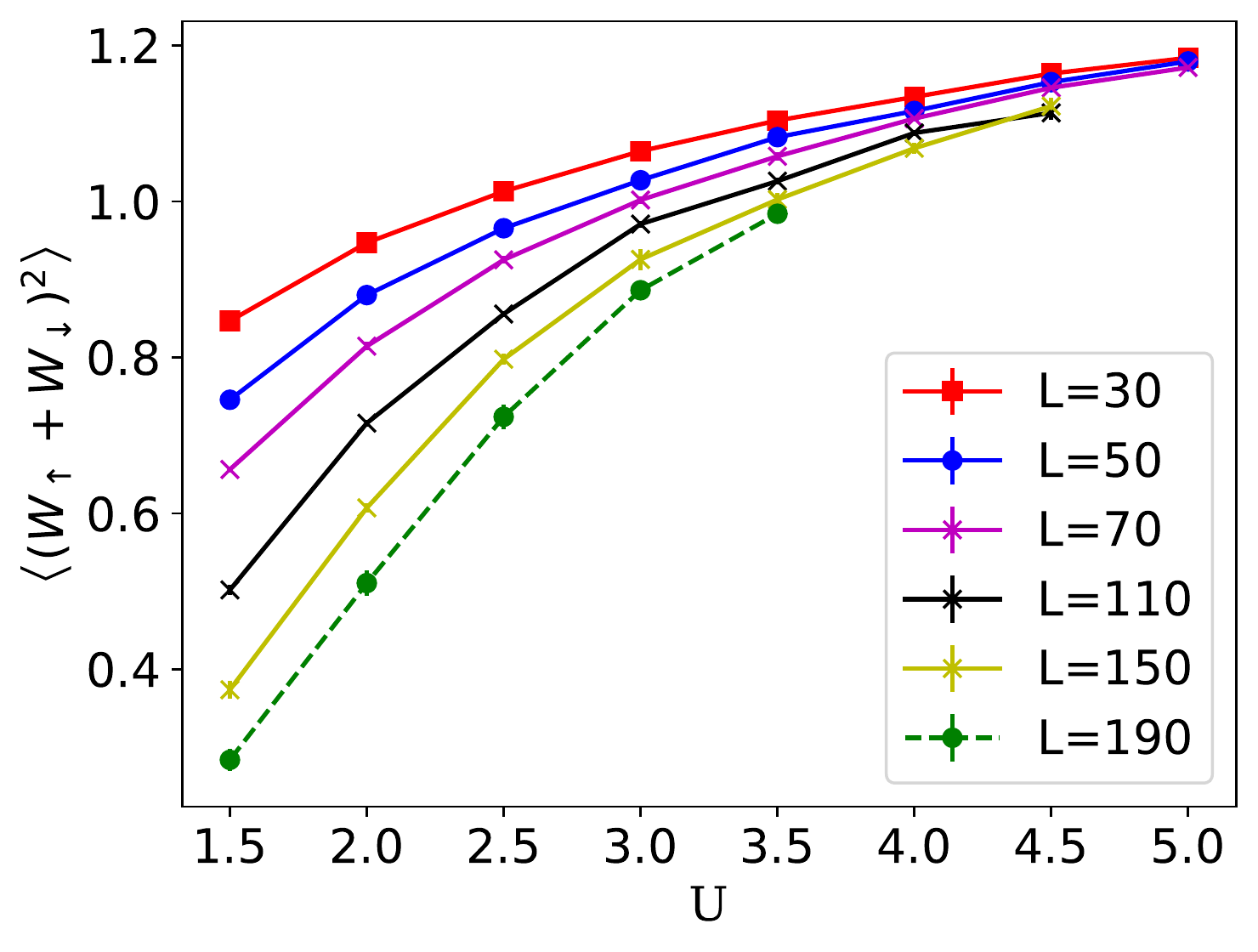} &
        \includegraphics[width=0.5 \columnwidth]{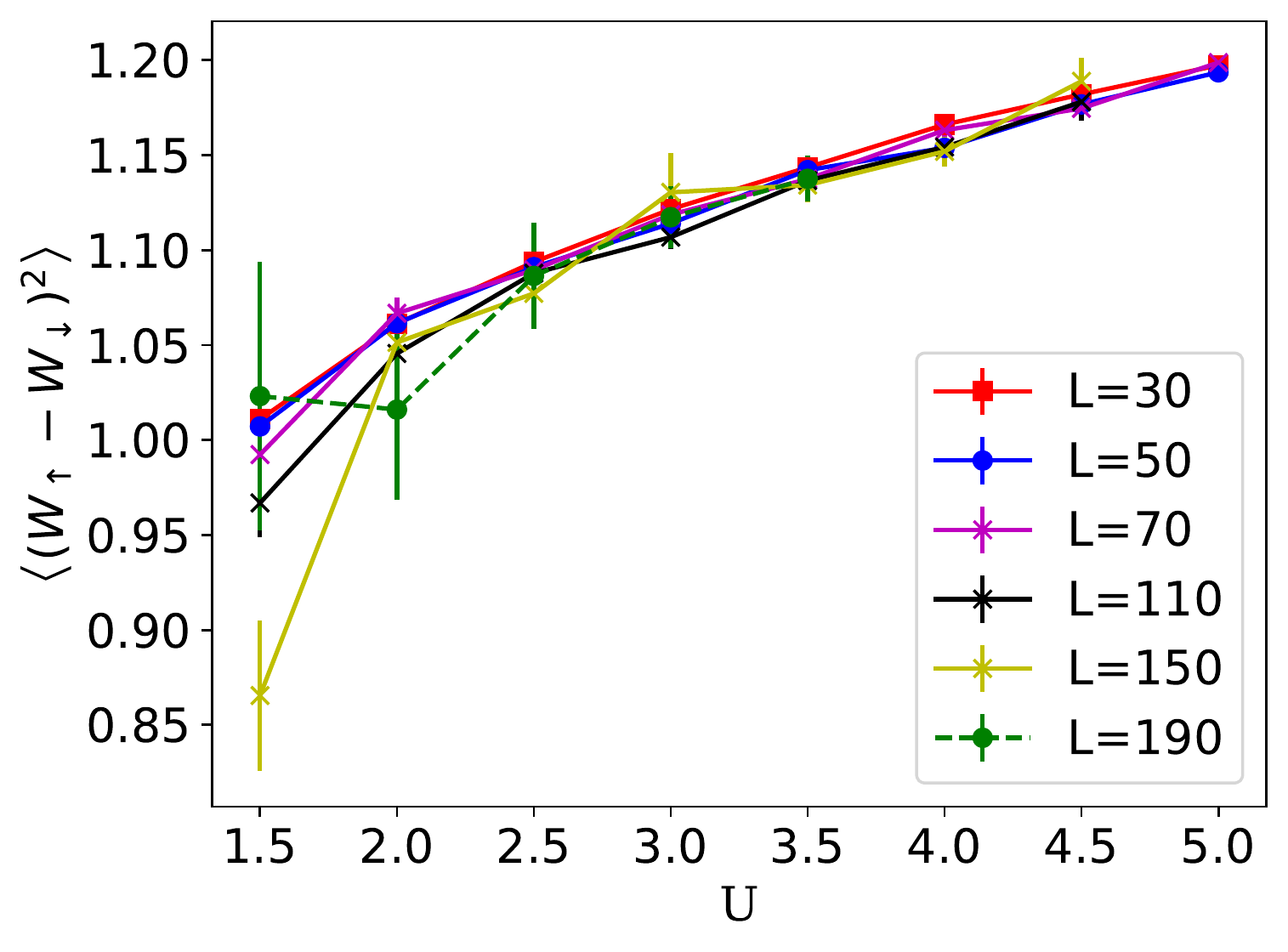}
        \end{tabular}
        \caption{Left: Winding numbers squared in the counter-flow  channel for $V = 2$; Right: Winding numbers squared in the pair-flow channel for $V= 2$. The convergence in the pair flow channel is seen as well as the ongoing renormalization flow in the counter-flow channel, suggesting (at least mesoscopic) pairing fluctuations. Whether they persist in the thermodynamic limit (cf. the right panel of Fig.~\ref{fig:V2_app}, and both panels of Fig.~\ref{fig:V2_U2_FSS} ) cannot be inferred, unless perhaps for the smallest values of $U$. 
         \label{fig:V2_spf_scf} }
\end{figure}

\begin{figure}[!htb]
        \begin{tabular}{ll}
        \includegraphics[width=0.5 \columnwidth]{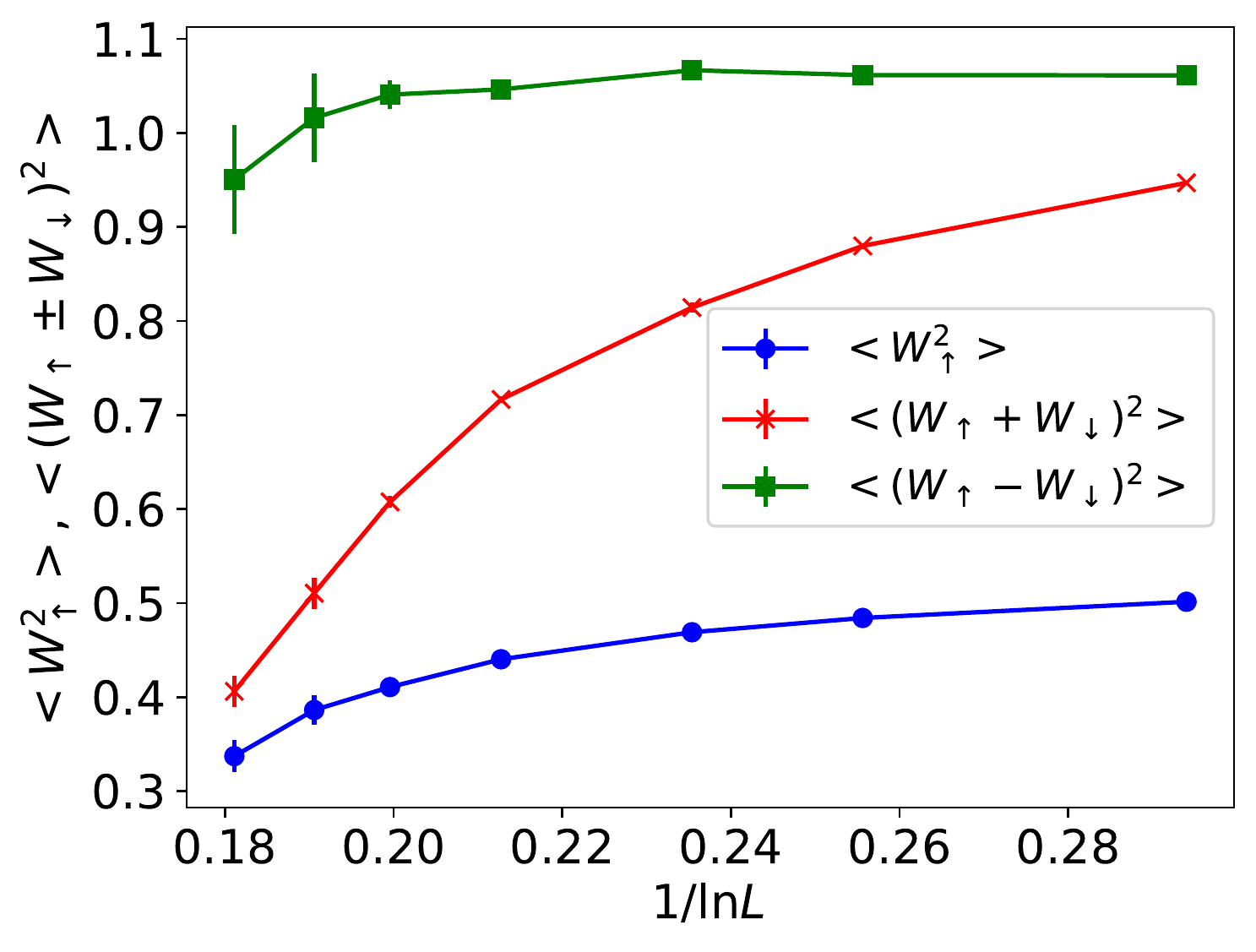} &
         \includegraphics[width=0.5 \columnwidth]{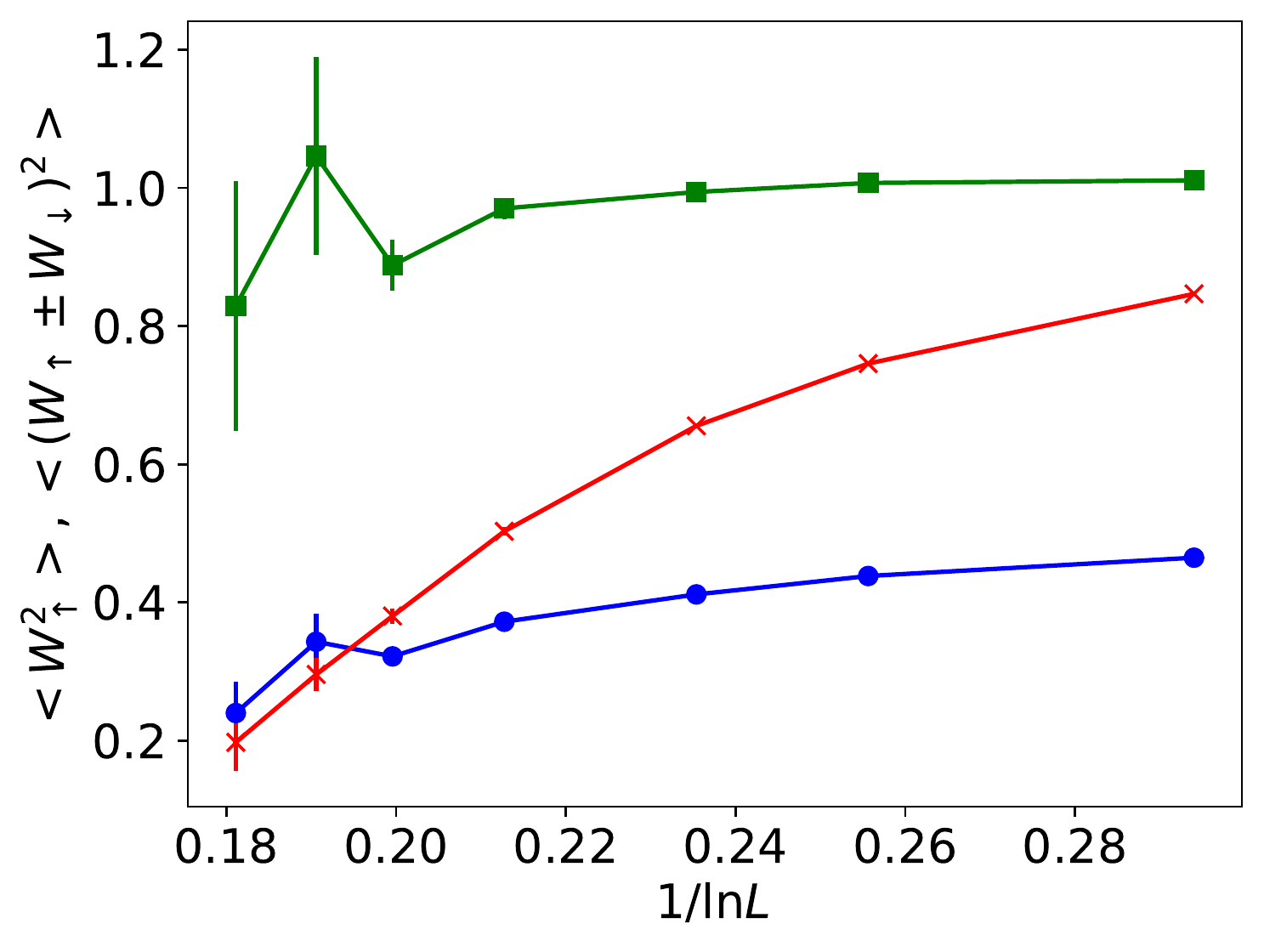} 
        \end{tabular}
        \caption{Left: Winding number squared for the spin-up particles (blue), in the counter-flow channel (red), and in the pair-flow channel (green) as a function of inverse system size with inverse temperature $\beta = L$ and system parameters $V=2$ and $U = 2$. The counter-flow channel shows a strong  and unstoppable renormalization to zero whereas the pair-flow channel stays finite and will approach 4 times the value of the winding squared of the spin-up particles in case of a thermodynamic super-flow. Note that we still have a very sizeable $Z(L=150) \approx 0.57$ (cf. the right panel of Fig.~\ref{fig:V2_app}).
      Right: Winding number squared for the spin-up particles (blue), in the counter-flow channel (red), and in the pair-flow channel (green) as a function of inverse system size with inverse temperature $\beta = L$ and system parameters $V=2$ and $U = 1.5$.
                   \label{fig:V2_U2_FSS} }
\end{figure}

For $V=2$ the renormalization flow in the fermionic channel is stronger than for $V=1$ but it is still weak, as can be seen in the right panel of Fig.~\ref{fig:V2_app}. It is therefore not possible to determine the fate of the fermionic phase diagram with certainty. The bosonic Mott transition can, as for $V=1$ be analyzed separately from the fermions due to the different length scales involved. If we apply the same formula $U^{\rm crit} ( V= 2) =  U^{\rm crit} ( V= 0) + \frac{1}{\pi} V^2$, knowing that it is beyond its validity regime $V \ll 1$, then we find $ U^{\rm crit} ( V= 2)  \approx 4.5$,
which is in fair agreement with the left panel of Fig.~\ref{fig:V2_app}. The flow for $U=4.5$ is indeed logarithmic, indicating proximity to the transition, but $U=4.5$ is in fact already on the insulating side. Looking again at the right panel of Fig.~\ref{fig:V2_app}, we see that $Z \sim 0.8$ on the scales of relevant flow in the bosonic channel; hence $4.3(2)$ is more likely for the transition point.
 
The powerlaw of the equal-time bosonic density matrix $\eta_b(V=2, U = 4) = 2.02(1)$ appears to be nearly constant with $L$ whereas it is flowing to zero with $L$ for $U=4.5$, as is shown in the left panel of Fig.~\ref{fig:V2_sf_bos_transition}.
The transition can also be studied from the scaled static structure factor, $\lim_{k \to 0} \frac{2 \pi S^{\rm bb}(k)}{k}$ shown in the right panel of Fig.~\ref{fig:V2_sf_bos_transition}, from which we extract $\eta_{\rm bb}$ in the limit $k \to 0$. Bosonization~\cite{MatheyWang2007} predicts the transition at $\eta_{\rm bb} = 2$, which is reflected in the data, see also the left panel of Fig.~\ref{fig:V2_Lutt}.

Fig.~\ref{fig:V2_spf_scf} suggests that there is a strong tendency towards pair-flow on mesoscopic system sizes. It is however impossible to monitor the flow for big enough system sizes, even where the lowest $Z$-factors are found, {\it i.e.}, for the lowest values of $U$ as can be seen in Fig.~\ref{fig:V2_U2_FSS} for $U=2$ and $U=1.5$. 
Whereas no CDW order can be seen for the largest $L$ (see Fig.~\ref{fig:V2_Lutt}), the extrapolation of the pairflow density to the thermodynamic limit is nevertheless dangerous, as shown in Fig.~\ref{fig:V2_U2_FSS} for $U=2$ and $U=1.5$ because the value of the superflow in the counterflow channel is still a quarter (half) of the one in the pair channel for $U=2 (U=1.5)$ for $L=150$. The superfluid density of the up fermions is also seen to be going down with $L$. In the pair flow phase this value should be exactly $1/4$ of the superflow density. We hence expect as a rule of thumb that $\langle W_{\uparrow}^2 \rangle$ should approach $\sim 0.26$ for $U=2$, which is one quarter of the value seen in the pairflow channel for the lowest system sizes and which appears roughly stable on the system sizes that we can simulate. Hence, the reduction seen in $\langle W_{\uparrow}^2 \rangle$ with $L$ is not necessarily a sign of insulating behavior at astronomically large $L$, but it cannot completely be ruled out either. The right panel of Fig.~\ref{fig:V2_Lutt} also shows that the spin gap is not fully developed on the length scales that we can simulate. Correlations in the charge sector decay algebraically, and slightly more rapid than for free fermions as long as the bosons are superfluid, but the value of $\eta_{\rm CDW}$ becomes less than 1 when the bosons turn insulating. 
Note that it is common for one-dimensional superconducting systems to show subdominant CDW algebraic correlations~\cite{Giamarchi2007}. The rather strong charge correlations can be understood based on the arguments presented in Sec.~\ref{sec:weak_coupling}, and it further seems reasonable that superfluid bosons suppress such correlations as uniformity in the system is beneficial for phase rigidity.


\subsection{Scan of phase diagram at $V=3$}

\begin{figure}[!htb]
        \begin{tabular}{ll}
        \includegraphics[width=0.5 \columnwidth]{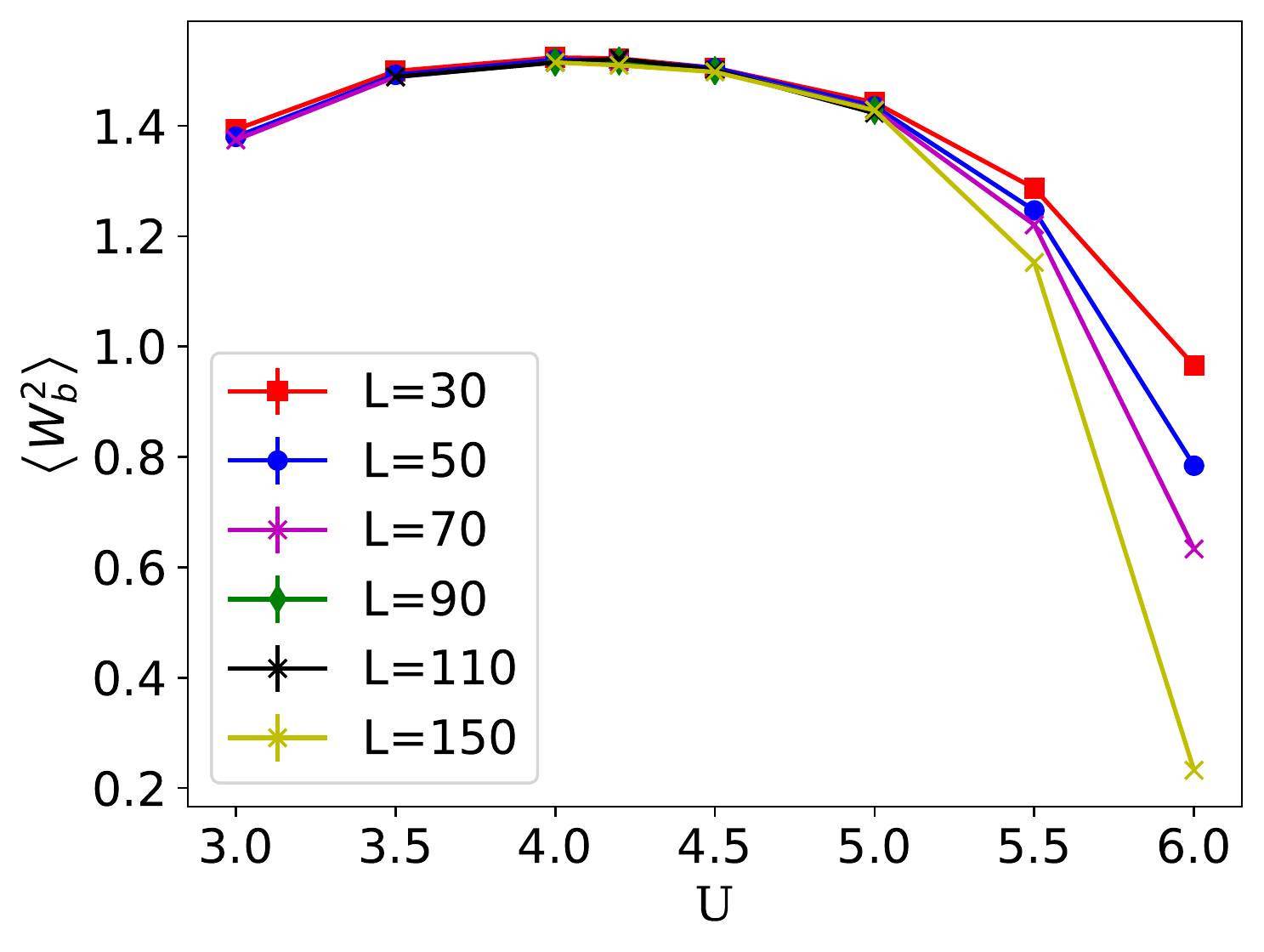} &
        \includegraphics[width=0.5 \columnwidth]{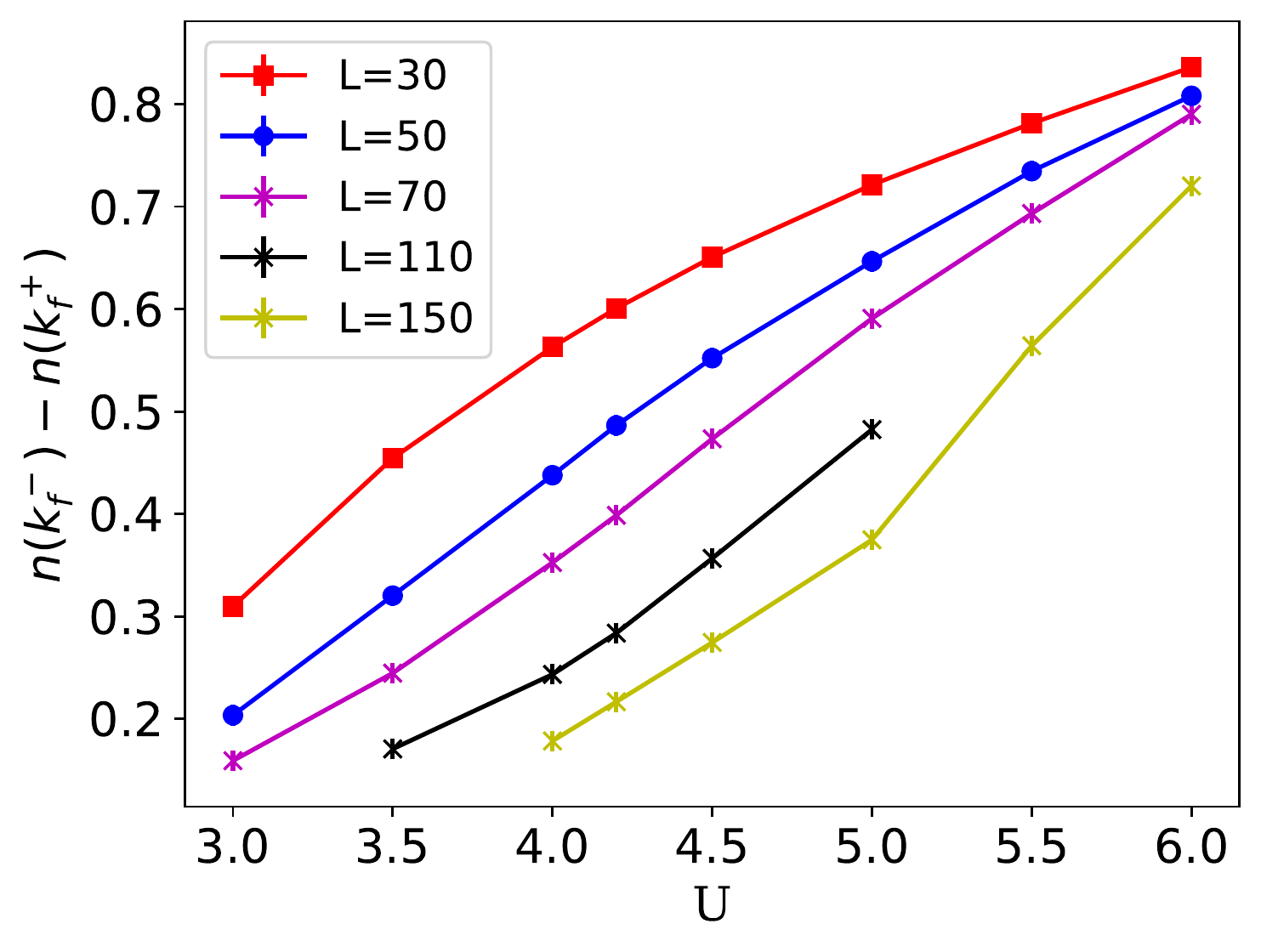}
        \end{tabular}
        \caption{Left: Bosonic winding number squared for $V= 3$.  Phase separation is seen for smaller values than $U \lesssim 3$; the bosons turn insulating for $U \gtrsim 5.3(2)$; Right : Jumps in the fermionic occupation number at $k_F$, showing a strong renormalization when the bosons are superfluid and a strong tendency towards free fermions deep in the bosonic insulating regime.
        \label{fig:V3_boson_jump}}

\end{figure}

\begin{figure}[!htb]
        \begin{tabular}{ll}
        \includegraphics[width=0.5 \columnwidth]{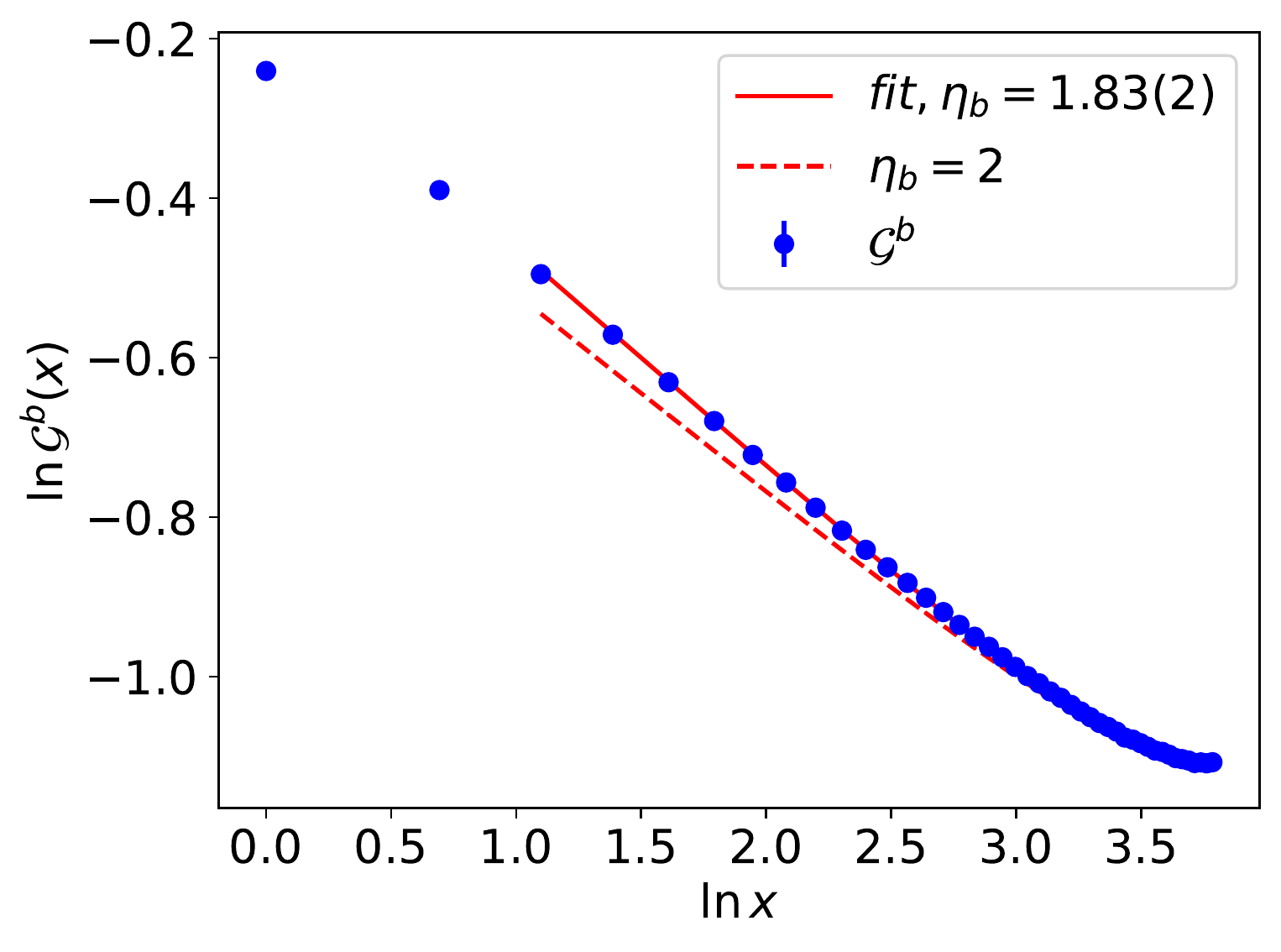} &
        \includegraphics[width=0.5 \columnwidth]{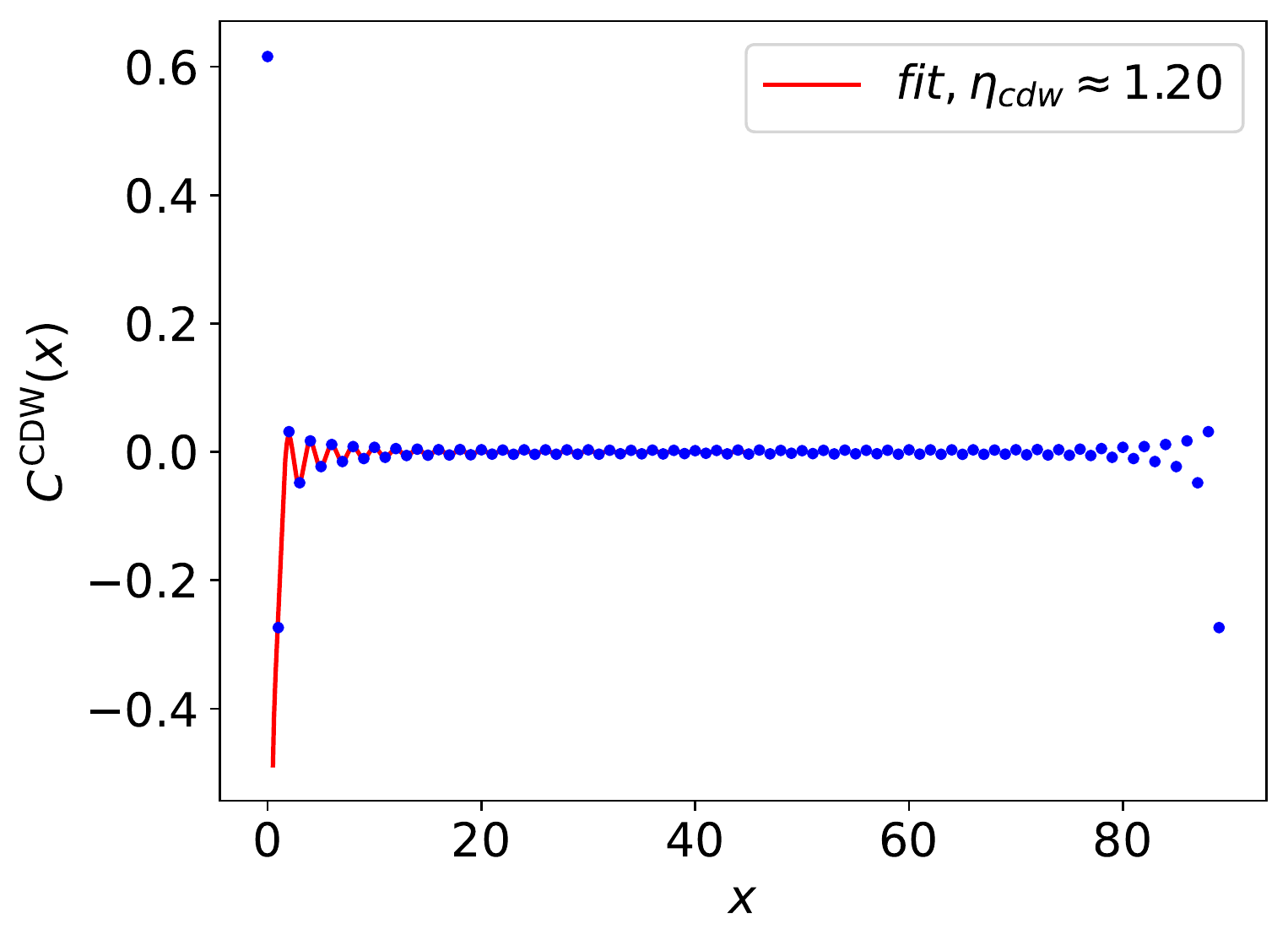}
        \end{tabular}
        \caption{Left: Single-particle bosonic density matrix showing a marginal superfluid for $V=3, U=5, L=\beta=90$. Right: CDW correlator for the same system. No signs of true long-range order are seen; the algebraic decay is faster than the one of free fermions and the amplitude of the correlator small. 
        \label{fig:V3_U5_L90_dm} }
\end{figure}

\begin{figure}[!htb]
        \begin{tabular}{ll}
        \includegraphics[width=0.5 \columnwidth]{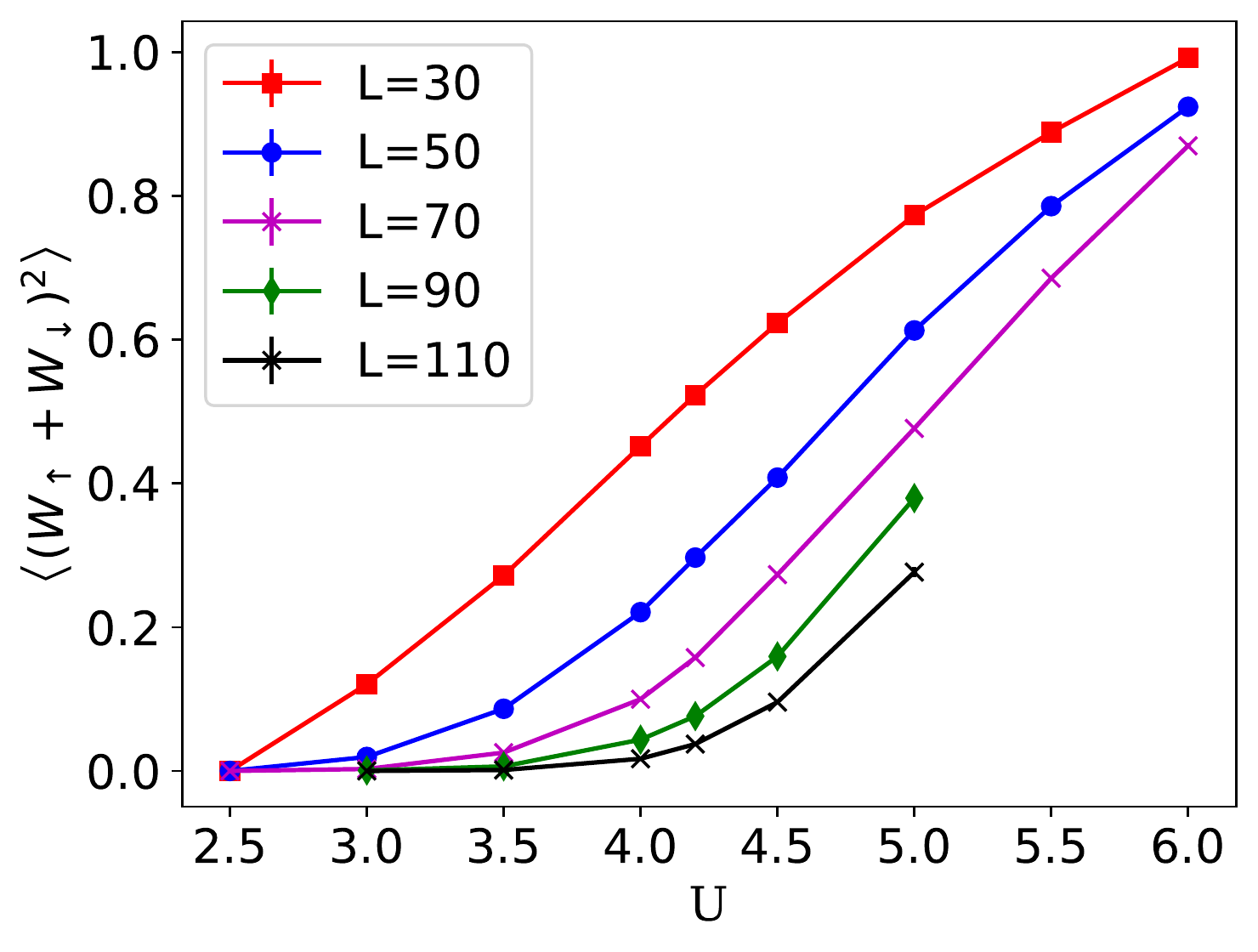} &
        \includegraphics[width=0.5 \columnwidth]{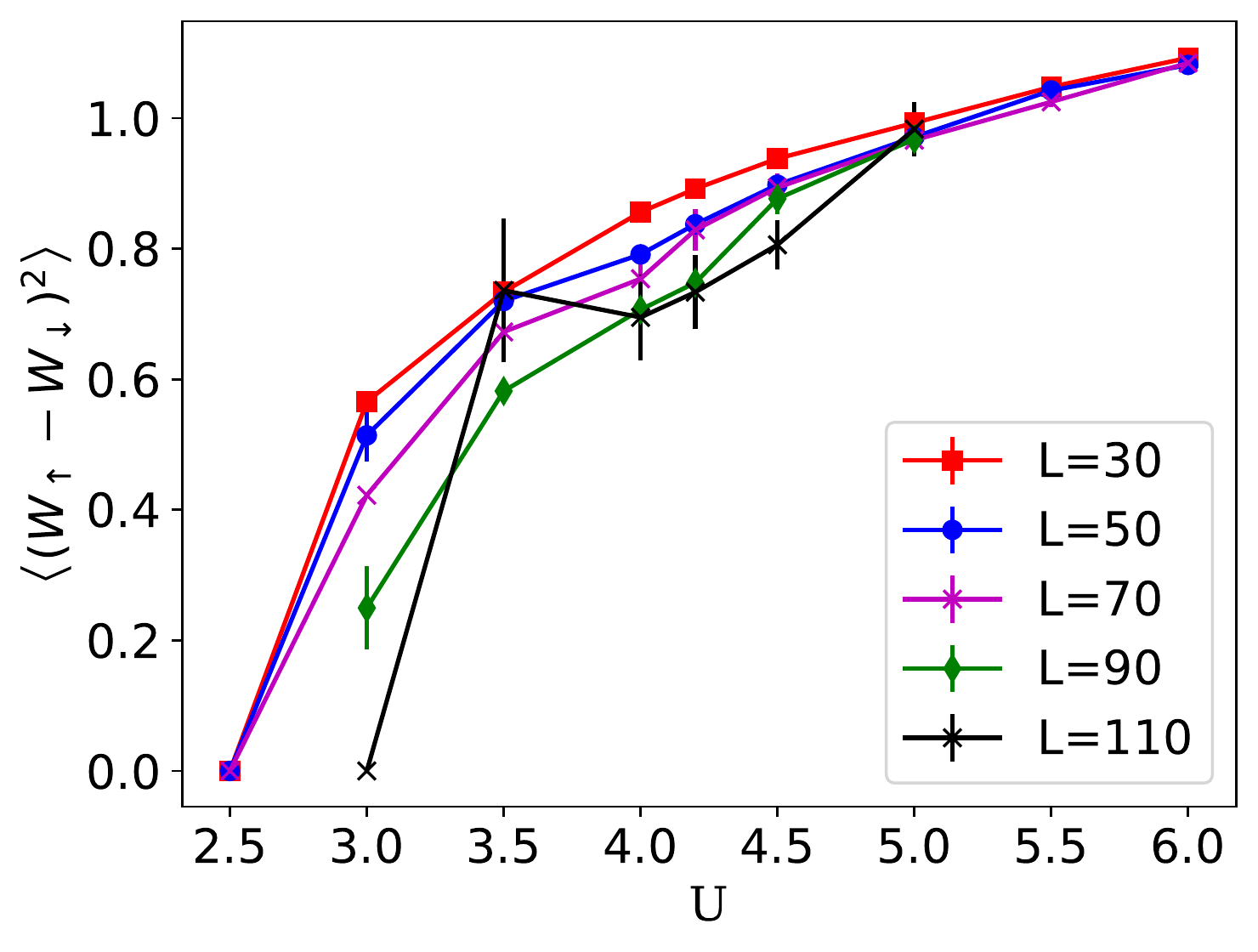}
        \end{tabular}
        \caption{Left: Winding number squared in the counter-flow channel at $V= 3$; Right: Same for the pair flow channel. Despite the noise, the data are suggestive of stable pair flow and it can be reached on length scales $L \sim 100$. 
          \label{fig:V3_spf_scf} }
\end{figure}

\begin{figure}[!htb]
        \begin{tabular}{ll}
        \includegraphics[width=0.5 \columnwidth]{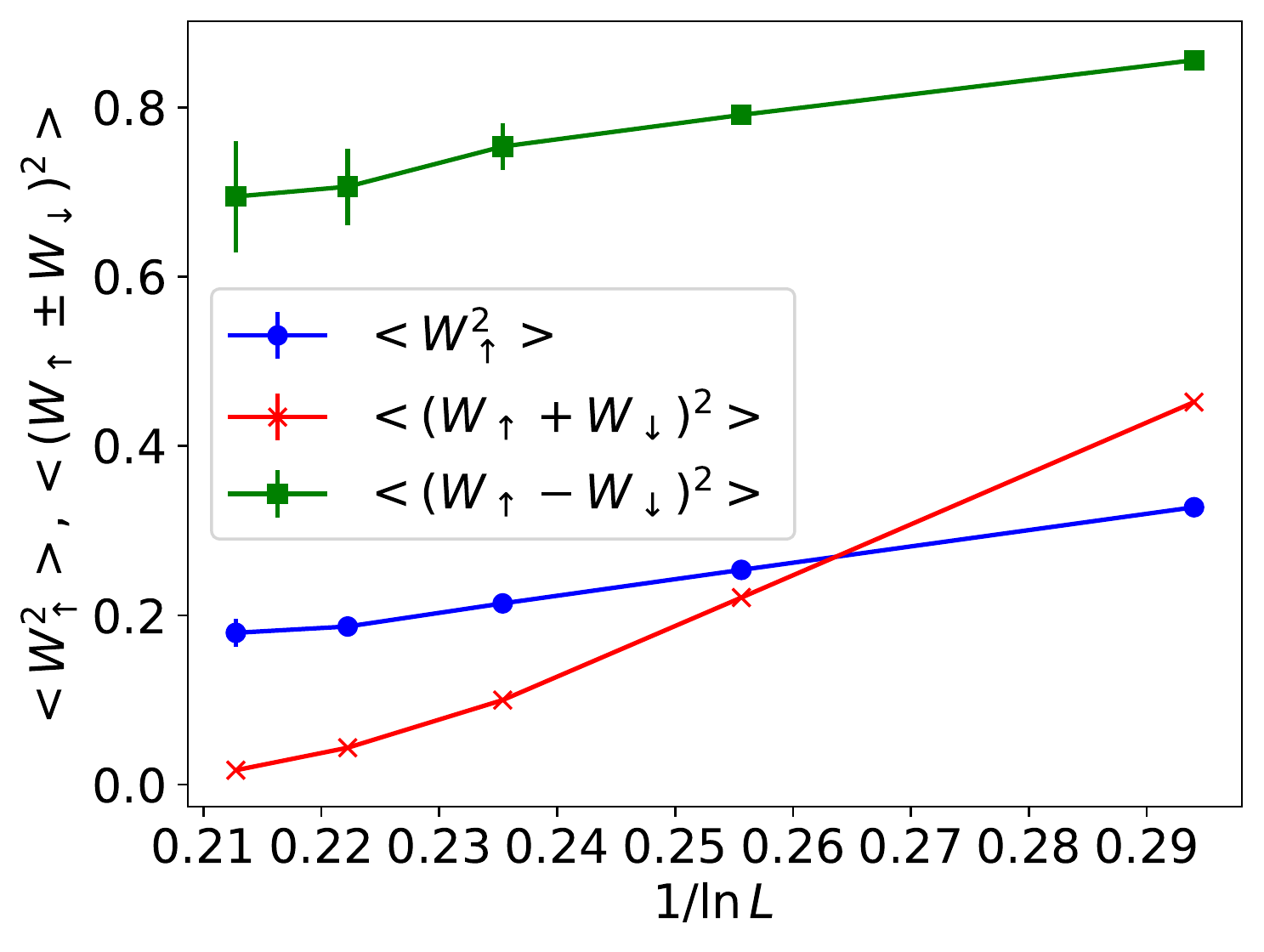} &
        \includegraphics[width=0.5 \columnwidth]{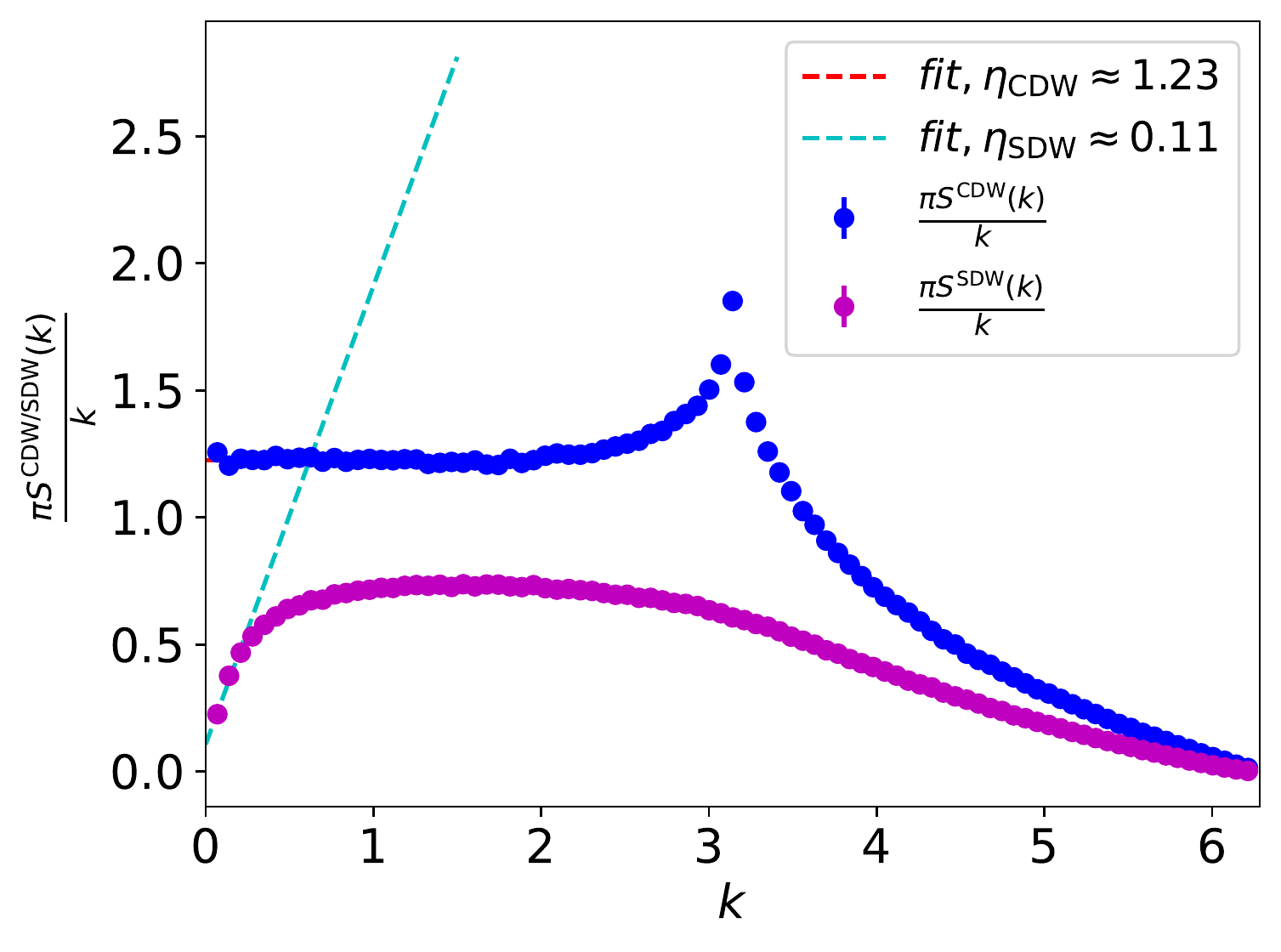}
        \end{tabular} 
        \caption{Left: Winding number squared for the spin-up particles (blue), in the counter-flow channel (red), and in the pair-flow channel (green) as a function of inverse system size with inverse temperature $\beta = L$ and system parameters $V=3$ and $U = 4$. Stable superflow is likely. 
        Right: Rescaled static structure factors for $V=3, U=4, L=\beta=90$. The low momentum analysis yields $\eta_{\rm CDW} = 1.22(1)$, in good agreement with Fig.~\ref{fig:V3_U5_L90_dm}. The spin gap is almost completely developed. The peak that develops at $k = \pi$ is present but weak. Only larger system sizes beyond reach can establish whether the peak value grows linearly with system size.
              \label{fig:V3_U4_FSS} 
              }
\end{figure}


In Fig.~\ref{fig:V3_boson_jump} we see that the transition from a bosonic superfluid to an insulator takes place for values of $U$ between $U=5$ and $U=5.5$. Phase separation is found for $U \lesssim 3$. In the bosonic superfluid regime the $Z$-factors for the fermions range from 0.1 to 0.5 at $L=\beta=110$. Surprisingly, the powerlaw of the bosonic density matrix at $U=5$ is such that $\eta_b =  1.81(2) < 2$ (which was given an explanation in Sec.~\ref{sec:bosonization}) while the system stays uniform, see Fig.~\ref{fig:V3_U5_L90_dm}. For $U=4$ this powerlaw was just $\eta_b = 2.01(5)$. The values $\eta_{\rm bb}$, extracted as before from the scaled bosonic static structure factor, are $\eta_{\rm bb} = 2.20(5) $ for $U=5, L=70$, and $\eta_{\rm bb} = 1.75(5) $ for $U=5.5, L=70$ and are compatible with a universal jump at $\eta_{\rm bb} = 2$ as predicted by bosonization~\cite{MatheyWang2007}. \\

From the winding numbers squared in the pair-flow and counter-flow channels, shown in Fig.~\ref{fig:V3_spf_scf}, we infer a strong tendency towards superconducting pair-flow correlations that are almost fully developed (cf. Fig.~\ref{fig:wind_superconductor} and the left panel of  Fig.~\ref{fig:V3_U4_FSS}). However, as can especially be appreciated from the data in the pair-flow channel in the range $3 \le U \le 5$ (see the right panel of Fig.~\ref{fig:V3_spf_scf}), the system is very hard to thermalize and simulate already for $L>50$. Autocorrelation times are extremely large, ranging from $10^3$ to $10^5$ for $L=70$ and rapidly increasing with $L$, which we attribute to the frustration caused by the induced interactions and the competition between paired and insulating phases.
We did not find evidence of gapped CDW order  anywhere in the phase diagram for $V=3$. The parameter $\eta_{\rm CDW}$ is above 1 everywhere in the bosonic superfluid regime: We find $\eta_{\rm CDW}(U=5, L=70) = 1.10(5)$, $\eta_{\rm CDW}(U=5.5, L=70) = 1.00(5)$, and $\eta_{\rm CDW}(U=6, L=70) = 0.93(5)$. 

Based on the system sizes that are accessible to the simulations, the system is for $V=3$ (i) in the phase separation regime for $U \lesssim 3$, (ii) a bosonic superfluid and uniform $s-$wave superconductor for $3 \lesssim U < 5.3(2)$, (iii) a bosonic Mott insulator for larger values of $U$. The fermions remain superconducting when $U$ remains close to the critical value of the bosonic Mott transition, and then cross over towards free fermions when the gap $\Delta \approx E_F$.  As $\eta_{\rm CDW} < 1$ in this regime a tendency towards algebraic charge density wave order is also observed.

\subsection{Scan of phase diagram at $V=4$}

\begin{figure}[!htb]
        \begin{tabular}{ll}
        \includegraphics[width=0.5 \columnwidth]{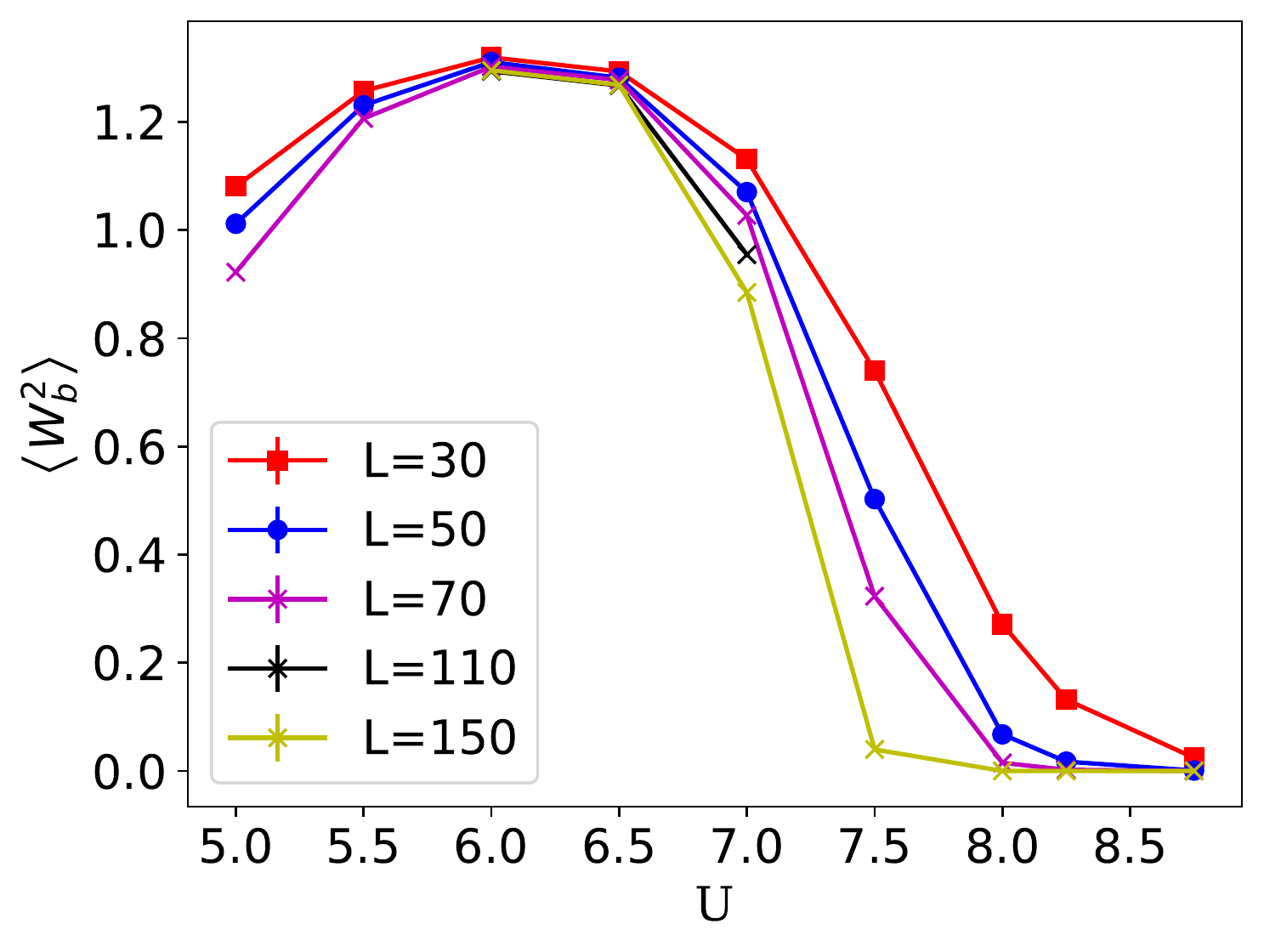} &
        \includegraphics[width=0.5 \columnwidth]{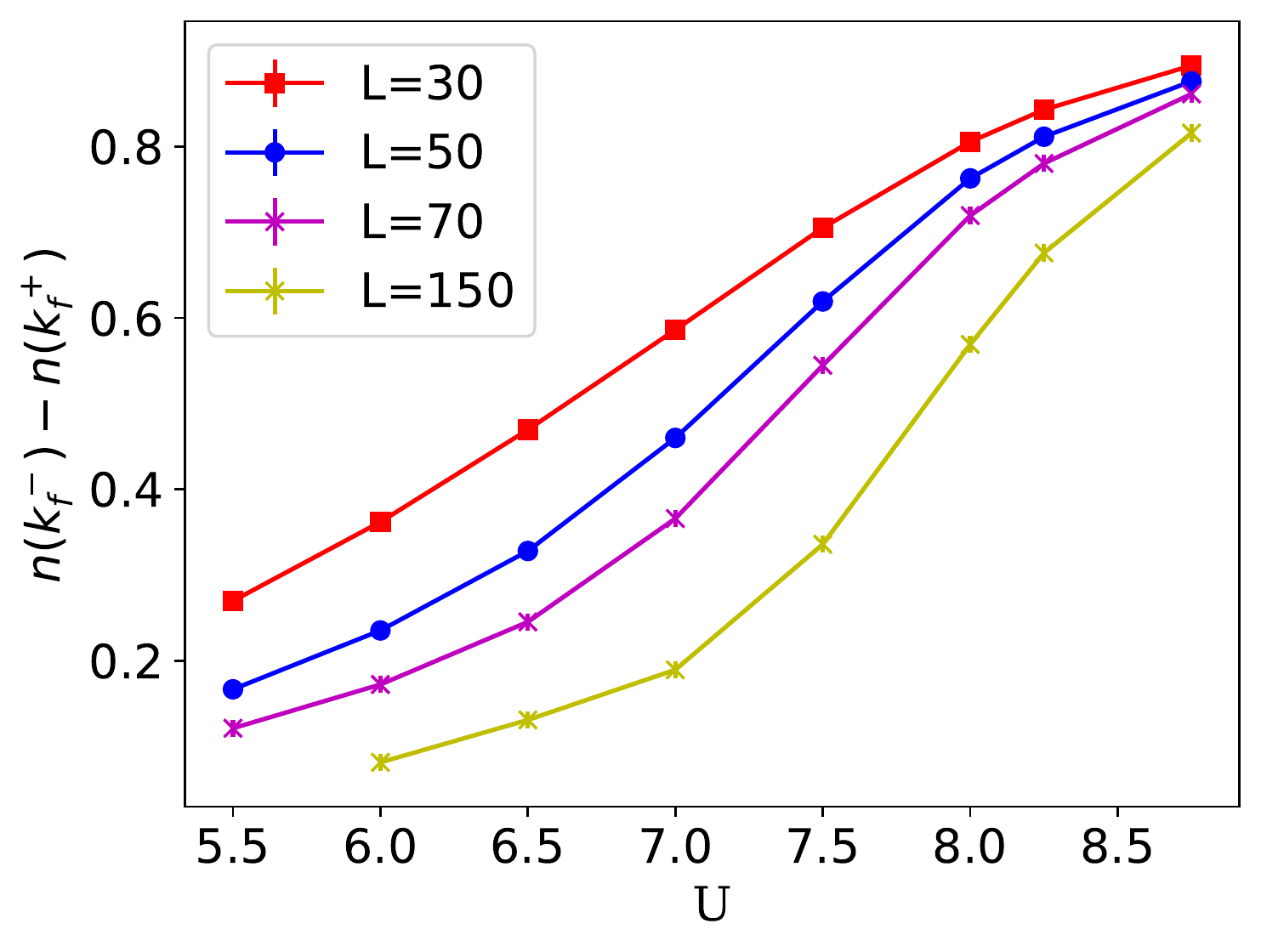}
        \end{tabular}
        \caption{Left: Bosonic winding number squared for $V = 4$. The bosons are (marginally) superfluid in the range $U = 5.5$ to $U=6.5$; for $U=7$ the flow indicates insulating behavior. Right : jumps in occupation number at $k_F$. As we see below, the low values of $Z(L) \le 0.2$ allow us to see strong competition between competing instabilities on the length scales accessible to our simulations.  
        \label{fig:V4_boson_jump}}
\end{figure}

\begin{figure}[!htb]
        \begin{tabular}{ll}
        \includegraphics[width=0.5 \columnwidth]{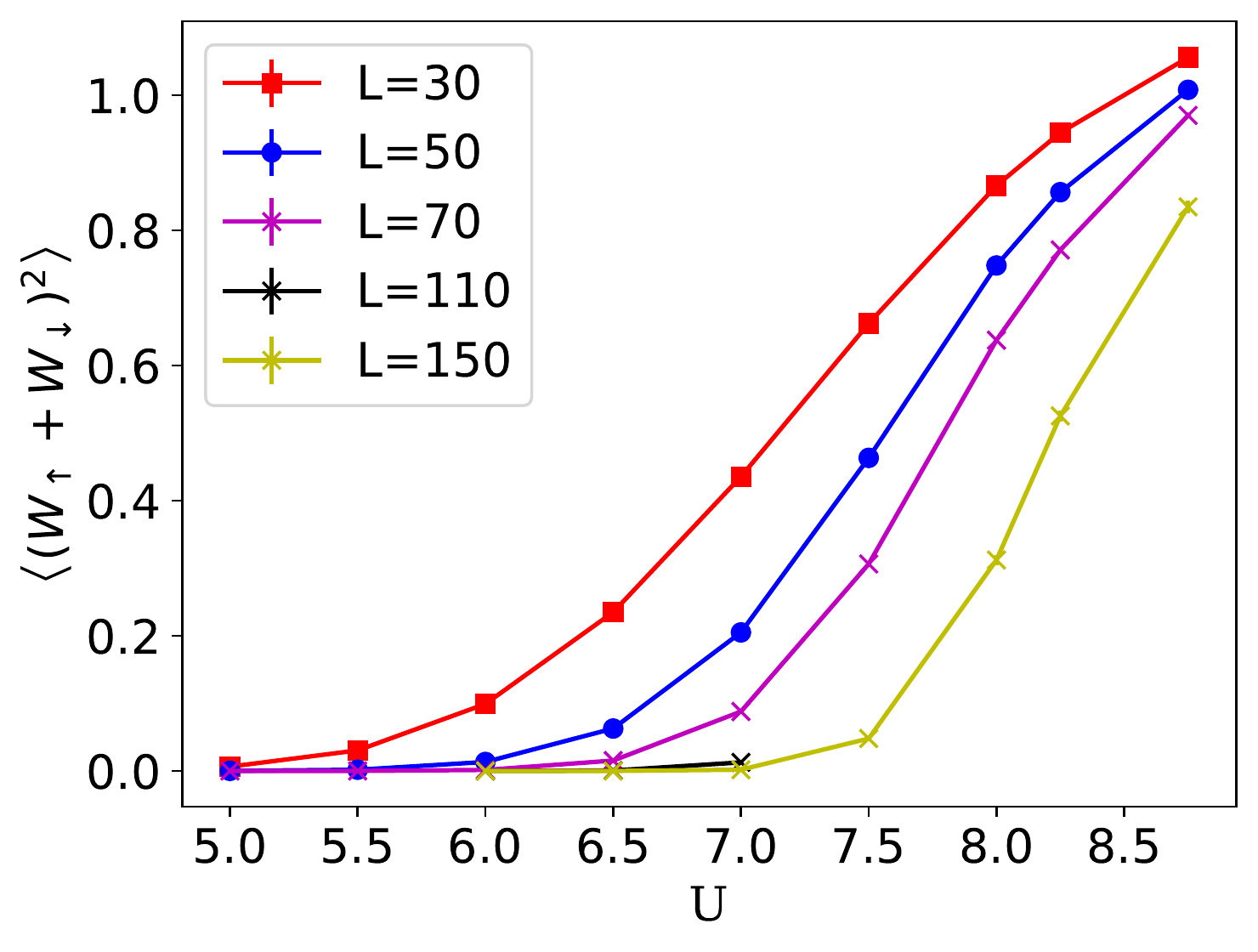} &
        \includegraphics[width=0.5 \columnwidth]{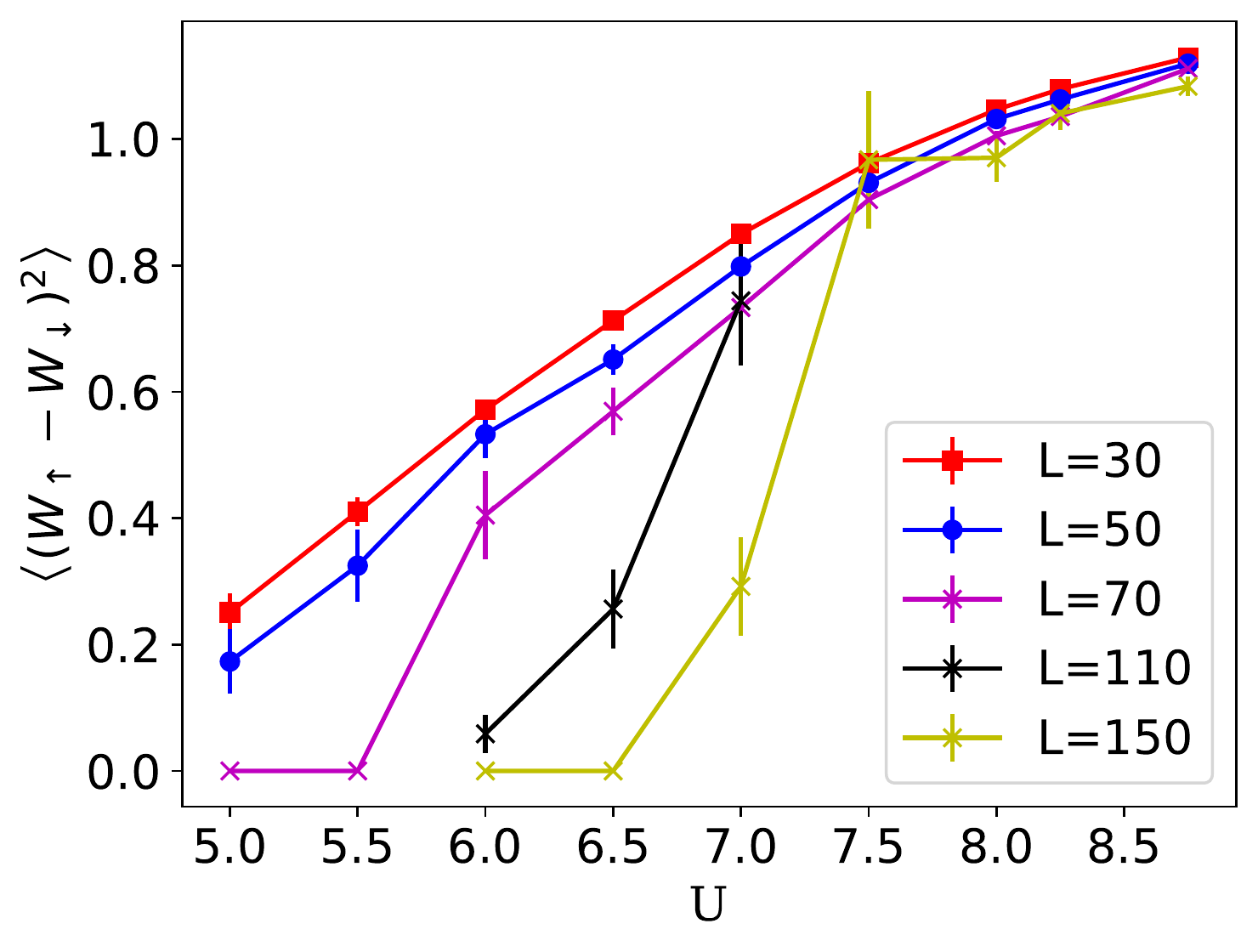}
        \end{tabular}
        \caption{Left: Winding numbers squared in the counterflow channel for $V = 4$. It shows the typical renormalization with system size to 0 that we have seen before for lower values of $V$, but here we reach (almost) 0 on accessible system sizes for all values of $U$ till the strong coupling estimate, $U \lesssim 2V$. Right: Winding numbers squared in the pairflow channel for $V = 4$.  Whereas the system sizes $L=30$ and $L=50$ can be simulated rather easily, this is not the case for bigger system sizes, but the data suggests that in the thermodynamic limit there will be no pair superflow for $U < 7.0(5)$. 
                \label{fig:V4_pair_scf}}
\end{figure}

\begin{figure}[!htb]
        \begin{tabular}{ll}
        \includegraphics[width=0.5 \columnwidth]{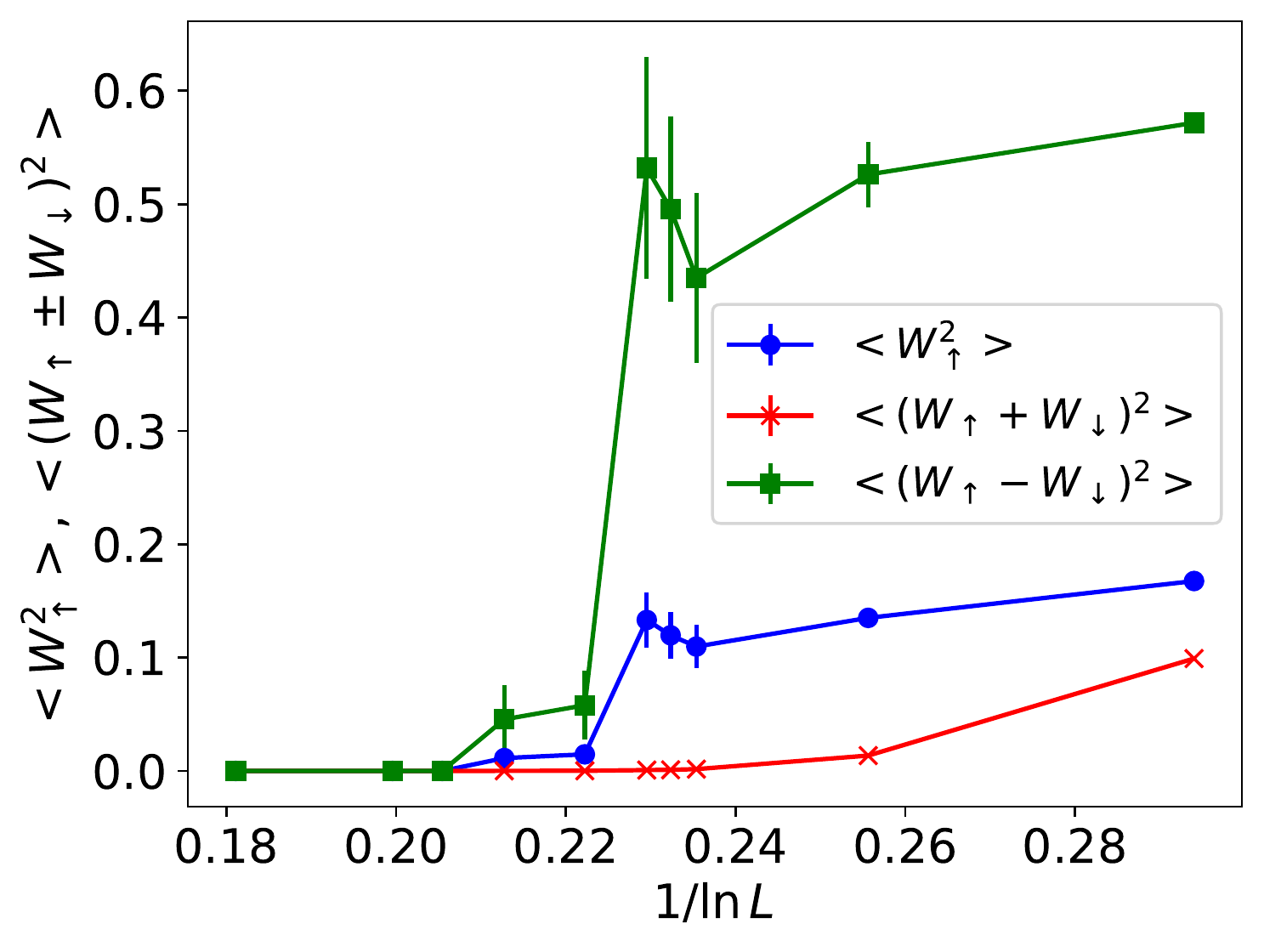} &
        \includegraphics[width=0.5 \columnwidth]{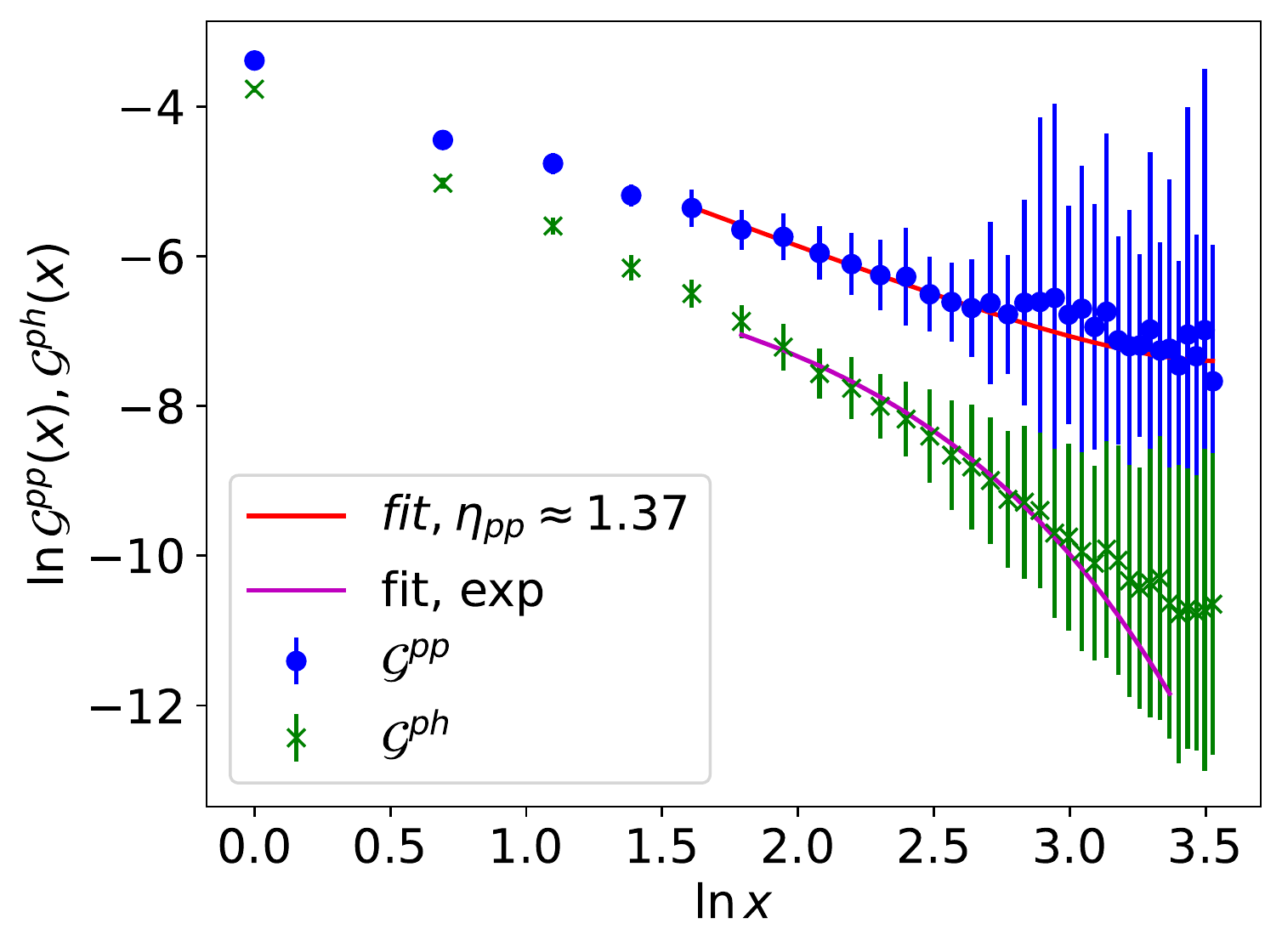} 
        \end{tabular}
        \caption{ \label{fig:V4_U6_FSS}Left: Finite size analysis in the relevant superflow channels for $V = 4, U = 6$.  At mesoscopic length scales $L = \beta < 70$ a tendency towards pairing is seen in the fermionic channels; however then jumps in the data indicate a transition towards an insulator. 
        Right: Powerlaw decay of the four-point correlator $\mathcal{G}^{pp}(x)$ (computed for hard-core bosons only) and exponential decay of the four-point correlator $\mathcal{G}^{ph}(x)$ for $V=4, U=7, L=70$ are indicative of pair superflow, cf. Fig.~\ref{fig:V4_pair_scf}.
        }
\end{figure}

\begin{figure}[!htb]
        \begin{tabular}{ll}
        \includegraphics[width=0.5 \columnwidth]{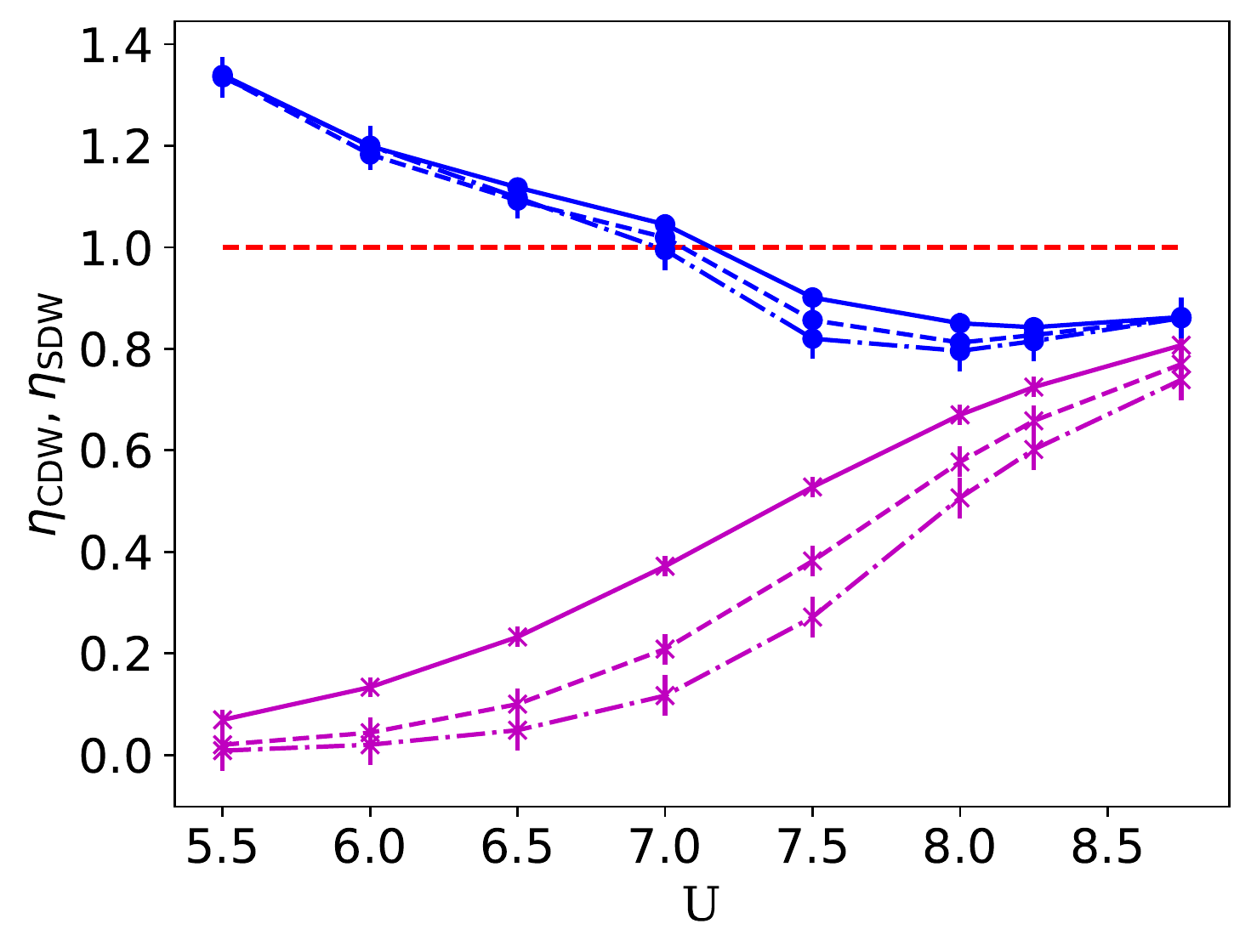} &
        \includegraphics[width=0.5 \columnwidth]{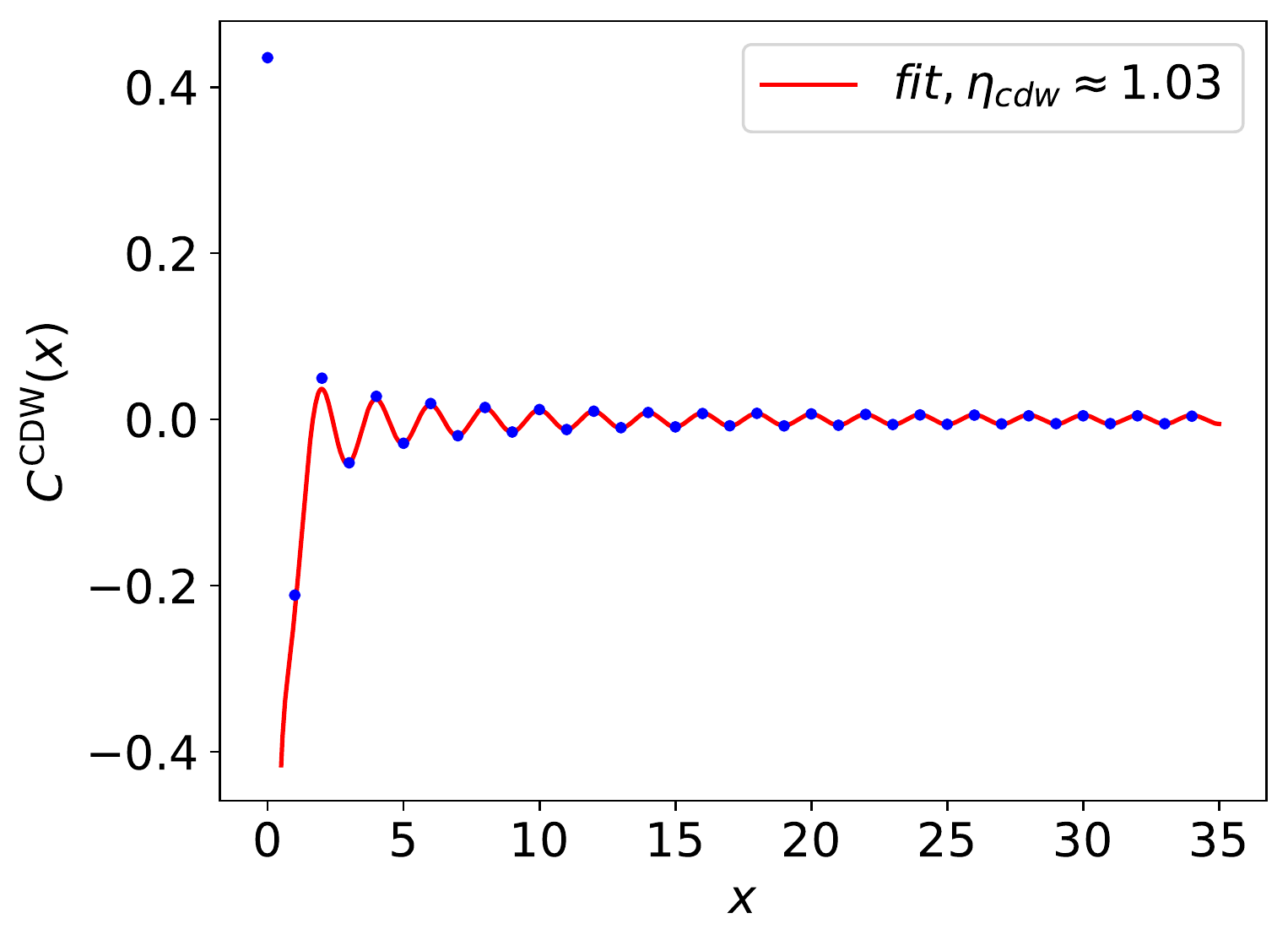}
        \end{tabular}
        \caption{Left: The quantities $\eta_{\rm CDW}$ (blue) and $\eta_{\rm SDW}$ (purple) for $V = 4$ and $L=30, 50, 70$ (top to bottom) obtained from their respective static structure factors.  Right: The CDW correlation function for $V=4, U = 6.5, L=\beta=70$. The fit value $\eta_{\rm CDW}=1.03(5)$ agrees within error bars with the value in the left panel.  
        \label{fig:V4_Lutt_cdw}}
\end{figure}

This intermediate regime is very challenging numerically. In the left panel of Fig.~\ref{fig:V4_boson_jump} we see constant bosonic winding numbers squared for $U=6$ and $U=6.5$. The decay of the single particle density matrix has however a powerlaw exponent that is less than 2, but $\eta_{\rm bb}$ is above 2 (and in fact very close to 2 for $U=6.5$). This is the behavior we previously referred to as a marginal superfluid, cf. Figs.~\ref{fig:wind_marginal_sf} and \ref{fig:V3_U5_L90_dm}.

From the $Z$-factors shown in the right panel of Fig.~\ref{fig:V4_boson_jump} we see essentially non-Fermi-liquid behavior up to $U=8.5$. Signs of phase separation were seen for $U \lesssim 5$: Specifically, for $U=5$ the bosonic superfluid density appears to flow to 0 while $\eta_{\rm bb}$ is considerably larger than 2. We will see such paradoxical behavior also for larger values of $V$, and interpret it as a sign of phase separation (see the discussion for larger values of $V$ below).
In addition, according to both the weak coupling estimate, $V^2/\pi$ and the strong coupling one $V/0.717$ we expect phase separation for $U=5$.

From the winding numbers squared in the pair-flow and the counter-flow channels shown in Fig.~\ref{fig:V4_pair_scf} we see a tendency towards stable pair flow for $U$-values in the range 7 to 8 on the system sizes that we can simulate, which are not enough to extrapolate to the thermodynamic limit because the $Z$-factors remain large.  We show in the right panel of Fig.~\ref{fig:V4_U6_FSS} the powerlaw decay of the four-point correlator $\mathcal{G}^{pp}(x)$ and the simultaneous exponential decay of the four-point correlator $\mathcal{G}^{ph}(x)$ for $U=7, L=70$, which is indicative of a pair superflow. The weakness of the signal is in line with the weak superflow properties that we also observed based on the analysis of the winding numbers.
For larger values of $U \ge 8.5$ the fermions cross over to quasi-free fermions, as we have seen before.

For smaller values of $U \le 7.0(5)$ the pair-flow looks stable on mesoscopic length scales but then abruptly jumps to 0, and the data are very noisy. In many cases, for instance for $U=6$, we see an unphysical increase in the flow with $L$ before jumping to 0 at larger length scales, see Fig.~\ref{fig:V4_U6_FSS}. This noisy behavior (due to autocorrelation times that exceed $10^6$) is in fact often encountered in our simulations (cf. the discussion for $V=3$ before). When this happens, one has to wonder if the data at larger system sizes (here for $L \ge 90$) have properly thermalized, but we are confident that this is so here: We saw a value of (near ) 0 for $U = 6, L = 90$ also in a simulation protocol in which the temperature was systematically lowered at fixed $L$, and also in a different protocol where we started from a configuration with stable superflow found at larger $U$ after which we reduced the value of $U$. We therefore consider this system to be an insulator for the fermions. 

In the left panel Fig.~\ref{fig:V4_Lutt_cdw} we show the extrapolated values of $\eta_{\rm CDW}$ and $\eta_{\rm SDW}$ obtained from the linear behavior of the static correlation functions $S^{\rm CDW/SDW}(q)$ for small $q$.
It is seen that the spin gap is fully developed for $5 < U < 7$ at $L \sim 100$. For larger values of $U$ the behavior crosses over to the one of free fermions, which sets in around $U=9-10$.  Similar as for lower values of $V$, we see that $\eta_{\rm CDW}$ drops below 1 when bosonic superfluidity is lost, indicating Luttinger liquid behavior with somewhat stronger charge fluctuations on top of a bosonic Mott insulator. We saw no conclusive evidence of lattice symmetry breaking for any value of $U$ at $V=4$ on the system sizes that we can simulate.


\subsection{Scan of the phase diagram at $V=5$}

\begin{figure}[!htb]
        \begin{tabular}{ll}
        \includegraphics[width=0.5 \columnwidth]{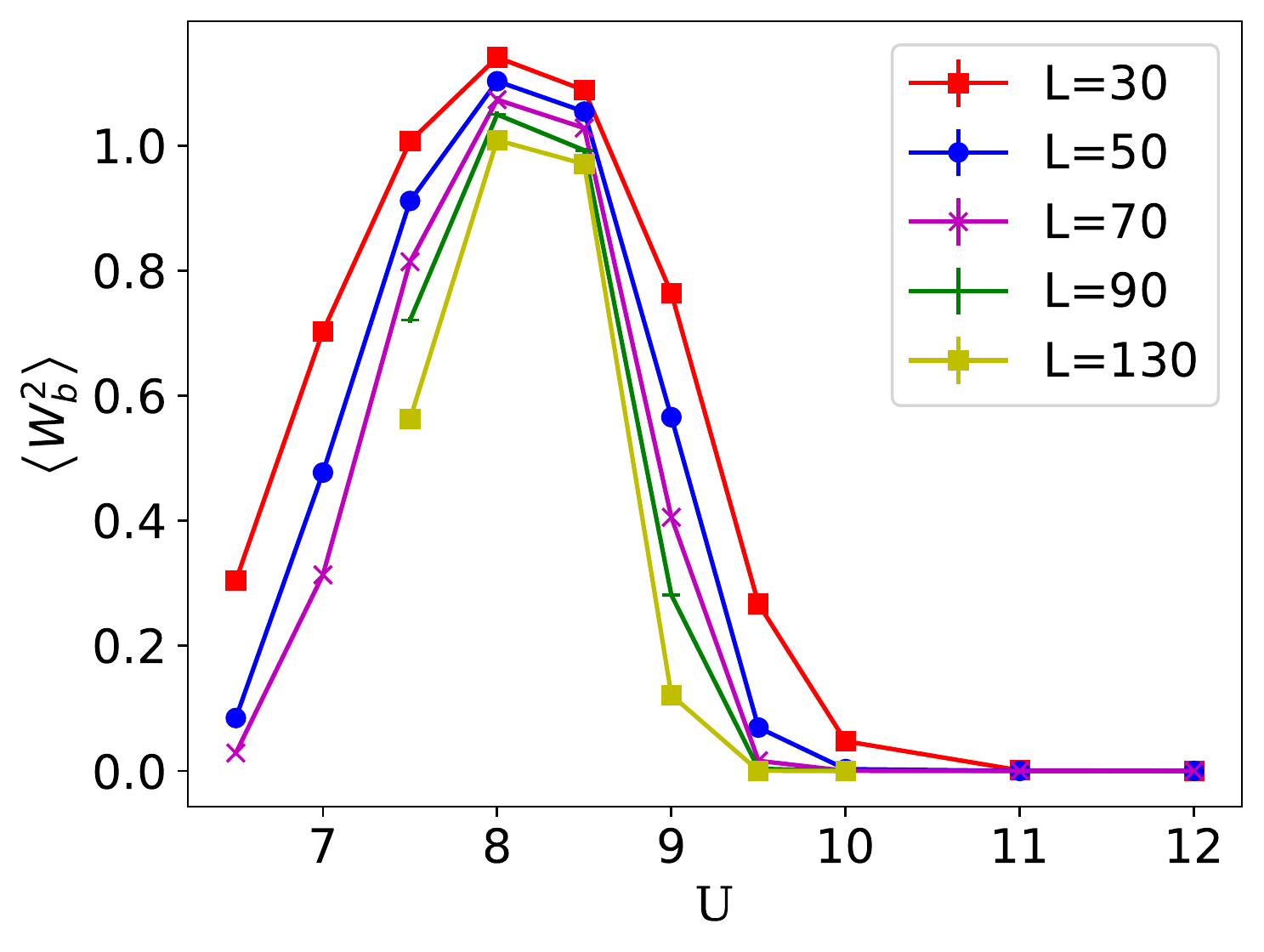} &
        \includegraphics[width=0.5 \columnwidth]{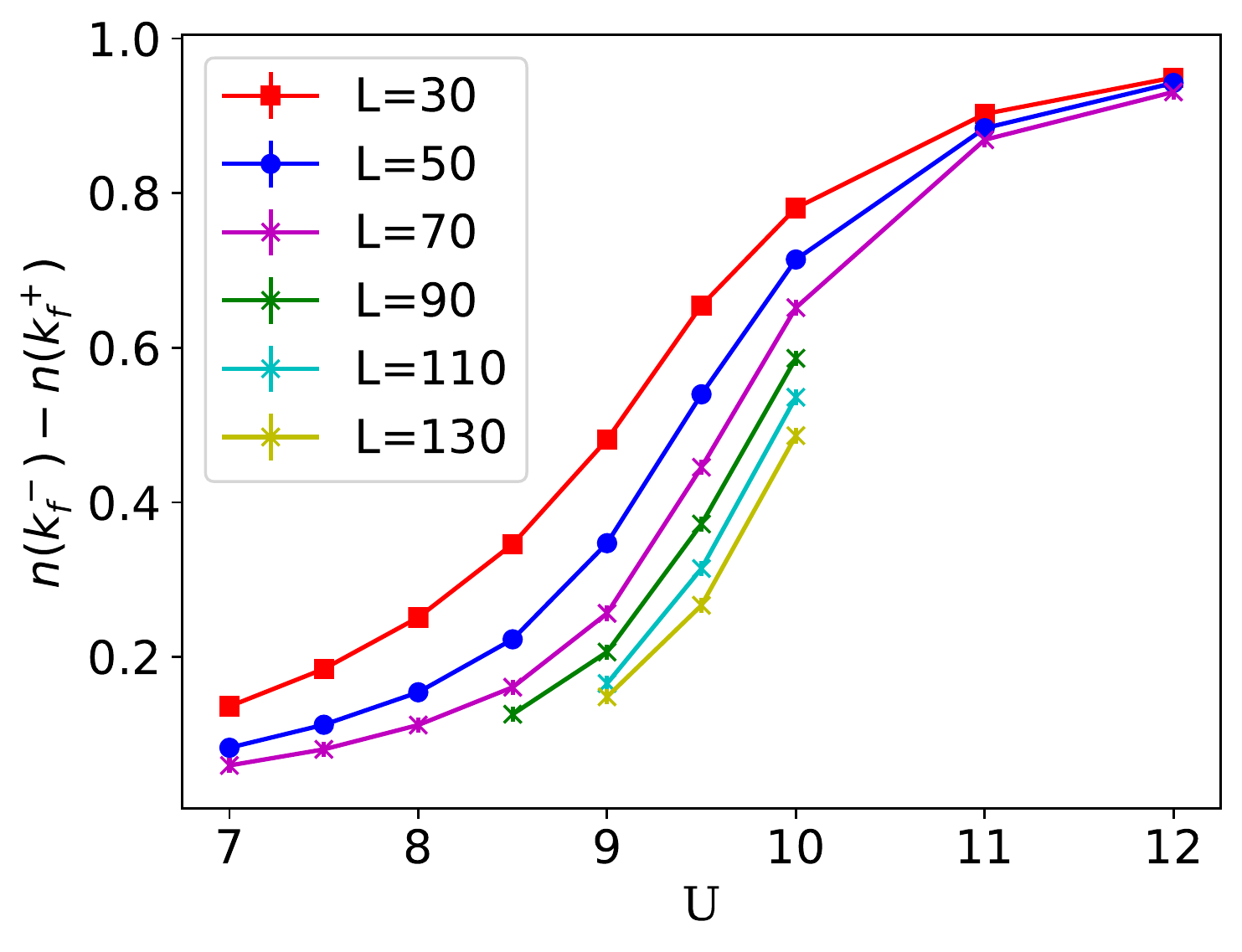}
        \end{tabular}
        \caption{Left: Bosonic winding number squared for $V= 5$. Superflow is thermodynamically stable for $U=8$ and $U=8.5$; and $U=7.5$ appears to be very close to the transition but on the insulating side. Right: Jumps in occupation number at $k_F$. The low values we see are promising for being close to the thermodynamic limit. 
        \label{fig:V5_boson_jump}}

\end{figure}

\begin{figure}[!htb]
        \begin{tabular}{ll}
        \includegraphics[width=0.5 \columnwidth]{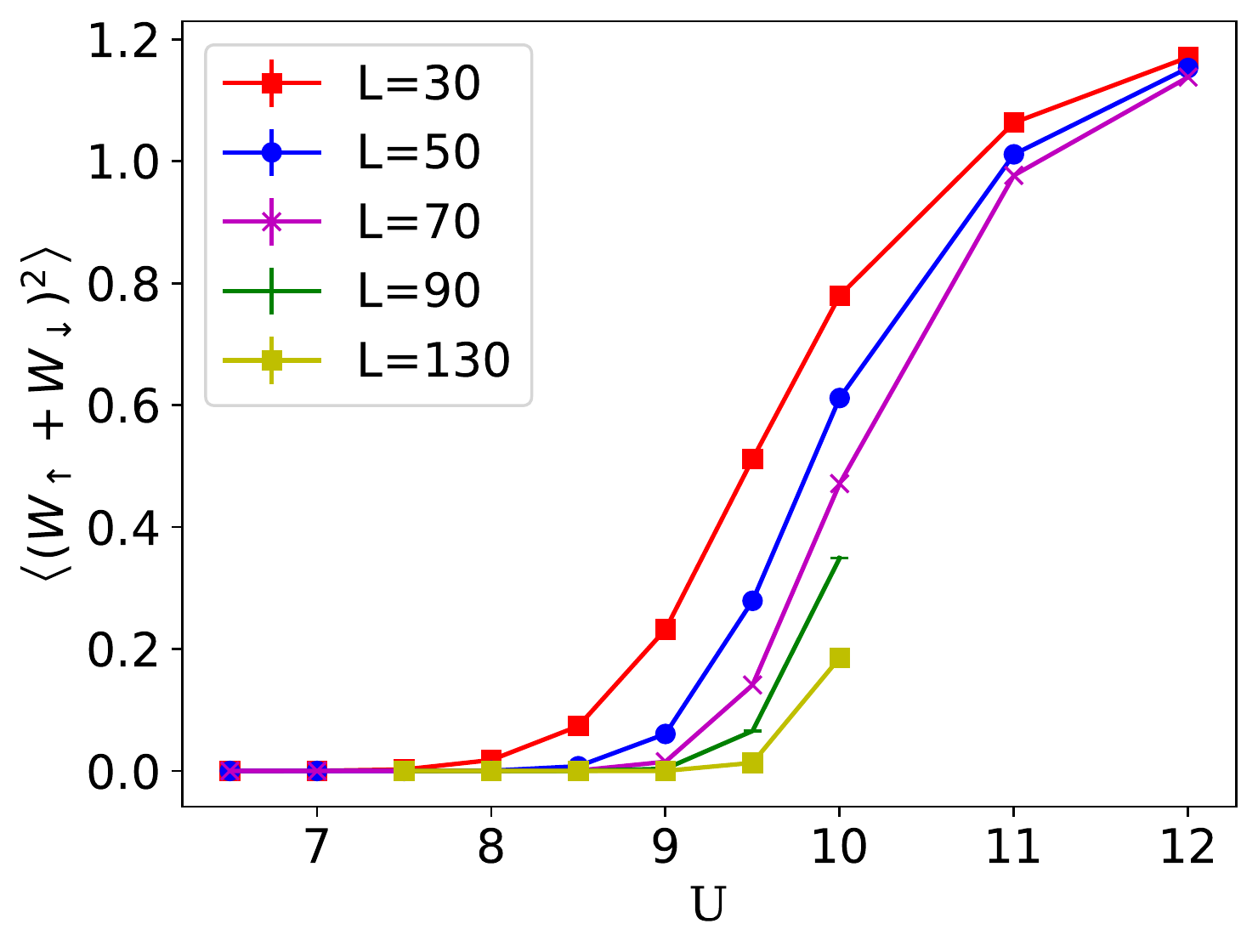} &
        \includegraphics[width=0.5 \columnwidth]{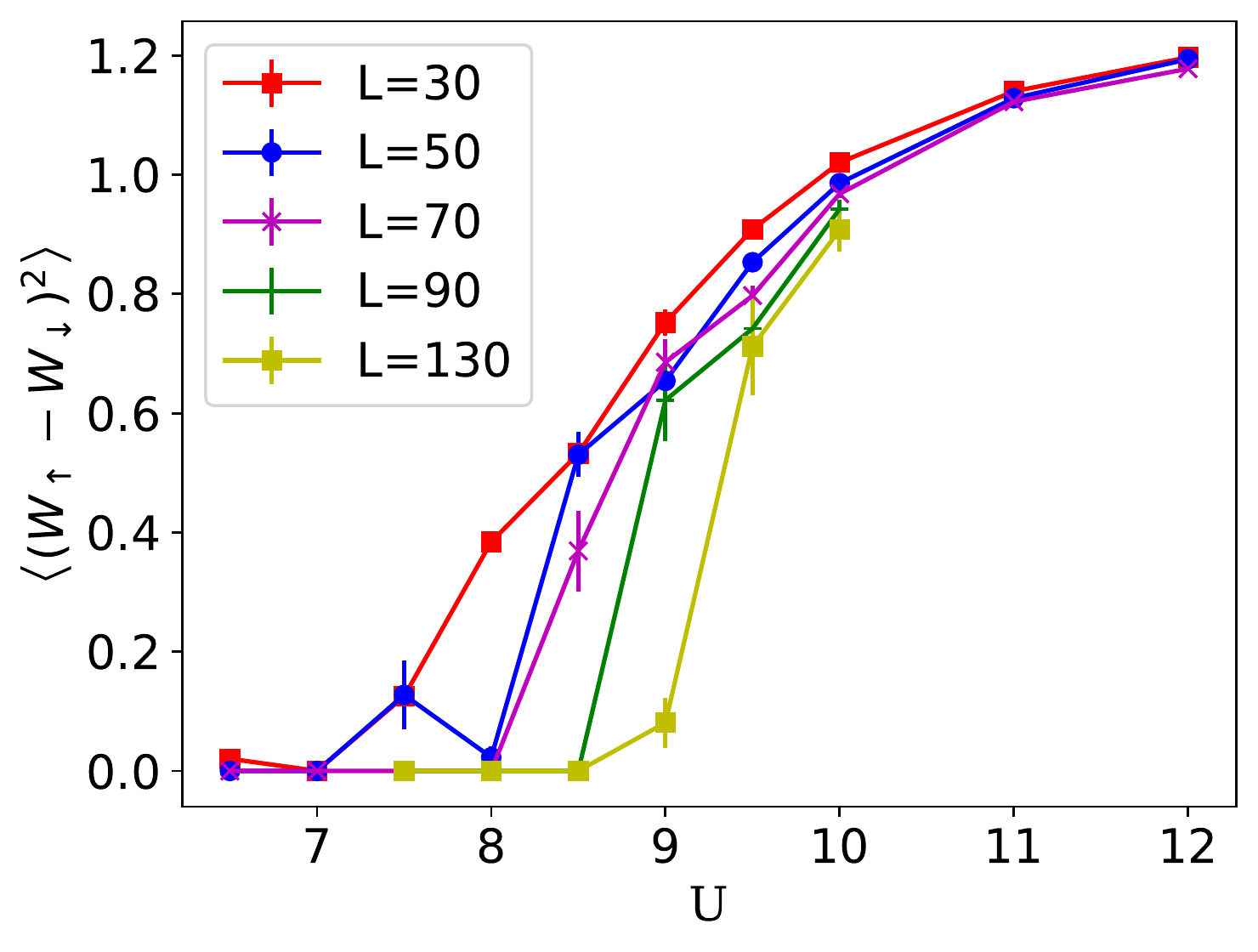}
        \end{tabular}
        \caption{Left: Winding number squared in the counter flow channel at $V= 5$; Right: Same for the pair flow channel. One-dimensional superconducting order is stable at least on mesoscopic length scales for $9.5 \le U \le 10$, but it is unclear what happens on longer length scales. Insulating behavior is certain for $U < 9$.
        \label{fig:V5_spf_scf} }
\end{figure}

       \begin{figure}[!htb]
        \begin{tabular}{ll}
        \includegraphics[width=0.5 \columnwidth]{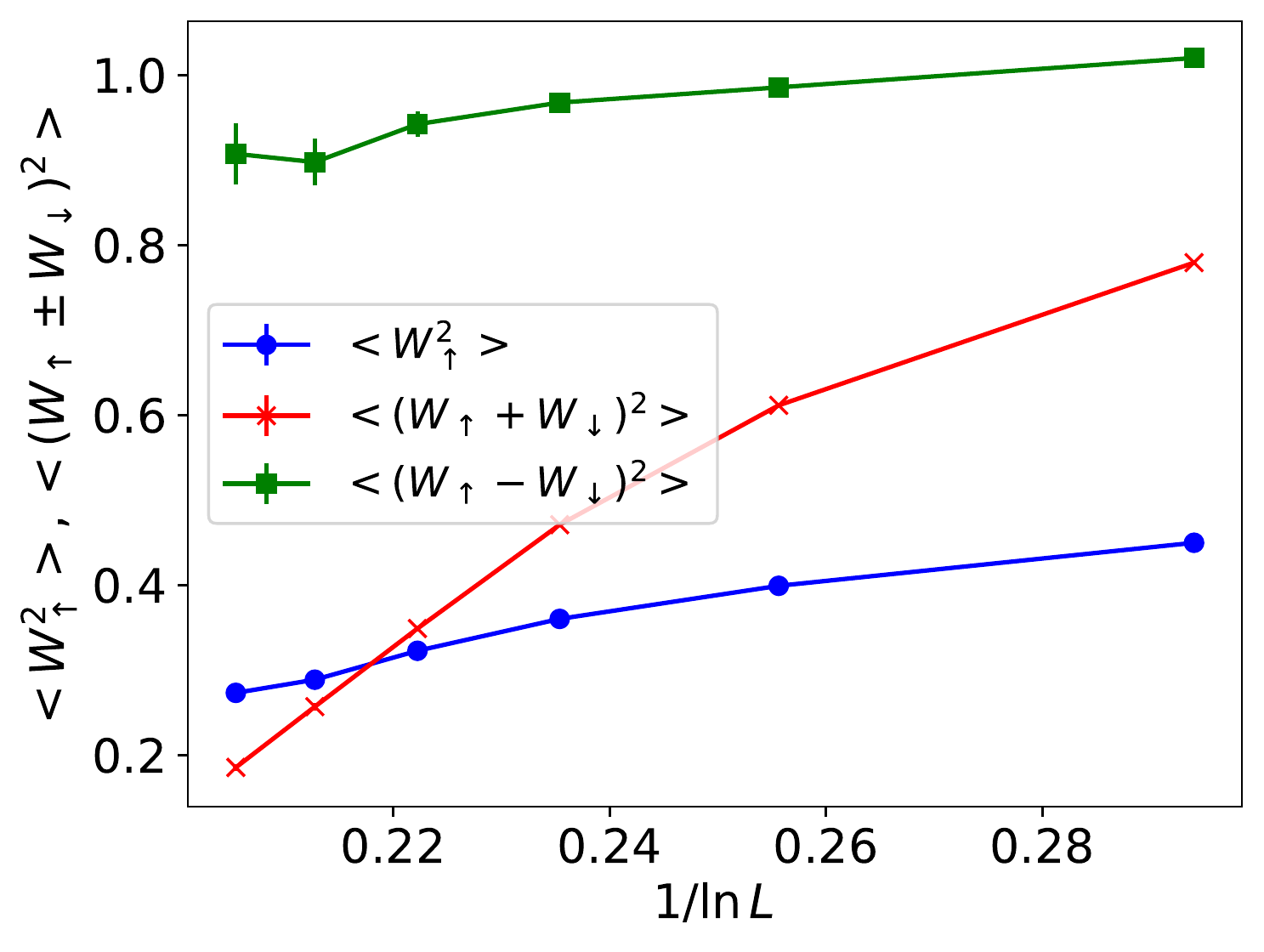} &
        \includegraphics[width=0.5 \columnwidth]{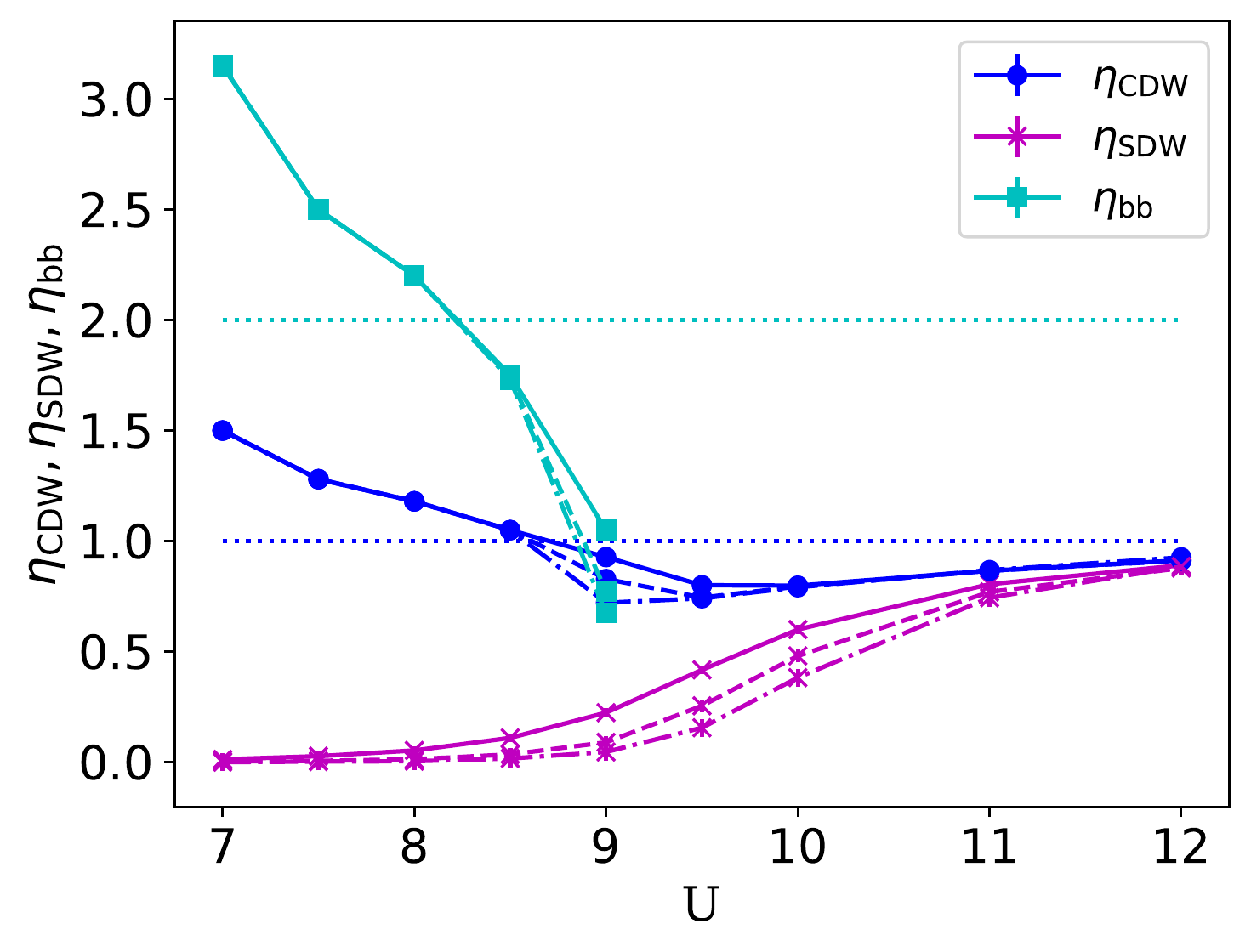} 
        
        \end{tabular}
        \caption{Left: Winding number squared for the spin-up particles (blue), in the counter-flow channel (red), and in the pair-flow channel (green) as a function of inverse system size with inverse temperature $\beta = L$ and system parameters $V=5$ and $U = 10$. 
        Right: The quantities $\eta_{\rm CDW}$, $\eta_{\rm SDW}$, and $\eta_{\rm bb}$ for $V = 5$ obtained from an analysis of the low momentum behavior of the CDW, SDW, and bosonic static structure factors for $L = 30, 50, 70$ (top to bottom). The spin gap is fully developed at $L = 100$ for $U < 10$.  Bosonization predicts a transition from a superfluid to a Mott insulator when $\eta_{\rm bb} = 2$ (cyan dotted line); simultaneously, $\eta_{\rm CDW}$ drops below 1 (blue dotted line).
              \label{fig:V5_U10_FSS} }
\end{figure}

\begin{figure}[!htb]
        \begin{tabular}{ll}
        \includegraphics[width=0.5 \columnwidth]{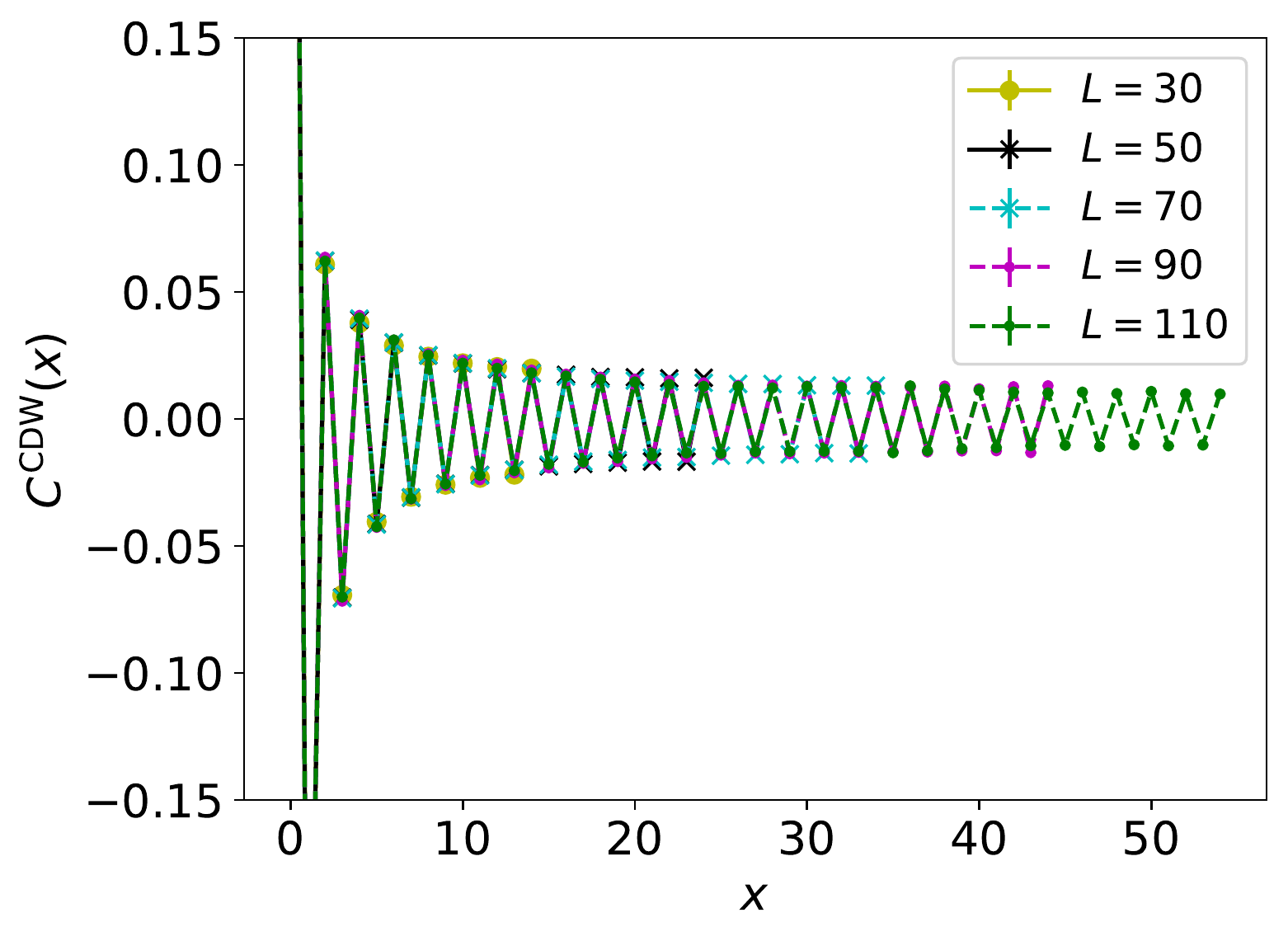} &
        \includegraphics[width=0.5 \columnwidth]{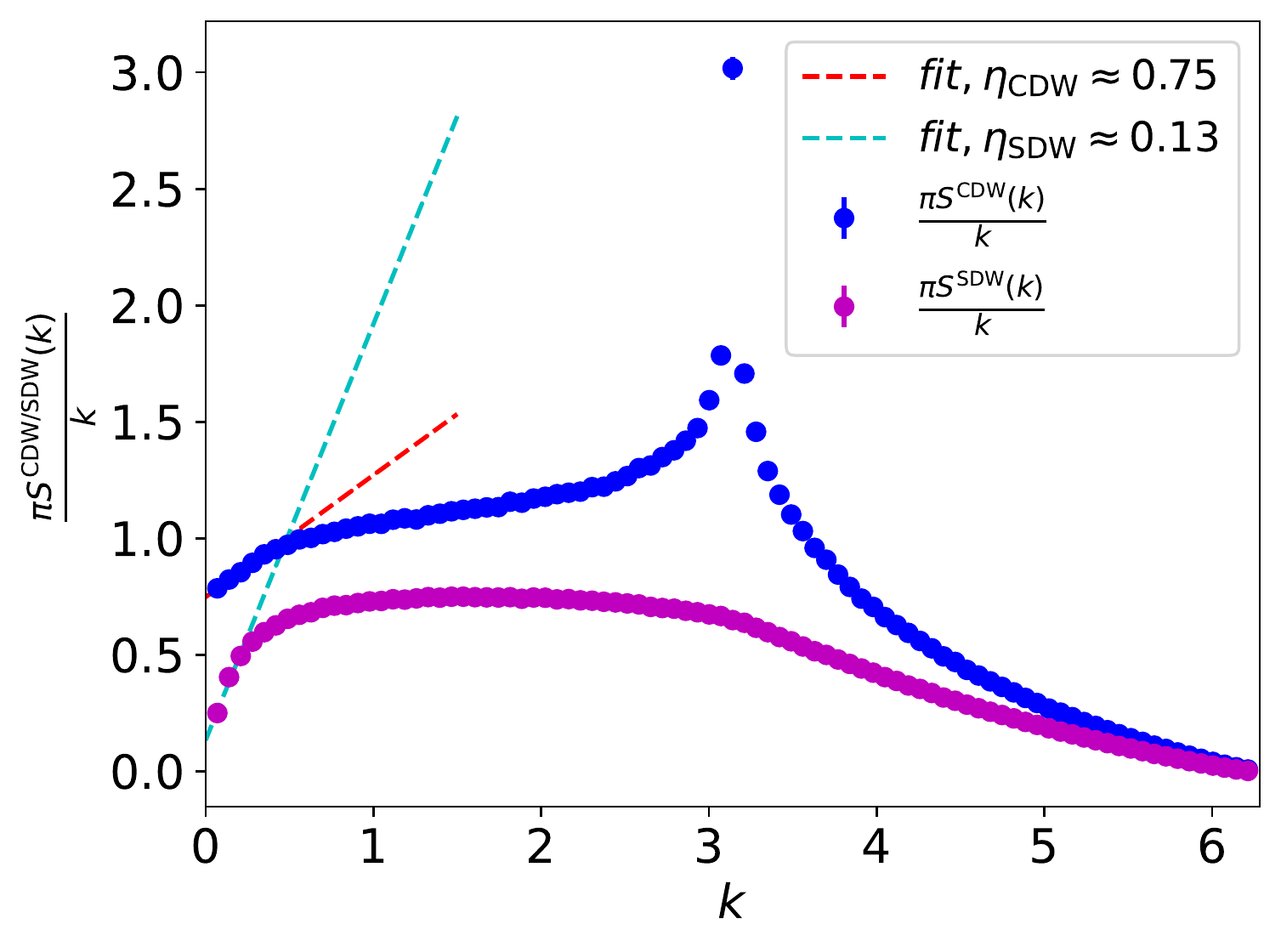}
        \end{tabular}
        \caption{Left: The CDW correlation function for $V=5$, $U=9.5$  and various system sizes and inverse temperatures chosen as $L = \beta$ showing convergence at short distances with system size and asymptotic powerlaw decay with exponent $\eta_{\rm CDW} \approx 0.74(2)$, indicative of  algebraic CDW correlations. 
        Right: Rescaled static structure factors in the CDW and SDW channels for $V=5, U=9.5, L=\beta=90$ from which $\eta_{\rm CDW}$ and $\eta_{\rm SDW}$ in the right panel of Fig.~\ref{fig:V5_U10_FSS} are extracted and which are compatible with the data shown on the left. The peak that develops at $k = \pi$ in the CDW static structure factor might be an indication of lattice symmetry breaking developing at longer length scales that are out of reach however. The SDW correlations are close to being fully gapped.
                \label{fig:V5_Lutt_cdw}}
\end{figure}

The strong coupling argument predicts  phase separation for $U \le V/0.717 \approx 7$ while the weak coupling argument sees phase separation up to $U \le V^2/\pi \approx 7.96$.
The bosonic winding number squared is plotted in the left panel of Fig.~\ref{fig:V5_boson_jump}.  We indeed see a renormalization to 0 for $U \le 7.5$ and stable superflow for $U \approx 8$ . The superflow is again marginal, as reflected in the decay of the  the single particle density matrix, $\mathcal{G}^b(x)$, which  decays already fast with an exponent $\eta_b \approx 1.3$ for $L=\beta=30$ and $U=8$, but barely faster for $L=\beta=90$ where  $\eta_b \approx 1.2$. We come back to the issue of bosonic superfluidity when discussing $\eta_{\rm bb}$.
The fermionic $Z$-factors are shown in the right panel of  Fig.~\ref{fig:V5_boson_jump}. The flow scales inversely with system size and we see a very strong renormalization everywhere except when $U \gtrsim 2V$, in agreement with the arguments of Sec.~\ref{sec:strong_coupling}.
The winding numbers squared in the counter-flow and pair-flow channels, shown in Fig.~\ref{fig:V5_spf_scf}, show the tendency towards free fermionic behavior for $U \ge 12$. For $9 \le U \le 10$ the counter-flow channel flows to zero with system size whereas the pair-flow seems to approach a constant on mesoscopic length scales, within error bars, indicating superconducting correlations. This is elaborated in the left panel of Fig.~\ref{fig:V5_U10_FSS} for $U=10$. 
For smaller values of $U \le 8.5$ any superflow disappears in the thermodynamic limit.  For $U=9$ and $U=9.5$ the signal is noisy; it indicates either thermodynamically stable superflow or mesoscopic flow  extending over hundreds of sites. It is, given the data, certain that the bosonic and the fermionic pair flow do not coexist, and likely that they are adjacent in the phase diagram.

We plot the quantities $\eta_{\rm CDW}$ and $\eta_{\rm SDW}$ as a function of $U$ and for $L= 30, 50, 70$ in the right panel of Fig.~\ref{fig:V5_U10_FSS} obtained from the low momentum analysis of the static CDW and SDW structure factors. In the region of putative bosonic superfluidity, we obtain values between 1 and $1.25$ (and a fully developed spin gap), indicating a decay faster than the one of free fermions. The fermions are hence paired because of the attractive induced interactions but a picture of molecules with charge density wave order does not apply.  For $U \ge 9$ the values of $\eta_{\rm CDW}$ are below 1. Note that the spin gap has not fully developed for $U \ge 9.5$ for $L=70$,  as can be seen in the lower curve of the right panel in Fig.~\ref{fig:V5_Lutt_cdw}. The spin gap develops however rather quickly for $U < 11$, and (exponentially) slowly for $U > 12$. 
Fig.~\ref{fig:V5_Lutt_cdw} also shows $\eta_{\rm bb}$, which is larger than 2 for $U=8$, in line with uniform bosonic superfluidity. For $U \ge 9$ the flow towards a bosonic Mott insulator is apparent. At $U=8.5$ the bosonic density-density correlation function has an exponent $\eta_{\rm bb} < 2$. It seems to be nearly constant on our system sizes (whereas it should flow to 0 in a Mott insulating phase). We attribute this to the (mesoscopic) superfluidity, which counters stronger density fluctuations; cf. also the behavior of $\eta_{\rm bb}$ for $U=4.5$ and $V=2$ in Fig.~\ref{fig:V2_Lutt} and for $U=14$ and $V=8$ in Fig.~\ref{fig:V8_Lutt_cdw}. Hence, within our system sizes, $U=8$ shows a bosonic superfluid and is very close to the phase separation line (and also very close to a Mott transition), whereas  $U=8.5$ is already on the Mott insulating side based on its value of $\eta_{\rm bb}$.

We saw the strongest CDW correlations for $U=9$ with $\eta_{\rm CDW} = 0.72(3)$ and $U=9.5$ with $\eta_{\rm CDW} = 0.75(3)$, both for $L = \beta = 70$,  as shown in Fig.~\ref{fig:V5_U10_FSS} and Fig.~\ref{fig:V5_Lutt_cdw}. These CDW correlations decay algebraically and for larger and smaller values of $U$ they decay as faster powerlaws; for $L=\beta=70$ we find $\eta_{\rm CDW} = 0.79(2)$ for $U=10$. We did not see any indication of lattice symmetry breaking for $V=5$ on the system sizes that we can simulate (cf. below for a discussion on this for $V=6$ and $V=8$).
There remains the small region between the strong coupling prediction for phase separation at $U \approx 7$ and the onset of bosonic superfluid for $U \le 8$. The data shows no sign  of superflow in any channel, and furthermore $\eta_{\rm CDW}$ and $\eta_{\rm bb}$ are large because of an upturn in their respective structure factors at low momentum. At the phase separation transition line the compressibility is divergent, with which the increase of $\eta_{\rm CDW}$ and $\eta_{\rm bb}$ is consistent. Most likely, on finite systems in the canonical ensemble, the data is already phase separated~\cite{Clay1999} and this leads us to the dotted cyan line in the phase diagram of Fig.\ref{fig:phasediagram}. The dotted cyan line is for $V=4$ and $V=5$ in fair agreement with the weak coupling prediction for phase separation.


\subsection{Scan of phase diagram at $V=6$}

\begin{figure}[!htb]
        \begin{tabular}{ll}
        \includegraphics[width=0.5 \columnwidth]{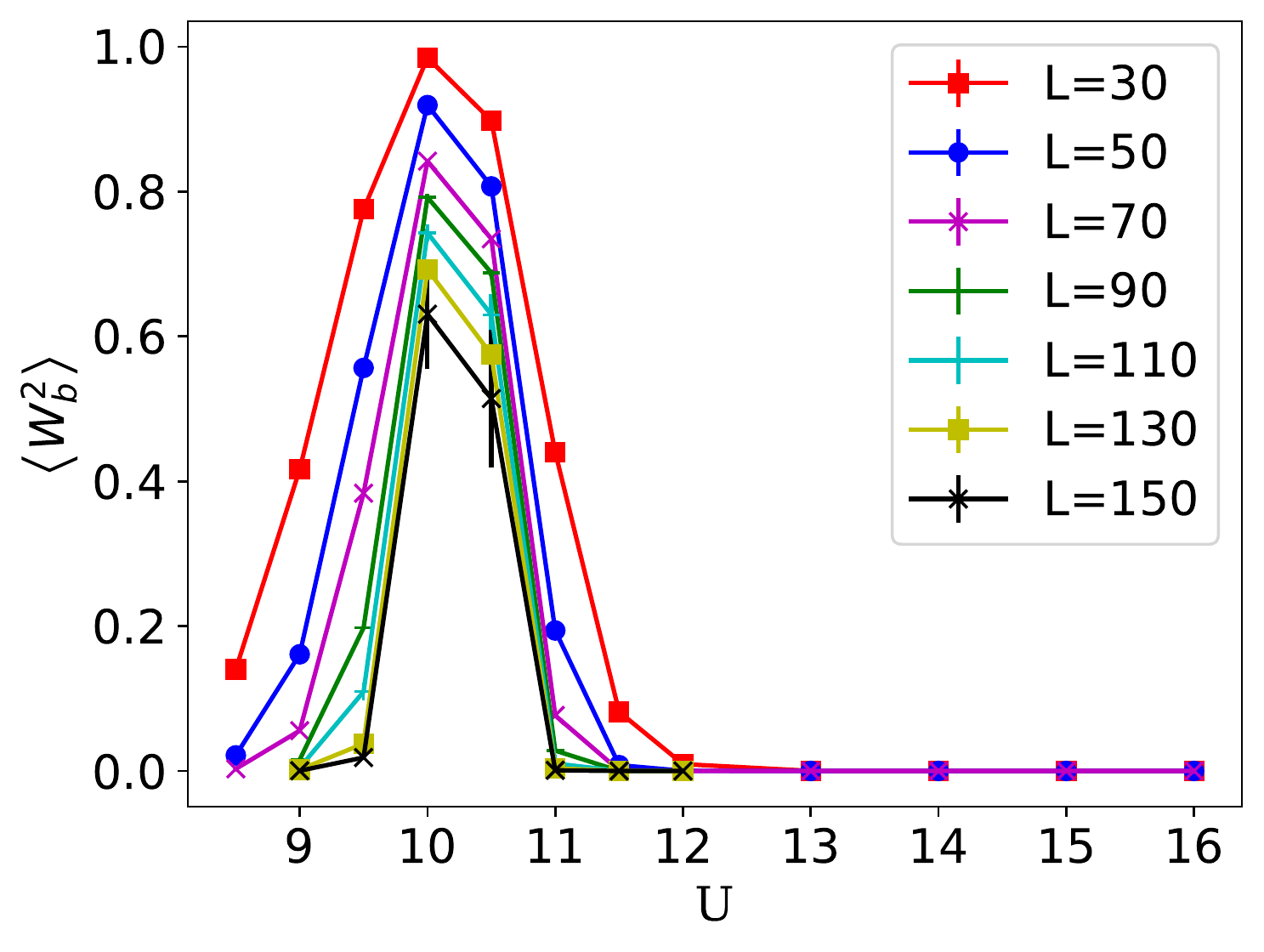} &
        \includegraphics[width=0.5 \columnwidth]{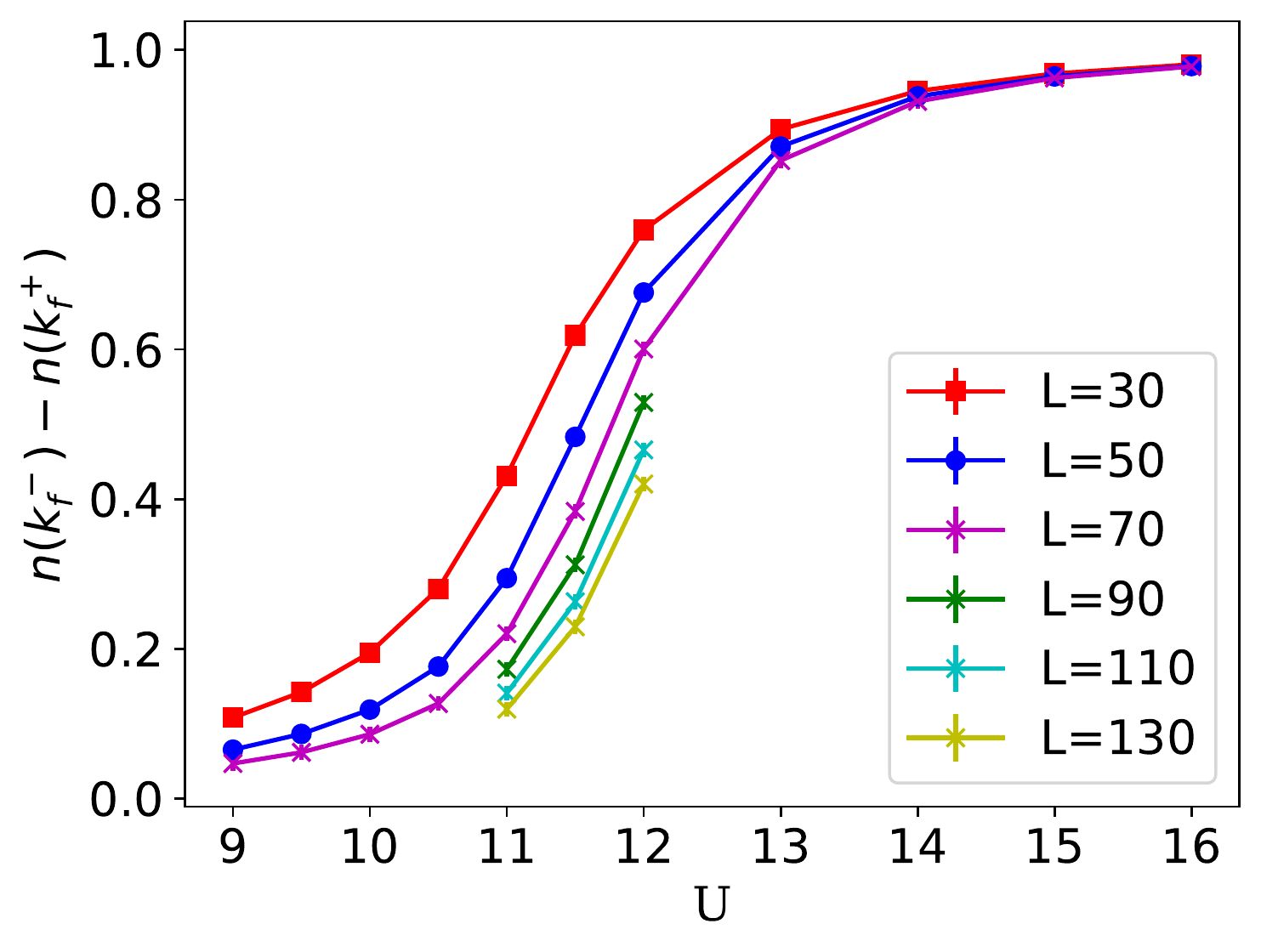}
        \end{tabular}
        \caption{Left: Bosonic winding number squared for $V= 6$.  Phase separation is seen for (at least) $U < 8.5$. The system flows towards an insulator in the thermodynamic regime despite the appearance of non-zero winding numbers at mesoscopic length scales in the range $U = 9.5$ to $U = 11$. Right: Jumps in occupation number at $k_F$, showing a strong tendency towards free fermions for $U \ge 13$, and non-Fermi liquid behavior  for $U < 12$. 
        \label{fig:V6_boson_jump}}
\end{figure}

\begin{figure}[!htb]
        \begin{tabular}{ll}
        \includegraphics[width=0.5 \columnwidth]{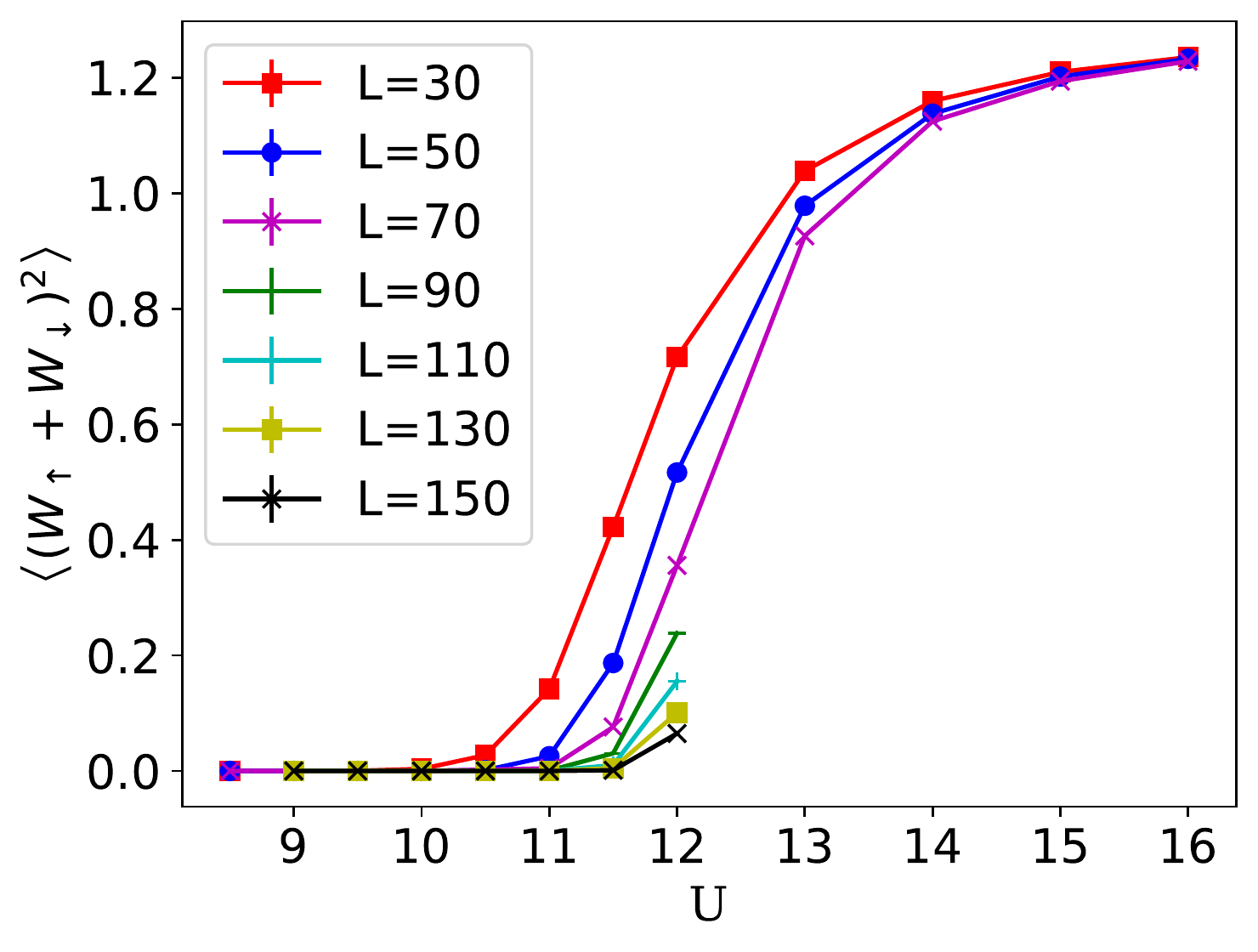} &
        \includegraphics[width=0.5 \columnwidth]{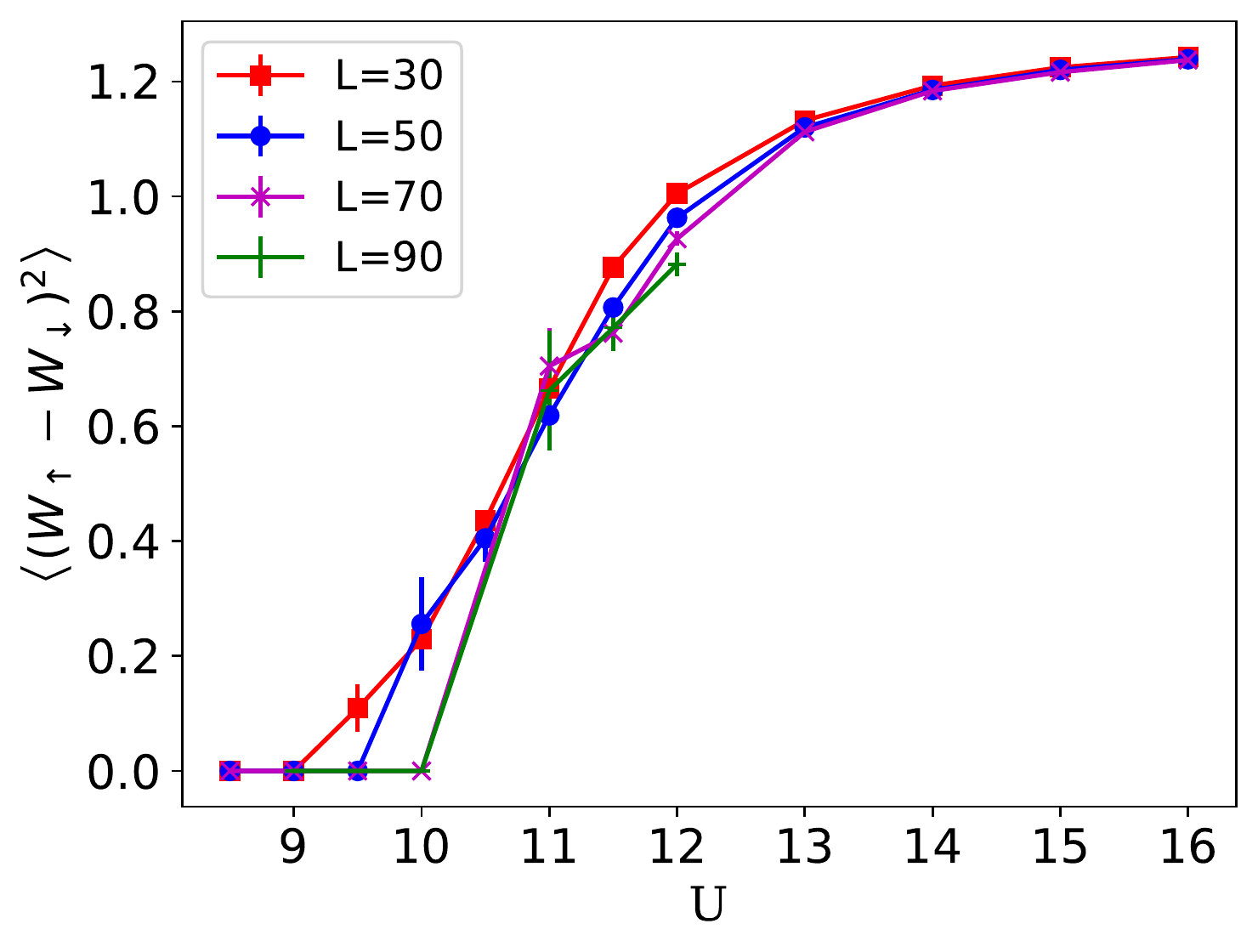}
        \end{tabular}
        \caption{Left: Winding number squared in the counter-flow channel at $V = 6$ as a function of $U$ for various system sizes and inverse temperatures with $L = \beta$. Right: Winding number squared  in the pair flow channel at $V = 6$ for the same system as in the left panel. For $U \ge 13$ the fermions are quasi-free on our system sizes; for lower values of $U$ the signal in the counter-flow channel flows to zero with system size. Insulating behavior in the pair flow channel is certainly seen for $U \le 10$. For $U=12$ it appears stable (see Fig.~\ref{fig:V6_U12_FSS}) and for  $U=11.5$ the signal is suggestive of stable pair flow as well. The error bars for the data in the pair-flow channel for $L> 100$ is too noisy over the range $9.5 \le U \le 11.5$ to show.
          \label{fig:V6_spf_scf} }

\end{figure}

        \begin{figure}[!htb]
        \begin{tabular}{ll}
        \includegraphics[width=0.5 \columnwidth]{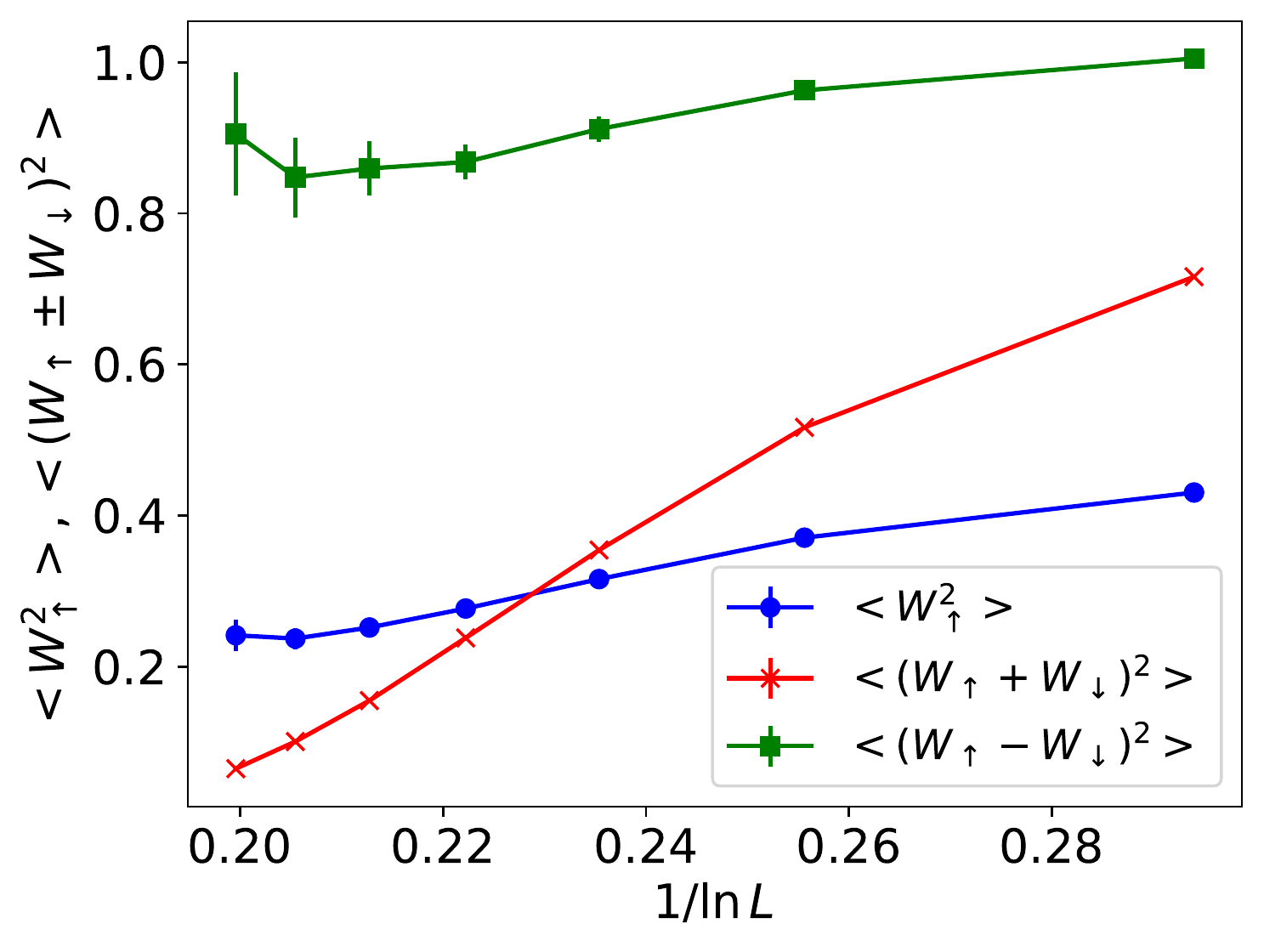} &
        \includegraphics[width=0.5 \columnwidth]{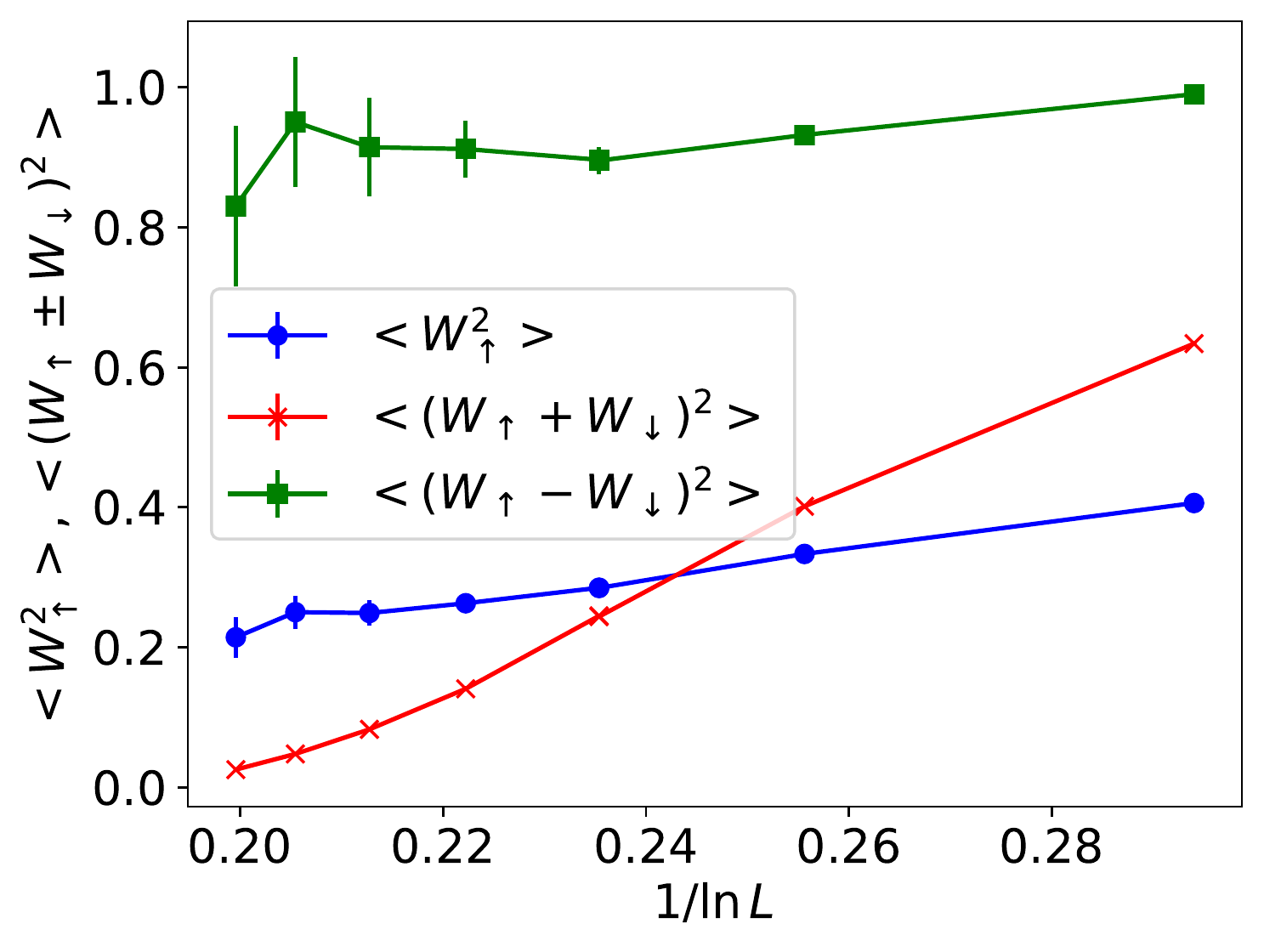}
        \end{tabular}
        \caption{Left: Winding number squared for the spin-up particles (blue), in the counter-flow channel (red), and in the pair-flow channel (green) as a function of inverse system size with inverse temperature $\beta = L$ and system parameters $V=6$ and $U = 12$. The counter-flow channel shows a strong  and unstoppable renormalization to zero whereas the pair-flow channel stays finite and will approach 4 times the value of the winding number squared of the spin-up particles in case of a thermodynamic super-flow. Note that the residue is still about $0.4$ on the largest system sizes (cf. Fig.~\ref{fig:V6_boson_jump}), so further renormalizations will happen on longer length scales.
        Right: Same but for $V=8$ and $U=16$, suggesting thermodynamically stable pair flow. This should be taken with some caution when taking into account that the residue is still $0.31(1)$ for $L=150$ (cf. Fig.~\ref{fig:V6_boson_jump}).
      \label{fig:V6_U12_FSS} }
\end{figure}

\begin{figure}[!htb]
        \begin{tabular}{ll}
        \includegraphics[width=0.5 \columnwidth]{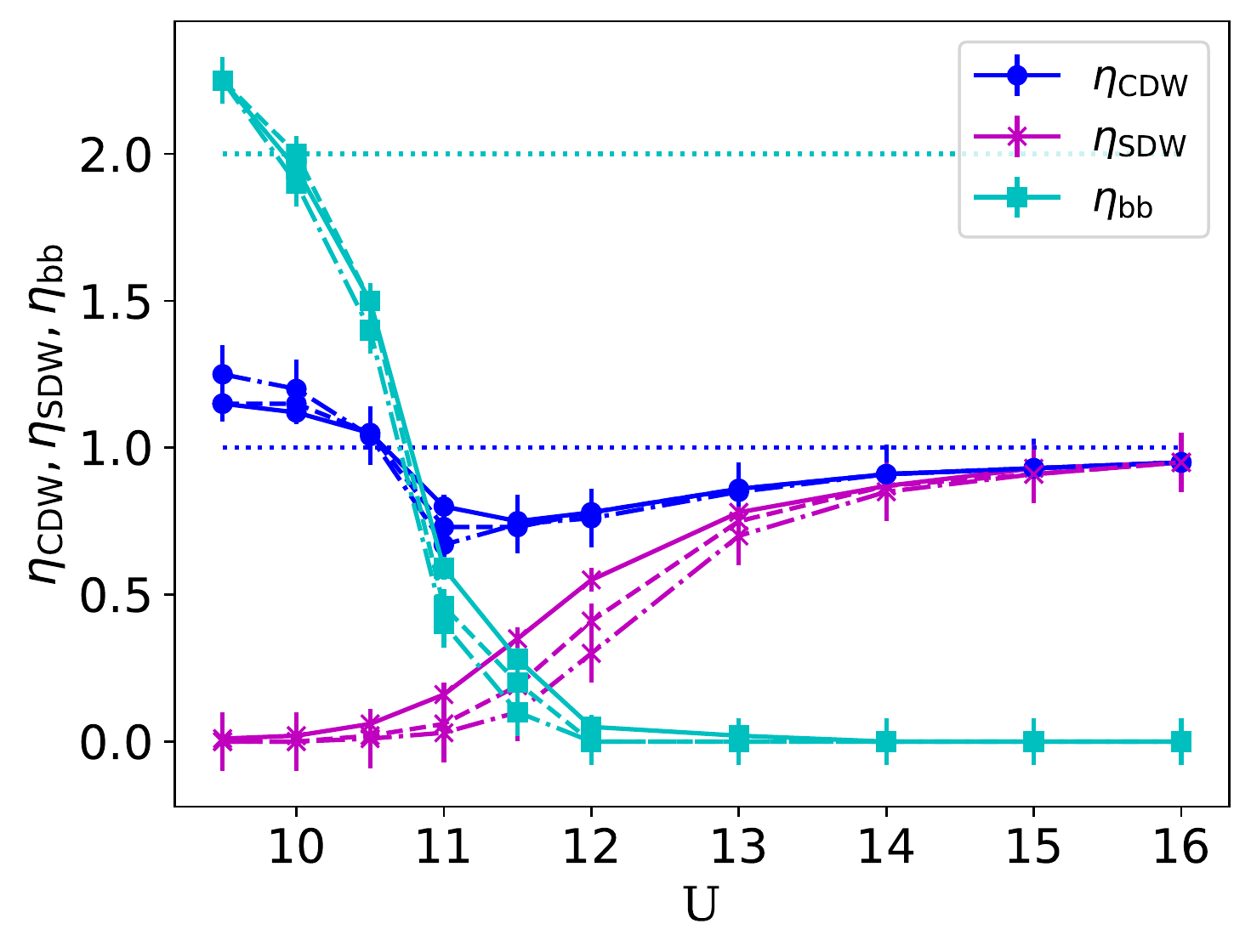} &
        \includegraphics[width=0.5 \columnwidth]{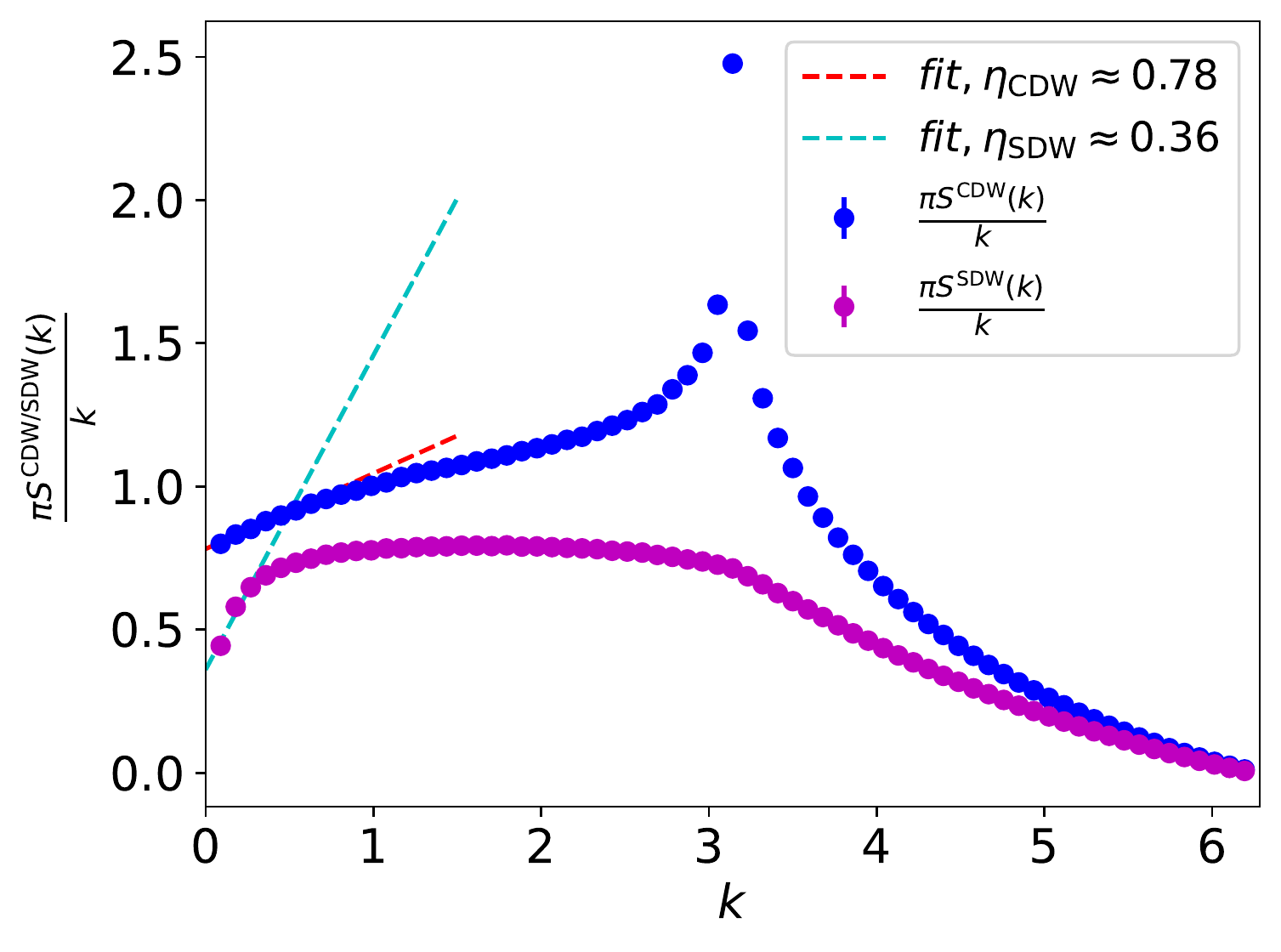}
        \end{tabular}
        \caption{Left: $\eta_{\rm CDW}$, $\eta_{\rm SDW}$, and $\eta_{\rm bb}$ as a function of $U$ for $V = 6$ computed at $L=\beta = 30, 50 , 70$ (solid, dashed, and dashed-dotted respectively) from extrapolation of the CDW, SDW and bosonic static structure factors for low momenta. When values for $\eta_{\rm CDW}$ below 1 are found, fits that involve lattice symmetry breaking are equally good but the value of the constant is very low and only larger system sizes could settle this issue. 
                Right: illustration of the scaled CDW and SDW static structure factors for $V=6, U=12$ in the CDW and SDW channels. The non-zero value below 1 seen in the scaled SDW correlator is certainly a finite size effect: it must approach 1 for systems with SU(2) symmetry, or go to 0 in case of a spin gap, which is the case here. There appears to be a $\delta-$peak developing at $k=\pi$ which also is suggestive of lattice symmetry breaking. For this to hold, one must show that this peak ($S^{\rm CDW}(k = \pi)$)  scales with $L$, and given the low values of the peak it is not possible to do so with certainty on the system sizes that we can reach (not shown). Cf. 
                 Fig.~\ref{fig:V8_U13_sf} for the same quantity for $V=8, U=2V=16$.
        \label{fig:V6_eta_cdw} }
\end{figure}

\begin{figure}[!htb]
        \begin{tabular}{ll}
        \includegraphics[width=0.5 \columnwidth]{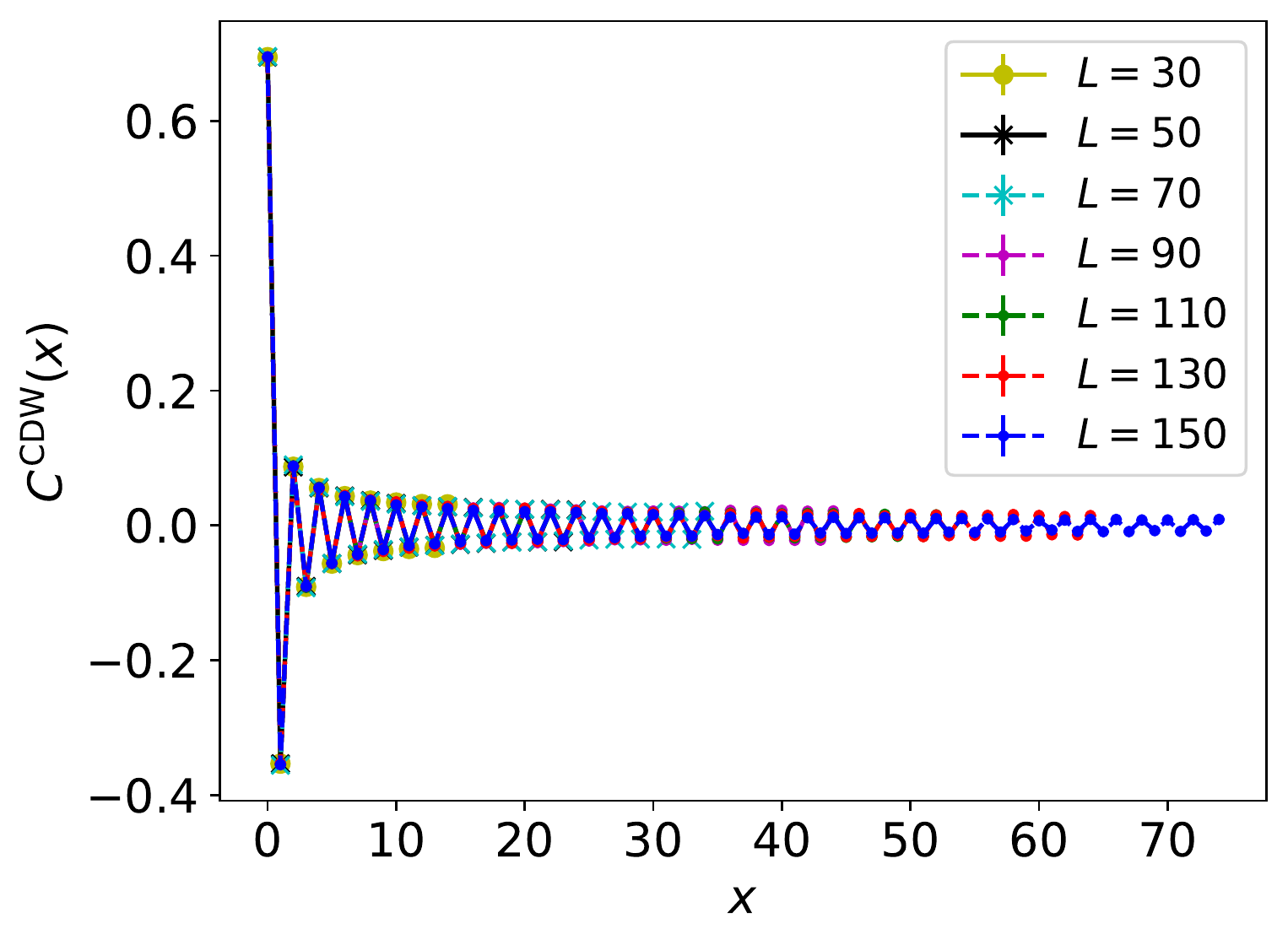} &
        \includegraphics[width=0.5 \columnwidth]{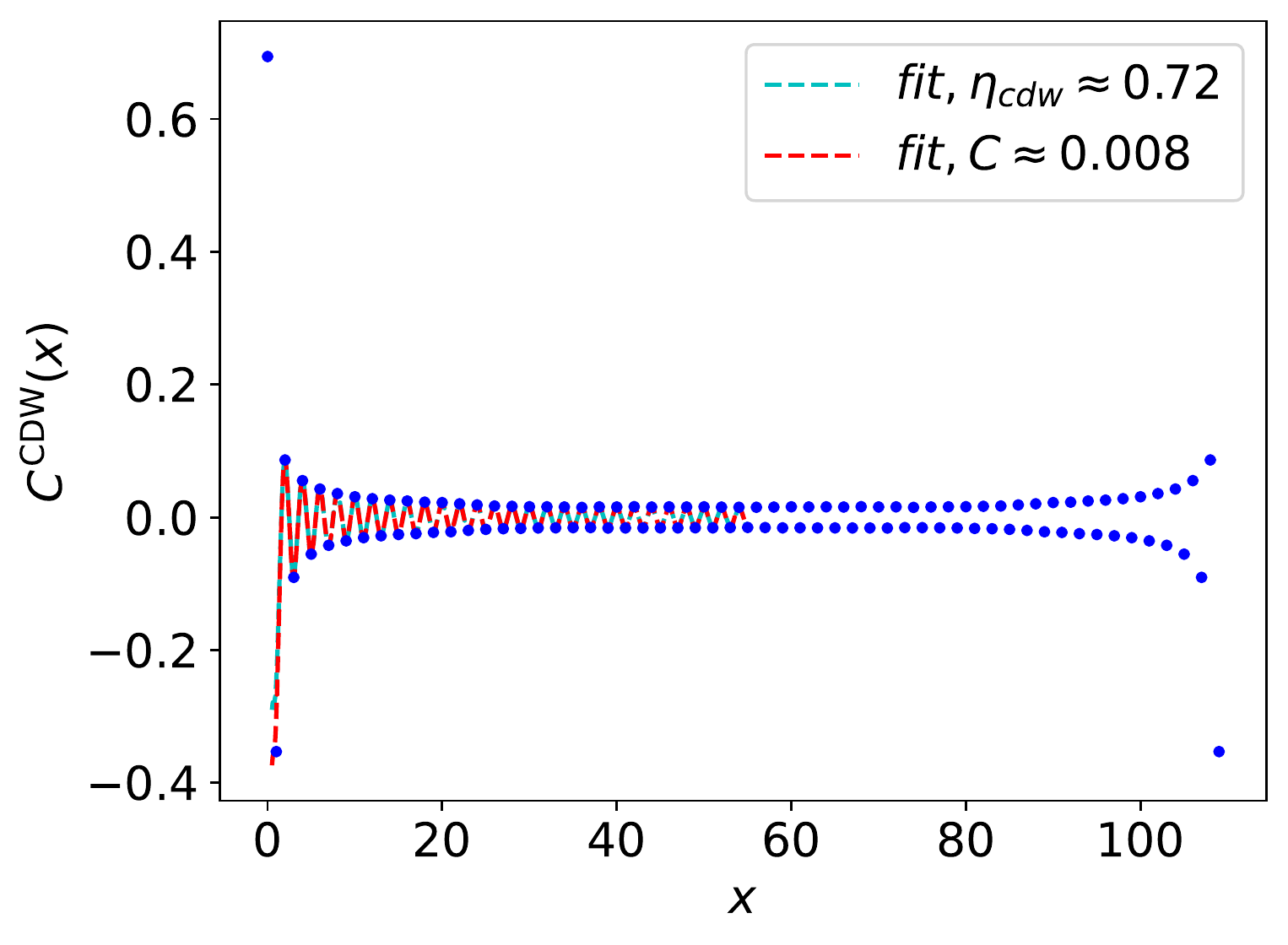}
        \end{tabular}
        \caption{Left: The CDW correlator for $V = 6, U = 11$ for different system sizes ($L = \beta = 30, 50, 70, 90, 110, 130, 150$) showing convergence at short distances. Right: The same CDW correlator for $L = \beta = 110$ is best fit with a power law; the alternative fit with a constant yields a constant $C$ that is too small in value according to our criterion comparing $C$ with $1/L$. Strong CDW correlations are however undeniable.
        \label{fig:V6_U11_cdw} }
        \end{figure}

        \begin{figure}[!htb]
        \begin{tabular}{ll}
        \includegraphics[width=0.5 \columnwidth]{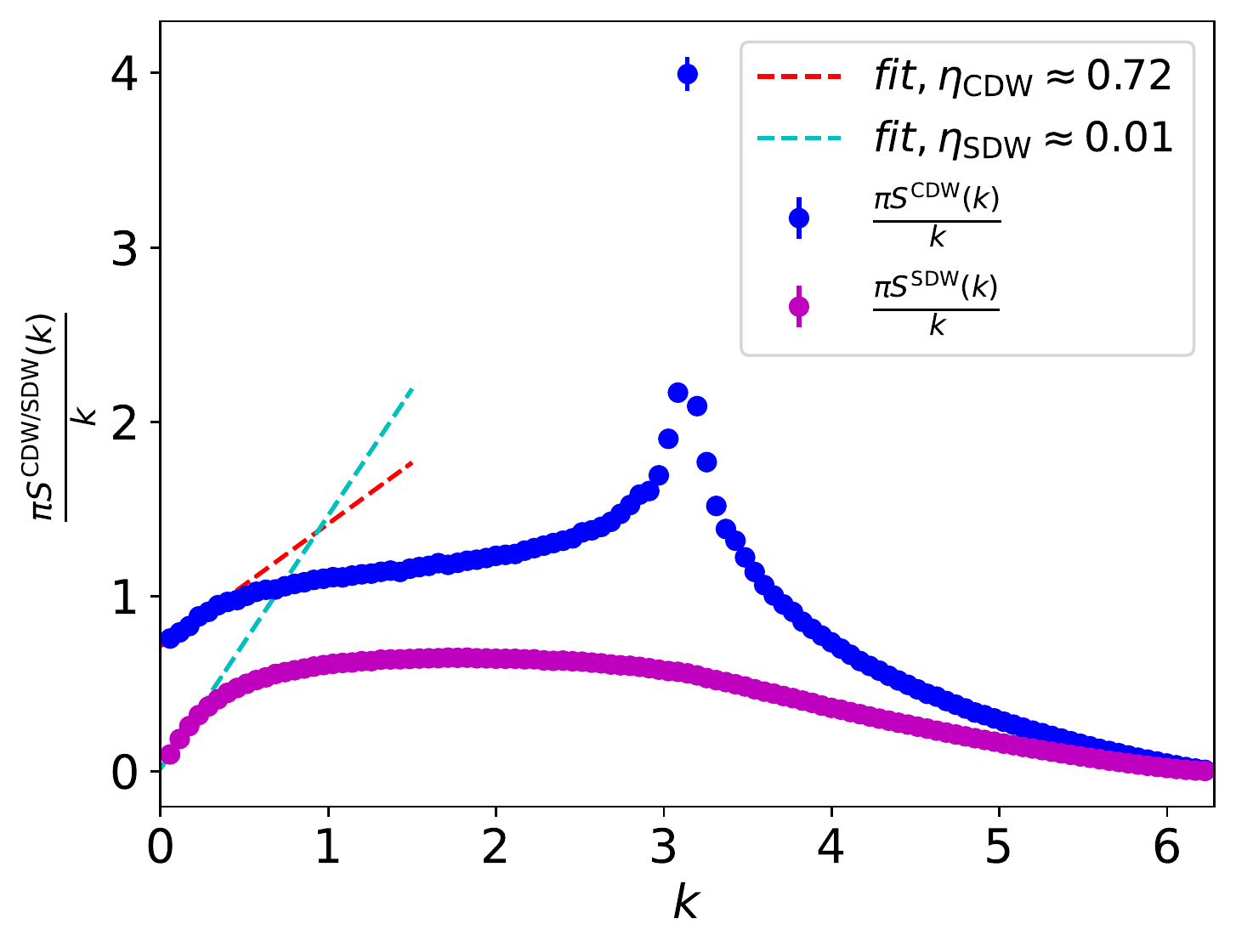} &
        \includegraphics[width=0.5 \columnwidth]{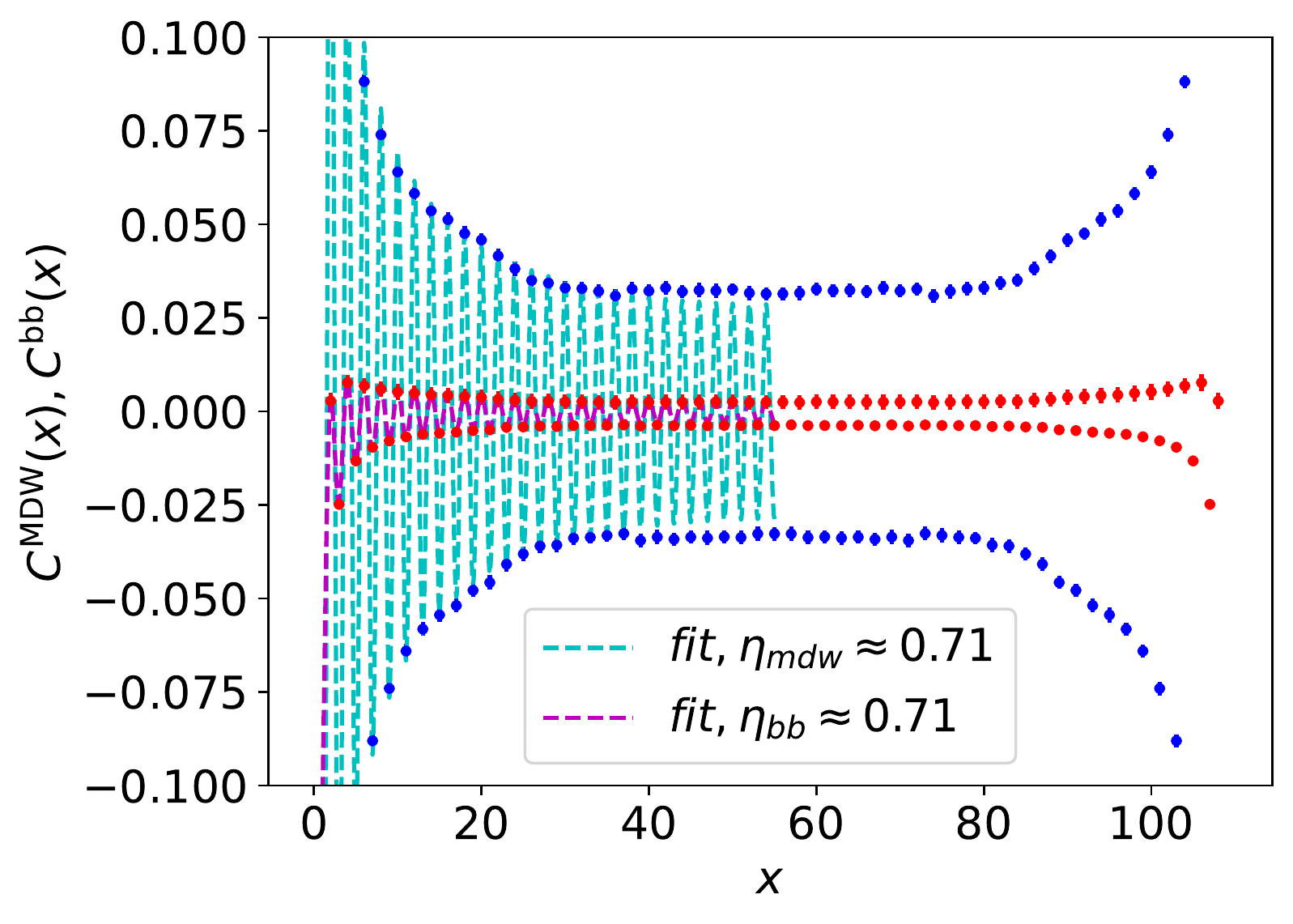}
        \end{tabular}
        \caption{Left: scaled static structure factor for the same system as in Fig.~\ref{fig:V6_U11_cdw} with $V=6, U=11, L=\beta=110$.   There appears to be a $\delta-$peak developing at $k=\pi$ which also is suggestive of lattice symmetry breaking. For this to hold, one must show that this peak ($S^{\rm CDW}(k = \pi)$) scales with $L$, and given the low values of the peak it is not possible to do so with certainty on the system sizes that we can reach (not shown). Right: The bosonic density-density ($C^{bb}(x)$) and MDW ($C^{\rm MDW}(x)$) correlation functions for the same system, showing a similar strong powerlaw, within error bars, for the the same asymptotic behavior as the CDW correlation function $(C^{\rm CDW}(x)$). 
        \label{fig:V6_U11_sf} }
        \end{figure}

For this value of $V=6$ the strong coupling analysis predicts phase separation to occur for $U \le V/0.717 \approx 8.36$. As we do not know if the strong coupling analysis applies, it could extend to higher values of $U$.

The bosonic winding number squared is plotted in the left panel of Fig.~\ref{fig:V6_boson_jump}. The flow stops for $U=9.5$ on a scale $L \sim 150$, and also for $U=10.5$ it will vanish. For $U=10$, the flow is reminiscent of the one of a critical point. The single particle density matrix $\mathcal{G}$ decays with an exponent $\eta_b \approx 1.1$ for $L=\beta=30$ and goes down very slowly with system size as is seen from $\eta_b \approx 1.0$ for $L=\beta=70$. The value of $\eta_{\rm bb}$ appears to be $1.98(2)$ for $L=70$ (see the left panel of Fig.~\ref{fig:V6_eta_cdw}). Whether any stable superflow for slightly different values of $U$ at $V=6$ can be found, is outside our resolution. We will consider $U=10$ a critical point; for the system sizes that we can simulate it behaves locally as a superfluid. 
The fermionic $Z$-factors are shown in the right panel of  Fig.~\ref{fig:V6_boson_jump}. We see a very strong renormalization towards non-Fermi liquid behavior everywhere except  for $U > 12$, which is in agreement with the argument given in Sec.~\ref{sec:strong_coupling}. For $U \le 11.5$ the low values of $Z(L) \le 0.2$ indicate that we have good chances of monitoring the entire flow on the system sizes that we can simulate. The fact that we still need to go to $L=150$ for $U=11$ and $U=11.5$ is however remarkable given that $U,V \gg t$.
The winding numbers squared in the counter-flow and pair-flow channels, shown in Fig.~\ref{fig:V6_spf_scf}, likewise show the tendency towards free fermionic behavior for $U \ge 14$. For $U = 12$ the counter-flow channel flows to zero with system size whereas the pair-flow seems to approach a nonzero value on mesoscopic length scales within error bars, indicating superconducting correlations; see the analysis in the left panel of Fig.~\ref{fig:V6_U12_FSS}. 
For smaller values of $U \le 10.5$ any superflow disappears in the thermodynamic limit.  For $U=11.5$ the signal is noisy; it indicates either thermodynamically stable superflow or strong correlations at least extending over hundreds of sites and be close to the transition. 
The induced pairing between the fermions cannot be caused by a soft bosonic mode.

We plot the quantity $\eta_{\rm CDW}$ as a function of $U$ and for $L=50, 70$ in the left panel of Fig.~\ref{fig:V6_eta_cdw} obtained from the low momentum analysis of the static CDW structure factor. In the region of the (local) bosonic superfluid, we obtain values between 1 and $1.2$ (and a fully developed spin gap), indicating a decay faster than the one of free fermions. The fermions are hence paired because of the attractive induced interactions but a picture of molecules with charge density wave order does not apply. As the bosons gap out for larger $L$ we expect $\eta_{\rm CDW}$ to diminish as well. For $U \ge 11$ the values of $\eta_{\rm CDW}$ are below 1. Whether true lattice symmetry breaking develops is impossible to say given the data, but unlikely as shown in Fig.~\ref{fig:V6_U11_cdw} and Fig.~\ref{fig:V6_U11_sf} for $U=11$, which has the lowest value of $\eta_{\rm CDW}$: It is equally well possible to fit the CDW data with a staggered constant at large distances as with a powerlaw over the cord function, but the value of this constant is so low (and lower than $1/L$ such that far bigger system sizes would be needed to answer this question). As shown in the right panel of Fig.~\ref{fig:V6_U11_sf} the $C^{\rm MDW}$ correlation function shows the same oscillating (alternating) pattern as $C^{\rm CDW}$ (and $C^{\rm bb}(x)$)  and has the largest amplitude. This strengthens the picture of a density wave consisting of alternating bosonic and fermionic molecules.

We now turn our attention to the region without superflow for $U = 8.5 - 9.5$. Care has to be taken with thermalization in this regime because in the absence of superflow properties we have no efficient updates, and the proximity of the phase separation further complicates the simulations.  The spin gap is fully developed, and the bosonic single particle density matrix decays exponentially on long enough length scales and the parameters $\eta_{\rm bb}$ and $\eta_{\rm CDW}$ are large. The behavior is similar to $V=4, U=5$ and $V=5, U = 7-8$, and, as discussed in the previous paragraph, this is interpreted as phase separation.

\subsection{Scan of phase diagram at $V=8$}

\begin{figure}[!htb]
        \begin{tabular}{ll}
        \includegraphics[width=0.5 \columnwidth]{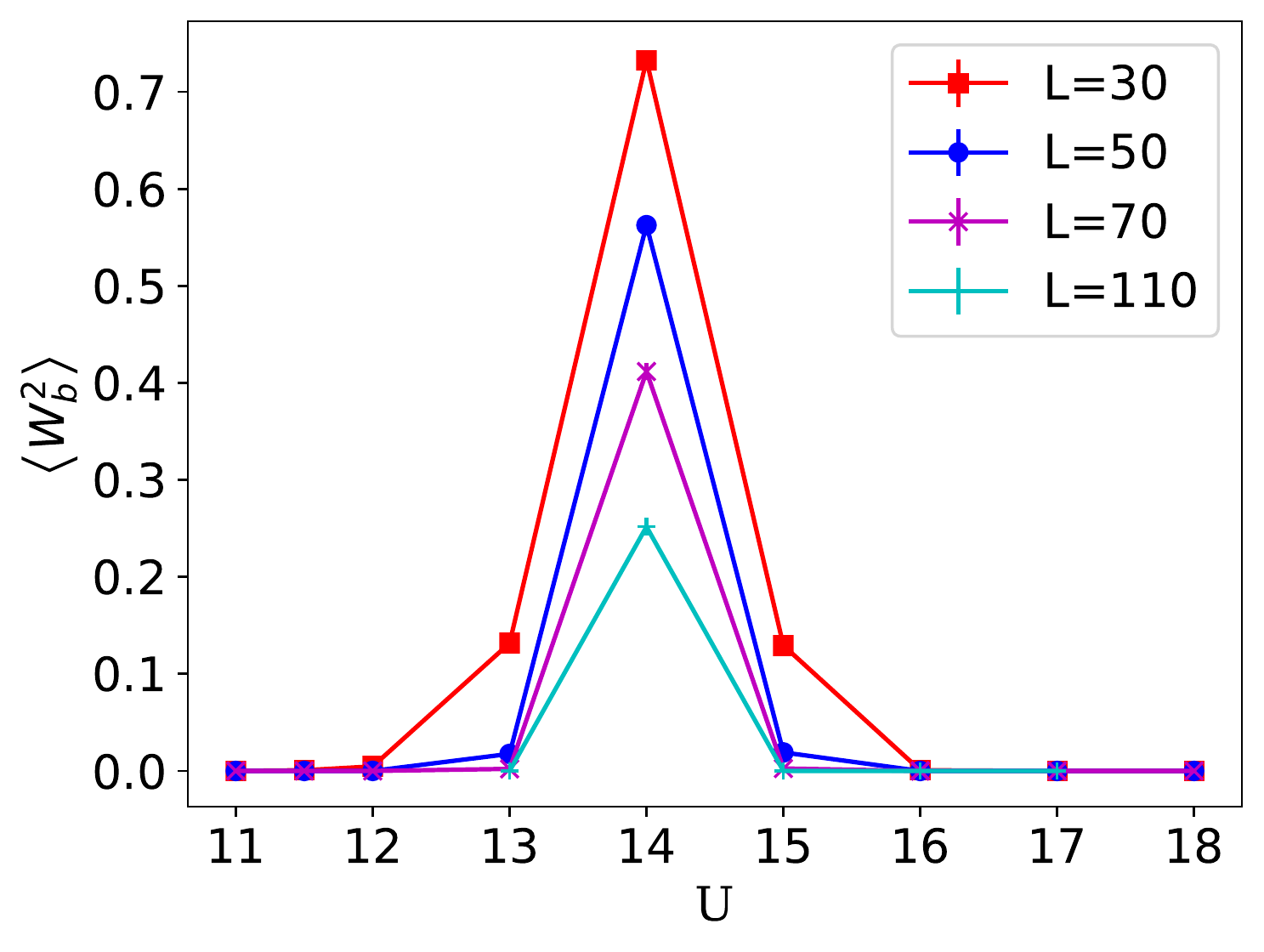} &
        \includegraphics[width=0.5 \columnwidth]{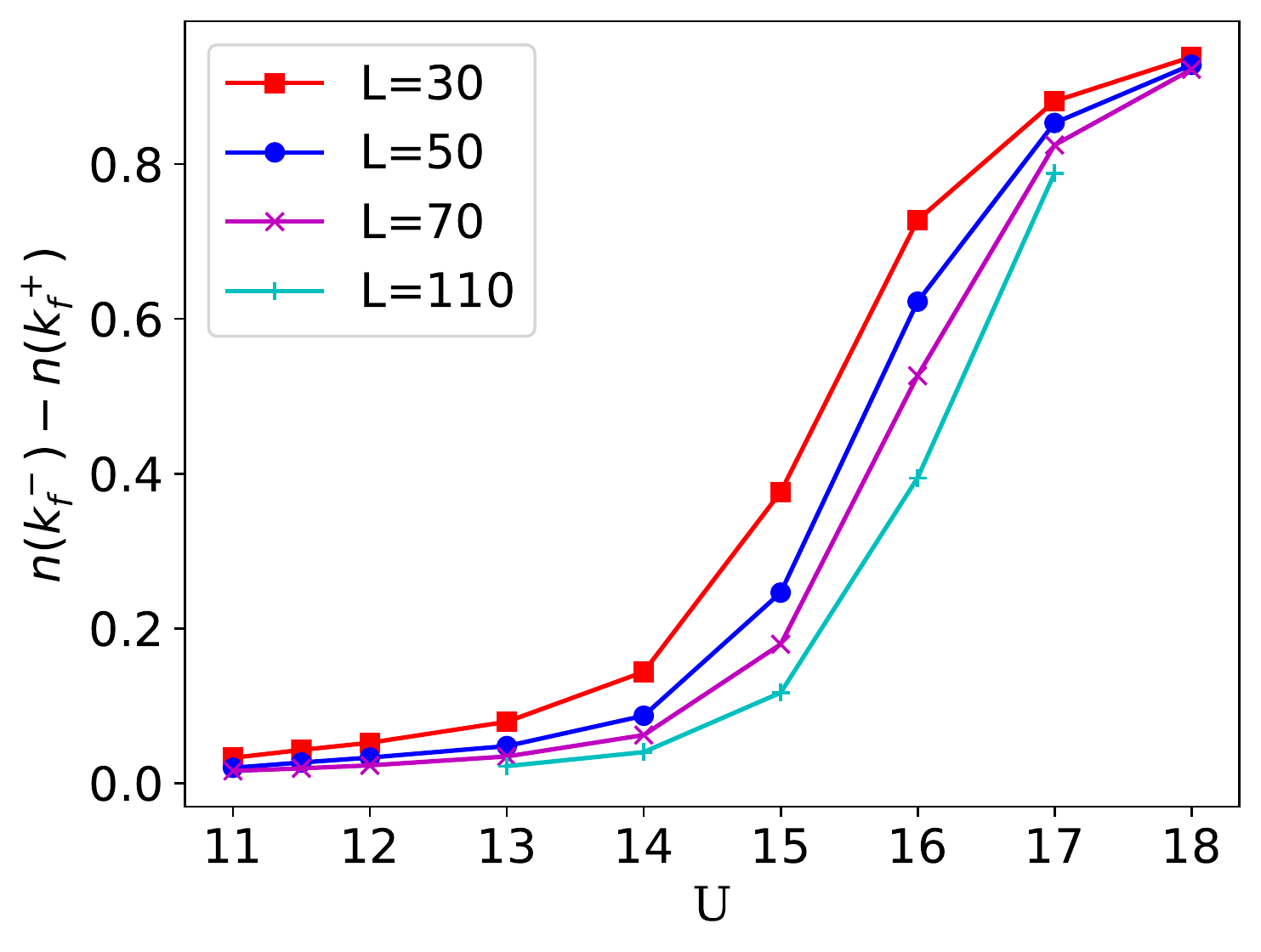}
        \end{tabular}
        \caption{Left: Bosonic winding number squared for $V= 8$.  Phase separation is seen for smaller values than $U \lesssim 12$. The system flows towards an insulator in the thermodynamic regime despite the appearance of winding number at mesoscopic length scales in the range $U \approx 14$. Right : Jumps in occupation number at $k_F$, showing a strong tendency towards free fermions for $U \ge 17$, and non-Fermi liquid behavior everywhere else.
        \label{fig:V8_boson_jump}}
\end{figure}

\begin{figure}[!htb]
        \begin{tabular}{ll}
        \includegraphics[width=0.5 \columnwidth]{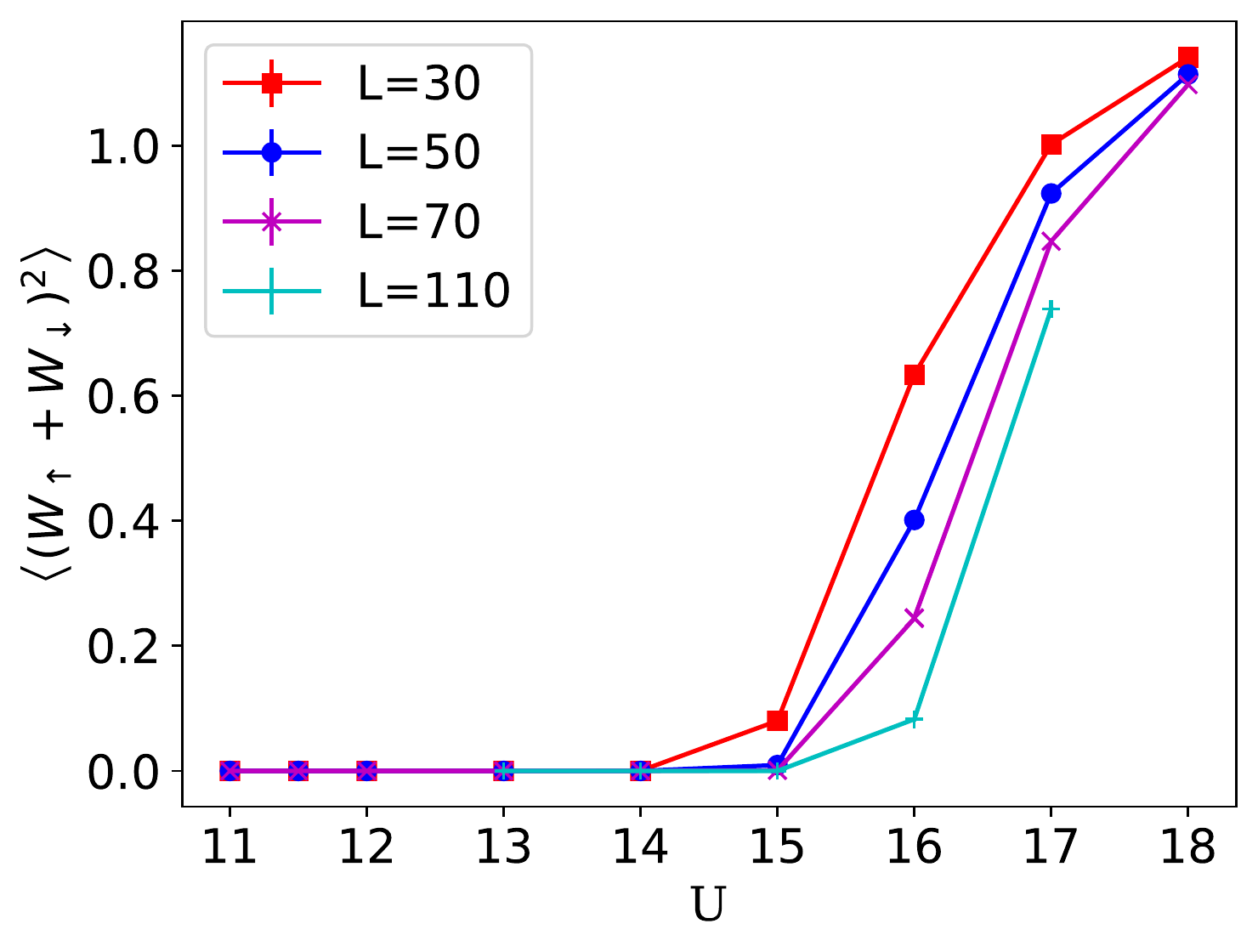} &
        \includegraphics[width=0.5 \columnwidth]{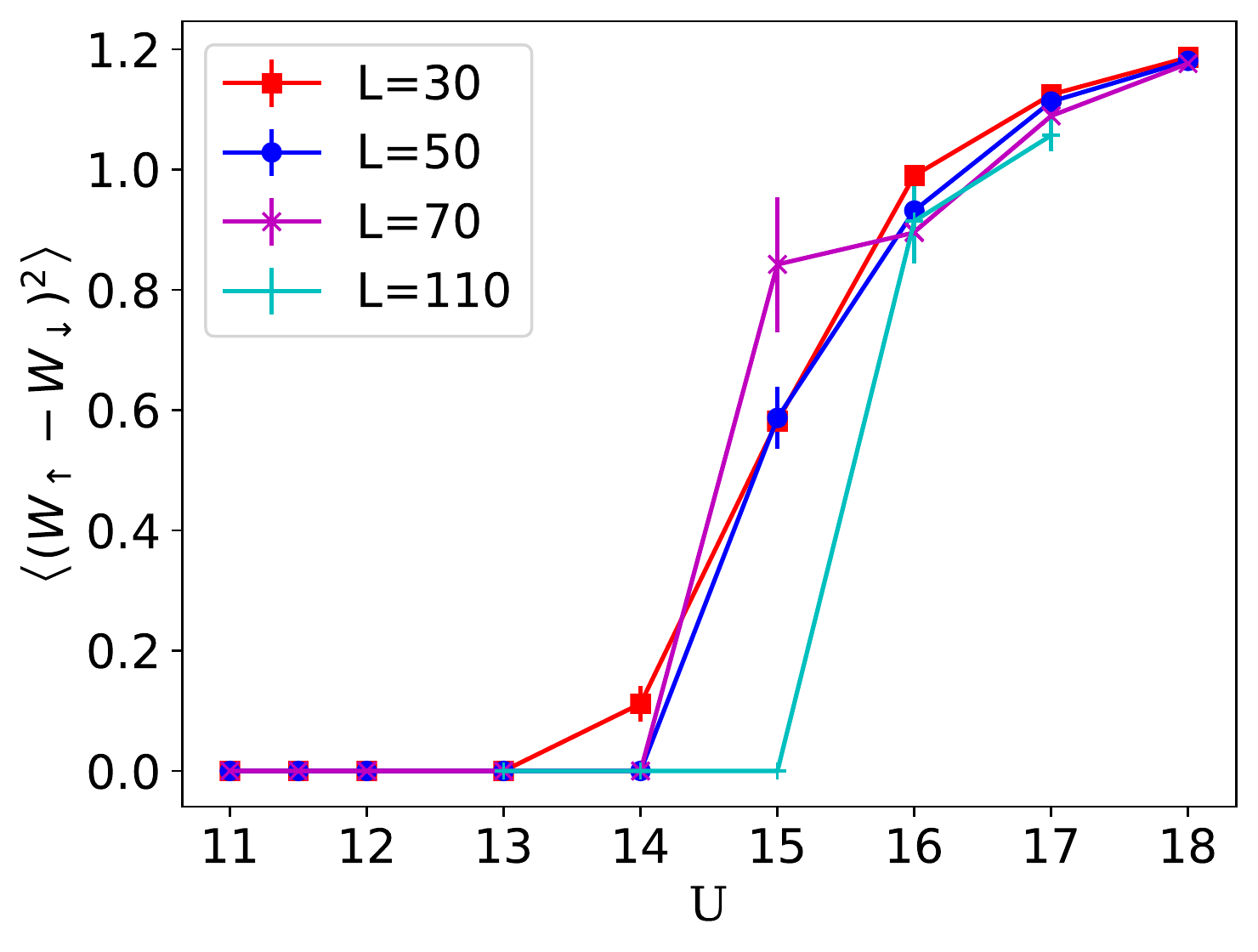}
        \end{tabular}
        \caption{Left: Winding numbers squared in the counter-flow channel for $V = 8$.  Right: Winding numbers squared in the pair-flow channel for $V = 8$. The width of the mesoscopic pair-superflow region seen around $U \approx 16$ is narrower than the corresponding region for $V=6, U=12$ and $V=5, U=10$, and situated closer to the asymptotic line $U=2V$; for $V \gtrsim 17$ the analysis of the various superflow quantities shows that the fermionic behavor is close to that of non-interacting ones on our length scales, in agreement with the right panel of Fig.~\ref{fig:V8_boson_jump}.
        \label{fig:V8_pair_scf}}
\end{figure}

\begin{figure}[!htb]
        \begin{tabular}{ll}
        \includegraphics[width=0.5 \columnwidth]{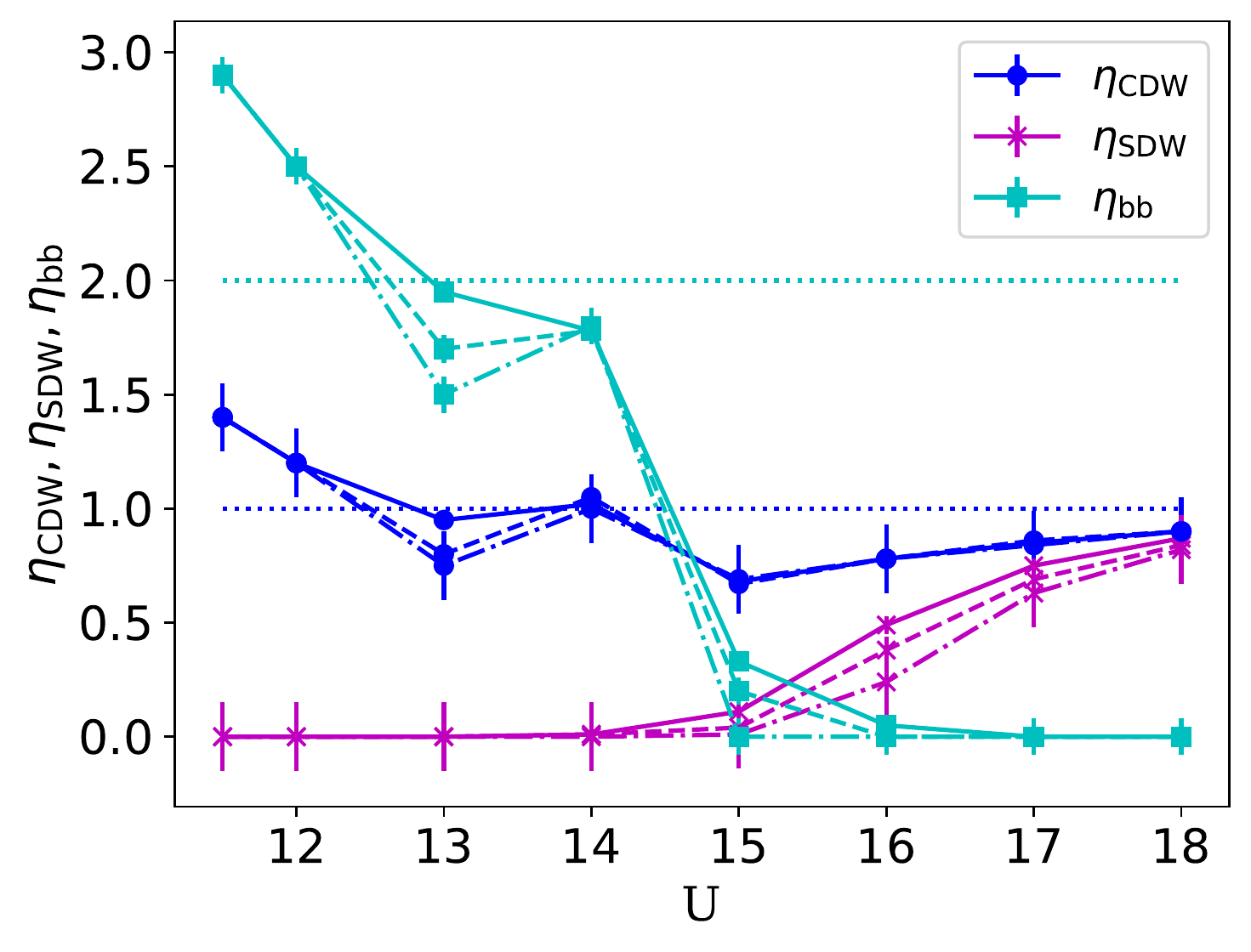} &
        \includegraphics[width=0.5 \columnwidth]{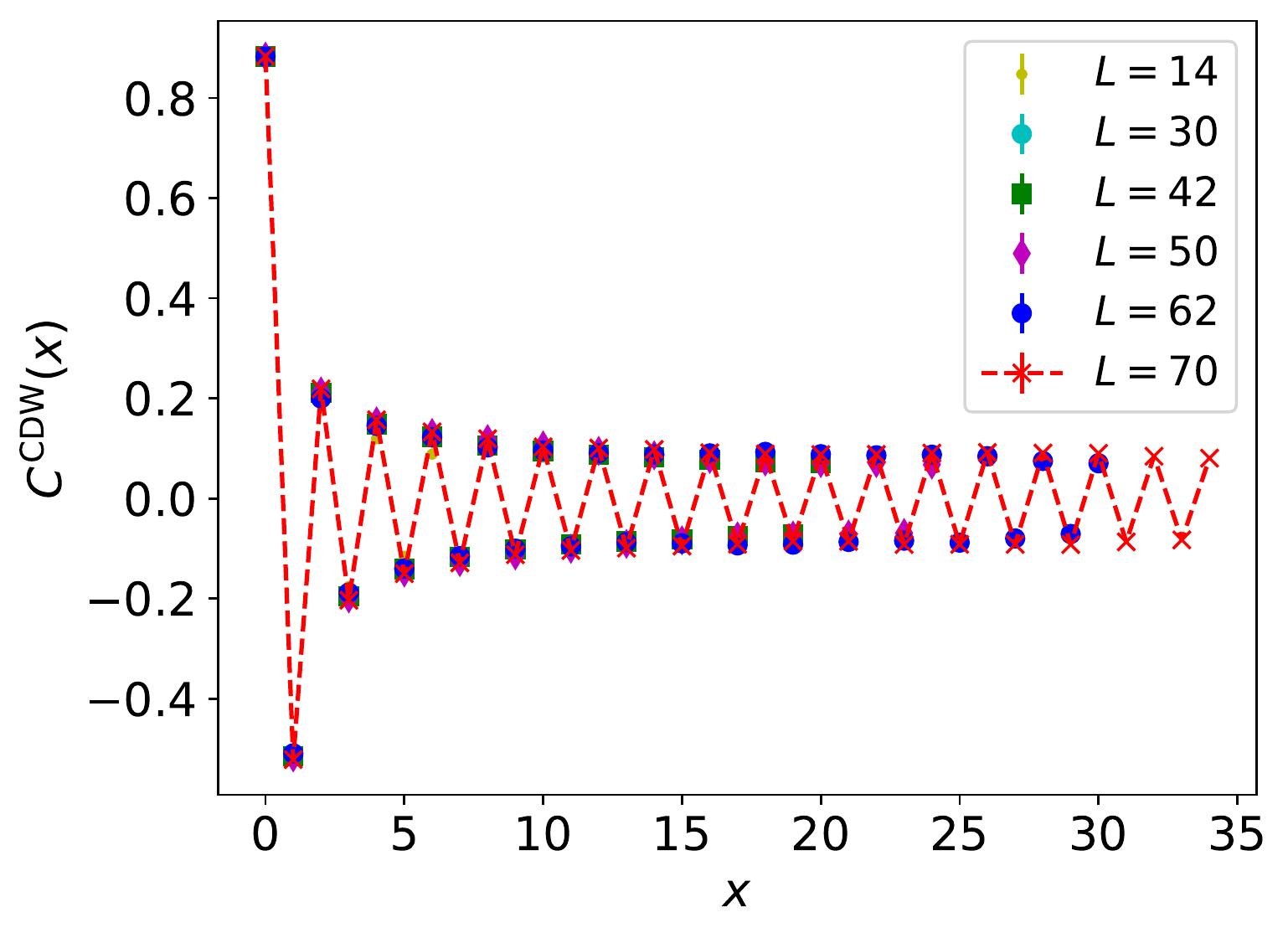}
        \end{tabular}
        \caption{Left: The quantities $\eta_{\rm CDW}$,  $\eta_{\rm SDW}$ and $\eta_{\rm bb}$ for $V = 8$ obtained from an analysis of the low momentum behavior of the CDW, SDW and bosonic static structure factors for $L = 30, 50, 70$ (top to bottom; solid, dashed and dashed dotted, respectively). The spin gap is fully developed at $L \sim 70$ for $U < 16$.  Right: The CDW correlator for $V = 8, U = 13$ for different system sizes ($L = \beta = 14, 30, 42, 50, 62, 70$) showing convergence at short distances.
                \label{fig:V8_Lutt_cdw}}
\end{figure}

\begin{figure}[!htb]
        \begin{tabular}{ll}
        \includegraphics[width=0.5 \columnwidth]{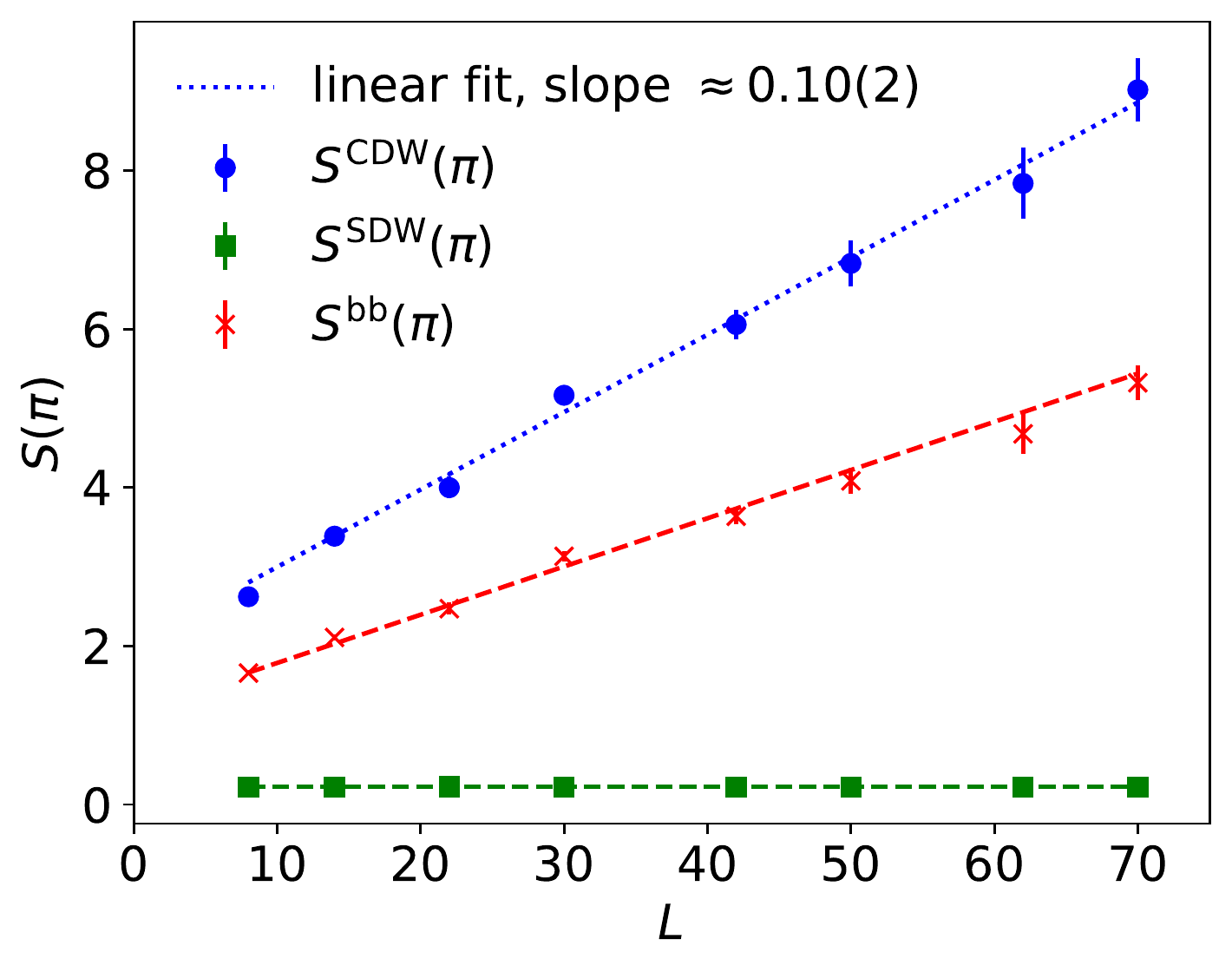} &
        \includegraphics[width=0.5 \columnwidth]{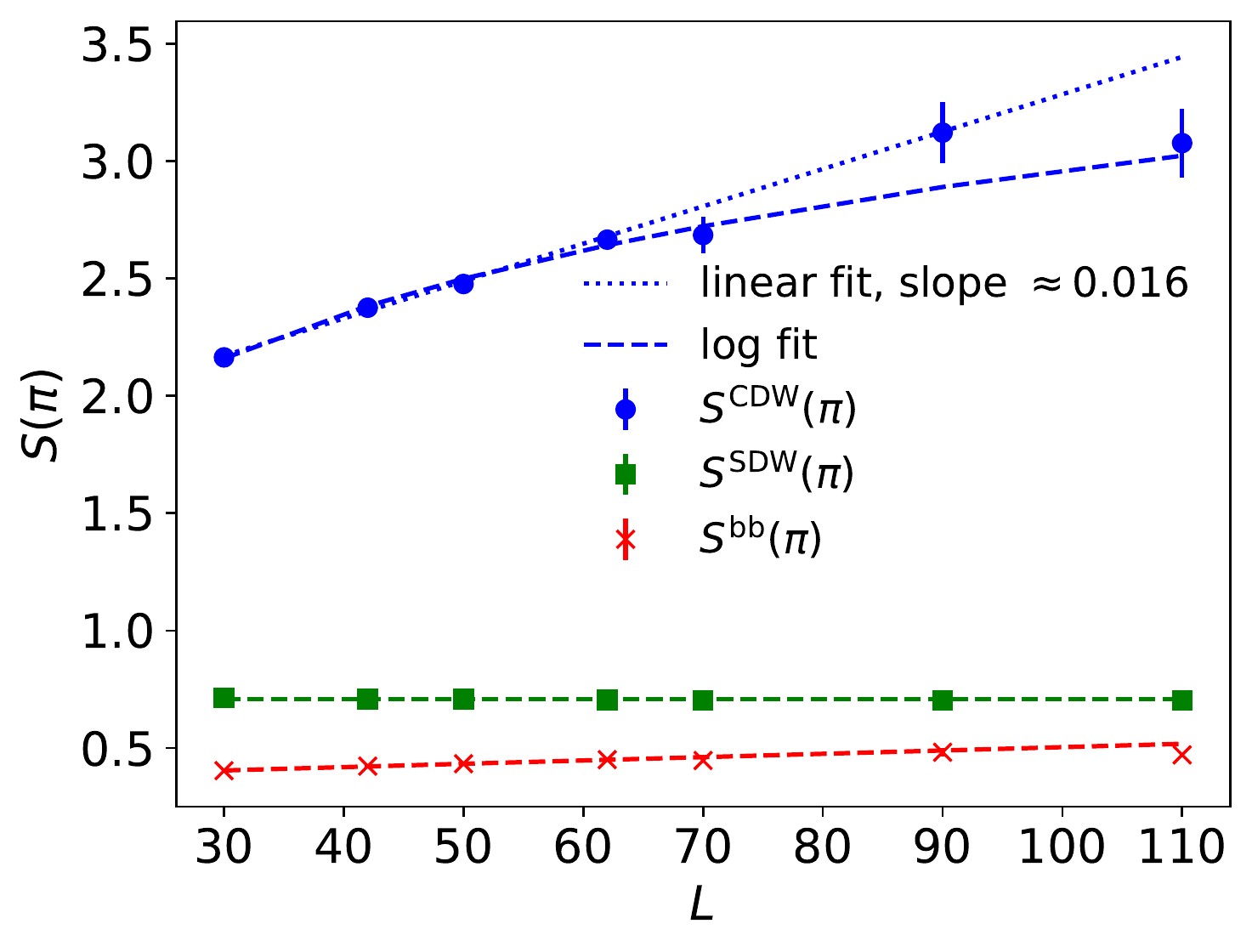}
        \end{tabular}
        \caption{Staggered structure factors for $V=8, U=13$ (left) and $V=8, U=16$ (right) in the bosonic, CDW and SDW channels. Spontaneous breaking of the lattice symmetry is seen in the linear scaling of this quantity with $L$ in the CDW and bosonic channels for $U=13$. The slope is compatible with the amplitude of the oscillations in the right panel of Fig.~\ref{fig:V8_Lutt_cdw}. For $U=16$ however, the scaling is most likely subextensive and no breaking of the lattice symmetry can be established, since the value of the slope in the linear fit is too low for our resolution and available system sizes. In such a scenario, a logarithmic fit is expected to take over at larger distances.
        \label{fig:V8_U13_sf} }
        \end{figure}

\begin{figure}[!htb]
        \begin{tabular}{ll}
        \includegraphics[width=0.5 \columnwidth]{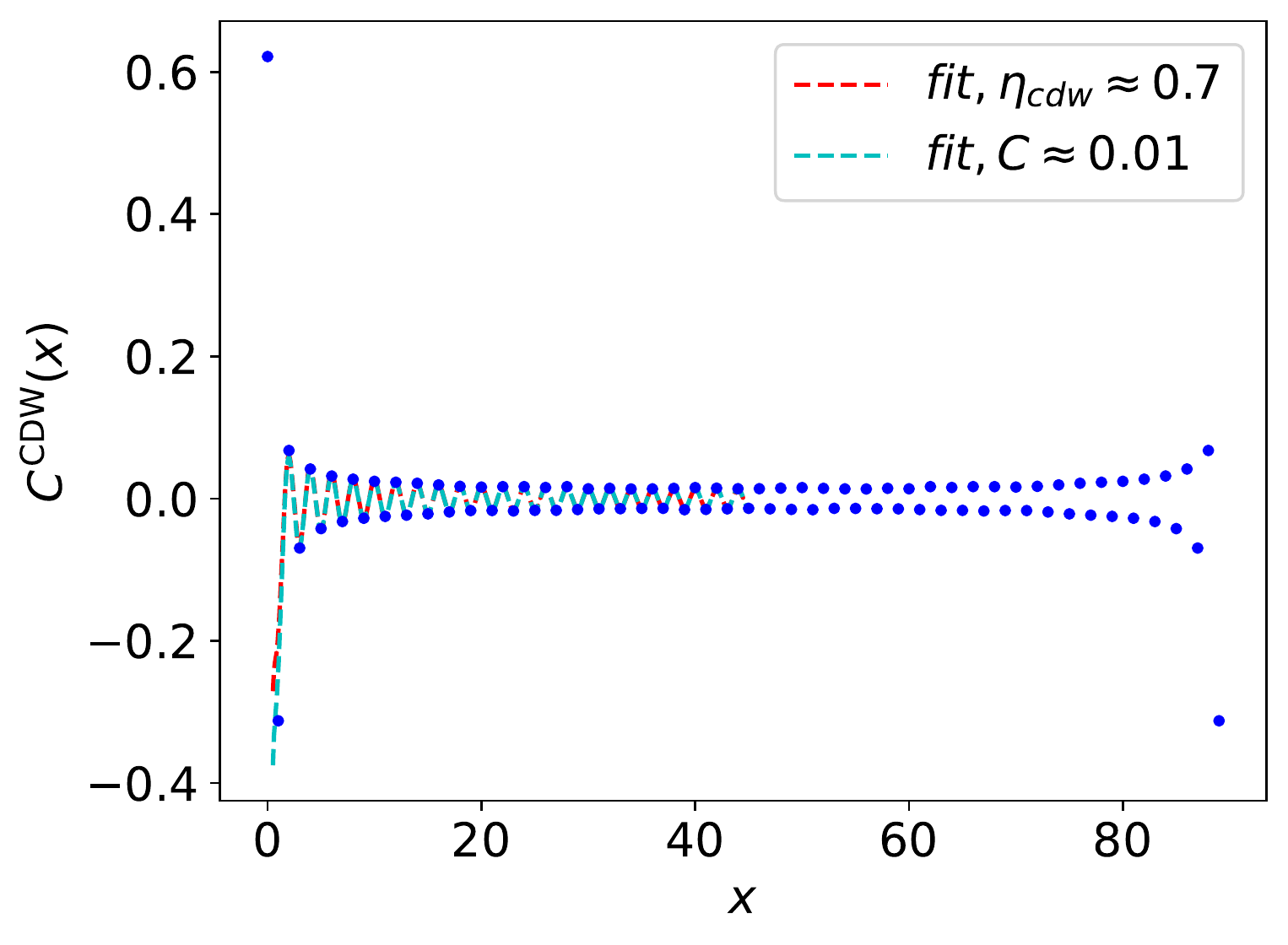} &
        \includegraphics[width=0.5 \columnwidth]{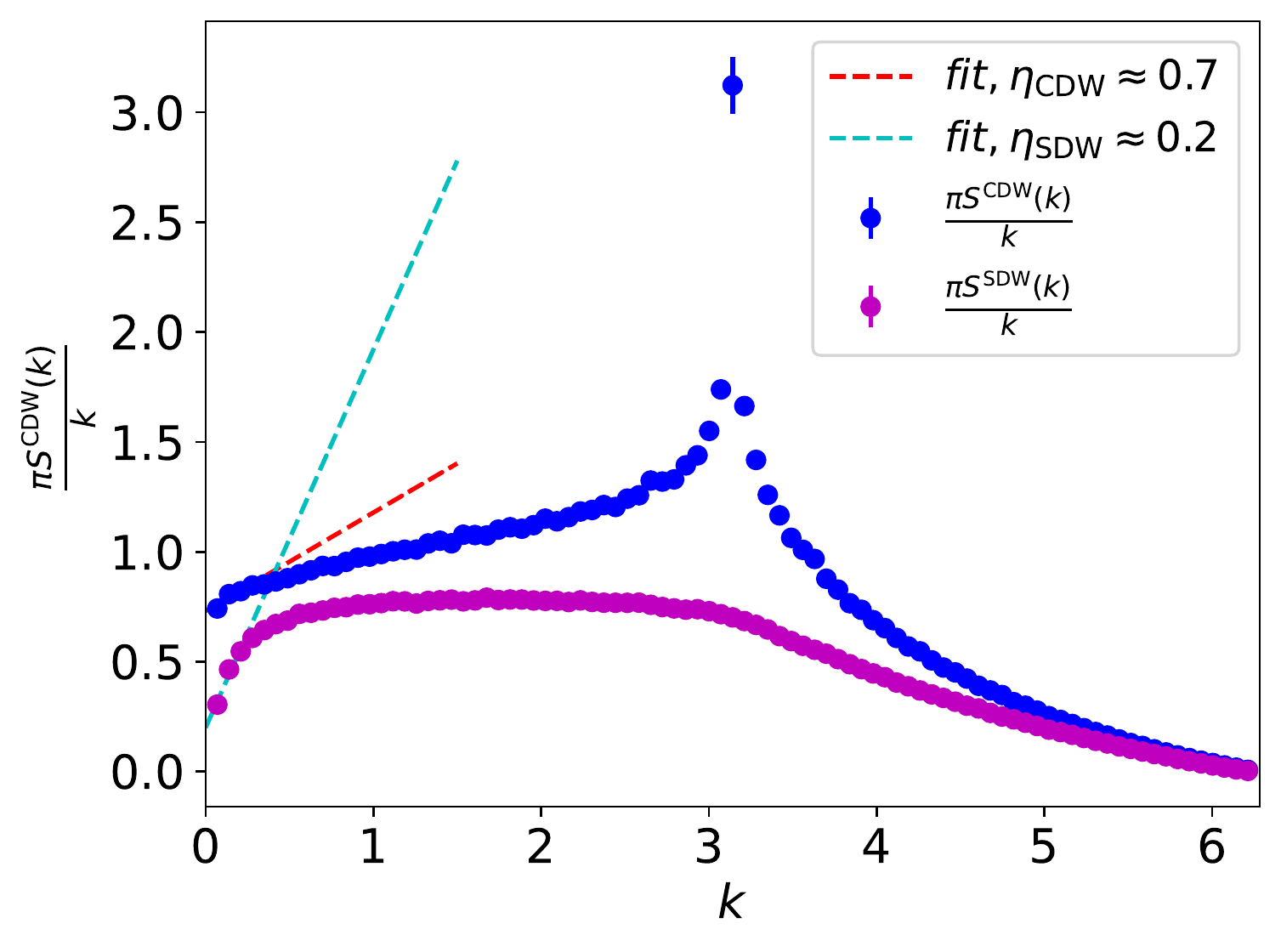}
        \end{tabular}
        \caption{Left: The CDW correlation function for $V=8, U = 16, L = \beta =90$ is asymptotically equally well fit by  a powerlaw as with a small constant alternating on even and odd sites. The value of $\eta_{\rm CDW}$ is seen to go down with $L$, starting from $\eta_{\rm CDW}(L = 30) \approx 0.81$. Error bars on $\eta_{\rm CDW}$ are dominated by systematic effects and estimated to be at least $5\%$.
        The alternative fit with a constant yields $C = 0.004(1)$, which is below our criterion value of lattice symmetry breaking. Hence, no spontaneous lattice symmetry breaking can be established, but strong CDW correlations remain undeniable.
        Right: The corresponding CDW and SDW scaled static structure factors, yielding a value of $\eta_{\rm CDW}$ in close agreement to the fit in the left panel. The peak that develops at $k = \pi$ is weak and in line with the left panel.
        \label{fig:V8_U16_cdw} }
        \end{figure}
                     
 According to the strong coupling estimates, phase separation is expected for $U \lesssim 11.15$, although, as discussed below, it extends probably to $U \approx 12$ (cf. also Fig.~\ref{fig:Lanczos_V8}).
  The bosonic superfluid properties are shown in the left panel of Fig.~\ref{fig:V8_boson_jump}. As one can expect for such strong values of $U$ and $V$ the bosons are in the thermodynamic limit always insulating. However, for $U=14$ non-zero winding numbers can be observed up to $\sim 500$ sites (after extrapolation, not shown), with the typical flow for one-dimensional systems.

In the right panel of Fig.~\ref{fig:V8_boson_jump} we see that the flow of the residue in the fermionic channel is completed for $U \le 15$. In Fig.~\ref{fig:V8_pair_scf} we observe that the pair- and counter-flow signals decay to zero with system size for $U=15$. However, very close to the $U = 2V = 16$ value, where our strong-coupling arguments favor a homogeneous system, indications of pair flow (at least on finite size systems) are seen. For $U \ge 18$ we are close to the behavior of decoupled, free fermions. In Fig.~\ref{fig:V8_Lutt_cdw} we show the dependence of $\eta_{\rm bb}, \eta_{\rm CDW},$ and $\eta_{\rm SDW}$ on $U$ for $L=30, 50, 70$ as obtained from the low momentum behavior of their respective static structure factors.
The curve for $\eta_{\rm bb}$ is non-monotonous. The most likely explanation is that for $U=14$ the mesoscopic bosonic superfluidity suppresses the density fluctuations, as we have seen for lower values of $V$.   The "fan" does not visibly open for $U=14$ although the value of $\eta_{\rm bb}$ is below 2 (because it is still rather close to 2), but we expect it to open for larger system sizes when the bosonic superfluidity is negligible. The suppression of charge fluctuations in the bosonic channel is then also reflected in $\eta_{\rm CDW}$. This parameter point ($U=14$ and $V=8$) demonstrates a recurrent theme of this paper, namely monitoring competing instabilities to more than 100 lattice sites.
 
For $U \ge 15$, we see strong indications of bosonic insulating behavior, similar to as what we have seen before in the regime $U \sim 2V$. For $U=13$ we observe a fast reduction of $\eta_{\rm bb}$ with $L$.
The spin gap is fully developed on our length scales up to the point when $U \sim 2V$, then shows a strong renormalization on the length scales that we can study for $U$-values in the range $15$ to $18$ and $\eta_{\rm SDW}$ approaches 1 from below in the quasi-free fermion regime for larger values of $U$. This is also similar to what we have seen before. 
For $U \ge 15$,  $\eta_{\rm CDW} < 1$ and it approaches 1 from below when further increasing $U$ as we approach the regime of quasi-free fermions. 

There are fingerprints of spontaneous lattice symmetry breaking for $U=13$  based on the linear scaling of the staggered static structure factors in the bosonic, CDW (and MDW) channels, as shown in Fig.~\ref{fig:V8_U13_sf}. The slope in the left panel of Fig.~\ref{fig:V8_U13_sf}) for the CDW channel is consistent with the value $C = 0.10(2)$ seen in the right panel of Fig.~\ref{fig:V8_Lutt_cdw}.
Although the bosonic structure factor also scales extensively with $L$, its slope is less steep than for the fermions. The spin density wave structure factors are obviously insensitive to such charge order.  Given the absence of any superflow, this data point would then be in an insulating phase characterized with alternating bosonic molecular pairs and fermionic molecular pairs, and is predicted by the strong coupling arguments of Sec.~\ref{sec:strong_coupling}. Nevertheless, the observed order is weak, and a final answer would require bigger system sizes. We were not able to let the simulations converge for those however.
For $U=16$, where the right panel of Fig.~\ref{fig:V6_U12_FSS} is suggestive of stable pair flow. In Fig.~\ref{fig:V8_U16_sf} the data for the CDW structure factor can equally well be fitted by a linear line as with a logarithm. Given the very low value of the linear slope, an interpretation of the data in terms of the absence of lattice symmetry breaking (and thus the logarithmic fit) is more likely however.  This can also be understood from the strong coupling arguments of Sec.~\ref{sec:strong_coupling}: For $U > 2V$ we expect a uniform $n=1$ bosonic Mott insulator, and $U=2V$ is the critical point. The strong coupling arguments have hence quantitative predictive power for the upper boundary at $V=8$.
In between $U=13$ and $U=16$ (not shown) the staggered structure factors are seen to diminish with increasing $U$, and one can not be sure about spontaneous lattice symmetry breaking.  For $\eta_{\rm CDW}$,  $U=14$ has a higher value than $U=13$, which we attributed to the mesoscopic bosonic superflow. For $U=15$ we find however the lowest $\eta_{\rm CDW}$.

The data in the regime $U=11.5 -12$ is again contradictory and misleading. The fermionic residues indicate that the absence of quasi-particles degrees of freedom, and the Luttinger parameters indicate a Luttinger liquid with $\eta_{\rm CDW} > 1$. The bosons are non-superfluid but have $\eta_{\rm bb} > 2$. As with similar conflicting results for $V=5$ and $V=6$, we think that these values of $U$ are  in the phase separation regime. The strong coupling arguments from Sec.~\ref{sec:strong_coupling} are for $V=8$ not quantitatively precise for the lower boundary, although the black dashed and cyan dotted lines will merge for somewhat larger values of $V$ in the phase diagram  Fig.~\ref{fig:phasediagram}: Lanczos results for very small system sizes and $V=20$ show very good agreement with the strong coupling predictions (cf Figs.~\ref{fig:Lanczos_V20},\ref{fig:Lanczos_V8}, and ~\ref{fig:Lanczos_V12}).

\section{Discussion and Conclusion} \label{sec:conclusion}
The simulation of multi-component systems, weakly induced interactions, competing instabilities at work on hundreds of lattice sites, and long autocorrelation times present a frontier in current computational physics. The system under investigation, a one-dimensional mixture of scalar bosons and $S=1/2$ fermions defined on the lattice and subject to Hubbard interactions, is perhaps the simplest model where such physics can be explored. 
We employed path-integral Monte Carlo simulations with worm-type updates to simulate this system. We have introduced worm-type operators for all single-particle Green functions, but restricted the two-worm operators to the particle-particle and particle-hole channels of the spin-up and spin-down fermions. 
Although the algorithm is efficient almost everywhere in the non-gapped part of phase diagram, there are cases where the autocorrelation times explode for reasons ill understood.
Despite the numerical challenges, the phase diagram shown in Fig.~\ref{fig:phasediagram} is brought under reasonable control. It is straightforward to understand that for $V \gg U$ the system phase separates, and that for $U \gg V$ the bosons impose a uniform $n=1$ Mott insulator on top of which the fermions remain quasi-free. In between, the bosons induce attractive interactions between the fermions, which must lead to a spin gap. We have systematically monitored the development of the spin gap and the fermionic residue as a function of $V, U$ and $L$, and used it as an indication of how close we are to the thermodynamic limit. The induced interactions can lead to phase separation, pair flow and/or molecular density wave structures. For not too high values of $V$, the bosons are found to be uniformly superfluid in between the phase separation and Mott insulator phases.  Bosonic superfluidity is the dominant energy scale, and this in turn suppresses fermionic charge fluctuations. The bosonic superfluid is most likely adjacent to the regime of phase separation by a first order transition. The Mott insulator is reached by a Kosterlitz-Thouless transition when $\eta_{\rm bb} = 2$ (this is the exponent governing the decay of the bosonic density-density correlation function), and this implies that $\eta_{\rm b}$ (the exponent governing the power law decay of the off-diagonal single particle, equal time bosonic density matrix is less than 2 at and close to the Mott transition), in line with the predictions from bosonization~\cite{MatheyWang2007}. We referred to $\eta_{b} < 2$ as a marginal superfluid (see Fig.~\ref{fig:wind_marginal_sf}). 
We find a bosonic superfluid (almost) up to $V=6, U = 10$, and even for $V=8, U=14$ there is mesoscopic bosonic superflow up to several hundreds of lattice sites. Superflow for such large values $U, V \gg t$  are a clear signature of the competition in the system: it must be that there is a delicate parameter regime where induced interactions  between the bosons mediated by the fermions lower the value of $U$ and prefer a homogeneous system. At least up to the system sizes that we can simulate (note that the "Cooper" pair size exceeds hundreds of lattice sites even for $V=2$), the signal in the counter and pair flow channel hints at pair flow for the fermions for low values of $V < 4$, with $\eta_{\rm CDW} > 1$ as long as the bosons are superfluid. For Mott-insulating bosons, we immediately arrive at $\eta_{\rm CDW} < 1$ throughout the phase diagram.
For $V=4$ and $V=5$ we observe a transition between pair flow and insulating fermionic behavior as a function of $U$ that is very hard to simulate and therefore hard to determine precisely. The  quantity $\eta_{\rm CDW}$ is not strongly affected by this. We expect for very large values of $V,U \gg t$ that over the range $ V / 0.717 < U < 2V$ a molecular charge density wave alternating with bosonic molecular pairs is formed (see Sec.~\ref{sec:strong_coupling}). We see some evidence for that for $V=8, U=13$ as shown in Fig.~\ref{fig:V8_U13_sf}, but increasing $U$ further leads to a reduction in the strength of the charge order correlations.

\begin{acknowledgements}
We thank Ludwig Mathey and Daw-Wei Wang for insightful conversations on their bosonization results.
We acknowledge support from FP7/ERC Consolidator Grant QSIMCORR, No. 771891, and the Deutsche Forschungsgemeinschaft (DFG, German Research Foundation) under Germany's Excellence Strategy -- EXC-2111 -- 390814868.  Numerical data for this paper is available under \url{https://github.com/LodePollet/QSIMCORR}.
Our simulations make use of the ALPS\-Core library~\cite{Gaenko17} for error evaluation.
 
\end{acknowledgements}


\bibliography{BFH.bib}

\end{document}